\documentclass[aps,prd,showpacs,floatfix,
onecolumn,
superscriptaddress,nofootinbib,preprintnumbers, 10pt, notitlepage]{revtex4-1} 

\makeatletter
\def\p@subsection{}
\makeatother

\usepackage{graphicx,amsmath,amssymb,lipsum,bm}

\usepackage[dvipsnames]{xcolor}
\definecolor{xlinkcolor}{rgb}{0.7752941176470588, 0.22078431372549023, 0.2262745098039215}

\usepackage{booktabs} 
\usepackage[normalem]{ulem}
\usepackage[utf8]{inputenc} 
\usepackage{mathtools}
\usepackage{aas_macros}
\usepackage{xspace}

\usepackage{ulem}
\usepackage{diagbox}

\usepackage[dvipsnames]{xcolor}
\usepackage{mathrsfs}

\usepackage[colorlinks=true,citecolor=xlinkcolor,linkcolor=xlinkcolor,urlcolor=xlinkcolor, backref=false,pdfborder={0 0 0}]{hyperref}

\newcommand{\be}{\begin{equation}}
\newcommand{\ee}{\end{equation}}
\newcommand{\beqa}{\begin{eqnarray}}
\newcommand{\eeqa}{\end{eqnarray}}

\newcommand{\md}{\mathrm{d}}
\newcommand\p{{\bm p}}
\renewcommand\k{{\bm k}}
\newcommand\q{\bm{q}}

\newcommand\G{\mathcal{G}_2}

\newcommand{\HH}{{\cal H}}

\newcommand{\MYhref}[3][black]{\href{#2}{\color{#1}{#3}}}%
\usepackage{xspace}
\newcommand{\cobra}{\MYhref{https://www.flickr.com/photos/198816819@N07/54686938553/in/dateposted-public/}{\textsc{cobra}}\xspace}

\def\d{\partial}
\newcommand{\bseq}{\begin{subequations}}
\newcommand{\eseq}{\end{subequations}}

\renewcommand{\ln}{\mathop{\rm ln}\nolimits}

\def\ltsima{$\; \buildrel < \over \sim \;$\xspace}
\def\gtsima{$\; \buildrel > \over \sim \;$\xspace}
\def\simlt{\lower.5ex\hbox{\ltsima}}
\def\simgt{\lower.5ex\hbox{\gtsima}}

\newcommand{\Mpch}{\, h^{-1}\mathrm{Mpc}\, }
\newcommand{\hMpc}{\, h\mathrm{Mpc}^{-1}\, }

\newcommand{\kmax}{\, k_{\rm max}\, }

\newcommand{\hn}{\hat{\bm n}}

\newcommand{\hk}{\hat{\bm k}}
\newcommand{\hq}{\hat{\bm q}}

\newcommand{\vq}{\bm q}
\newcommand{\vx}{\bm x}

\def\gsim{\raise0.3ex\hbox{$\;>$\kern-0.75em\raise-1.1ex\hbox{$\sim\;$}}}
\def\lsim{\raise0.3ex\hbox{$\;<$\kern-0.75em\raise-1.1ex\hbox{$\sim\;$}}}

\def\beqn#1{\begin{equation}\label{#1}}
\def\eeqn{\end{equation}}

\def\beqa#1{\begin{eqnarray}\label{#1}}
\def\eeqa{\end{eqnarray}}

\def\kmax{{k_\text{max}}}
\def\hMpc{h{\text{Mpc}}^{-1}}
\def\Mpch{h^{-1}{\text{Mpc}}}

\def\Z2{$\mathcal{Z_2}$}

\def\beq{\begin{eqnarray}}
\def\eeq{\end{eqnarray}}
\let\vec\mathbf

\newcommand {\ignore}[1]{}

\begin{document}

\preprint{MIT-CTP/5891}
\preprint{RBI-ThPhys-2025-23}

\title{One-Loop Galaxy Bispectrum:
Consistent Theory, 
Efficient Analysis with COBRA,
and Implications for Cosmological Parameters
}


\author{Thomas Bakx}
\email{t.j.m.bakx@uu.nl}
\affiliation{Institute for Theoretical Physics, Utrecht University, Utrecht, Princetonplein 5, NL}
\author{Mikhail M. Ivanov}
\email{ivanov99@mit.edu}
\affiliation{Center for Theoretical Physics -- a Leinweber Institute, Massachusetts Institute of Technology, 
Cambridge, MA 02139, USA}
\affiliation{The NSF AI Institute for Artificial Intelligence and Fundamental Interactions, Cambridge, MA 02139, USA}
\author{Oliver~H.\,E.~Philcox}
\email{ohep2@cantab.ac.uk}
\affiliation{Simons Society of Fellows, Simons Foundation, New York, NY 10010, USA}
\affiliation{Center for Theoretical Physics, Columbia University, New York, NY 10027, USA}
\affiliation{Leinweber Institute for Theoretical Physics at Stanford, 382 Via Pueblo, Stanford, CA 94305, USA}
\affiliation{Kavli Institute for Particle Astrophysics and Cosmology, 382 Via Pueblo, Stanford, CA 94305, USA}
\author{Zvonimir Vlah}
\email{zvlah@irb.hr}
\affiliation{Division of Theoretical Physics, Ruder Bošković Institute, 10000 Zagreb, Croatia}
\affiliation{Kavli Institute for Cosmology, University of Cambridge, Cambridge CB3 0HA, UK}
\affiliation{Department of Applied Mathematics and Theoretical Physics, University of Cambridge, Cambridge CB3 0WA, UK}

\begin{abstract} 
\noindent We present an efficient and accurate pipeline for the analysis of the redshift-space galaxy bispectrum multipoles at one-loop order in effective field theory (EFT). 
We provide a systematic theory derivation based on power counting, which features the first comprehensive treatment of stochastic EFT contributions -- these are found to  
significantly improve the match to data.
Our computational pipeline utilizes the \cobra technique that expands the linear matter power spectrum over a basis of principal components based on a singular value decomposition, allowing the cosmology dependence to be captured to sub-permille accuracy with just eight templates. This transforms the problem of computing the one-loop EFT bispectrum to a simple tensor multiplication, reducing the computation time to around a second per cosmology with negligible loss of accuracy. Using these tools, we study the cosmological information in the bispectrum by analyzing PTChallenge simulations, whose gigantic volume provides the most powerful test of the one-loop
EFT bispectrum so far. We find that the one-loop prediction provides an excellent match to the bispectrum data up to $k_{\rm max}=0.15\,h\mathrm{Mpc}^{-1}$, as evidenced by the precise recovery of the dark matter density $\omega_\text{cdm}$, Hubble constant $H_0$, and mass fluctuation amplitude 
$\sigma_8$ parameters, and the amplitude of equilateral primordial non-Gaussianity (PNG) $f_{\rm NL}^{\rm equil}$. 
Combined with the power spectrum, the \cobra-based one-loop bispectrum monopole and quadrupole yield tighter constraints than the tree-level bispectrum monopole, with the posteriors on  
$\omega_{\text{cdm}}$, $H_0$, and $\sigma_8$ shrinking by
44\%, 32\%, and 25\%, respectively.
We also find that the monopole
and quadrupole moments capture most of the cosmological 
information in
the redshift space bispectrum
in the $\Lambda$CDM model.
Our results suggest that the \cobra-based bispectrum analysis will be an important tool in the interpretation of data from ongoing redshift surveys such as DESI and Euclid. 
\end{abstract}

\maketitle

\section{Introduction}
\noindent The large-scale distribution of matter in the Universe has emerged as a powerful probe of the nature of dark matter and dark energy, the expansion history of the Universe, and the physics underlying its initial conditions. Current galaxy surveys such as Euclid \cite{Laureijs:2011gra} and DESI \cite{Aghamousa:2016zmz} (as well as proposed future experiments such as Spec-S5 \citep{Spec-S5:2025uom}) will collect millions of spectra of extragalactic objects such as galaxies and quasars, which trace the underlying dark matter distribution and contain a wealth of valuable cosmological information. 

The distribution of these tracers (which we refer to as galaxies from now on) is non-Gaussian due to non-linear gravitational interactions that govern the dynamics of the dark matter and baryons. However, on the typical scales probed by the aforementioned surveys these nonlinearities remain small and can be treated order-by-order in perturbation theory \cite[e.g.,][]{Scoccimarro:1995if,Bernardeau:2001qr}. Most modern prescriptions used to model the large-scale distribution of galaxies are based on the effective field theory of large-scale structure (hereafter EFT) \cite{Baumann:2010tm,Carrasco:2012cv} and its various formulations~\cite{Vlah:2015sea,Vlah:2015zda,Vlah:2018ygt,Senatore:2014vja,Lewandowski:2015ziq,Blas:2015qsi,Blas:2016sfa,Ivanov:2018gjr,Senatore:2014eva,Perko:2016puo,Desjacques:2016bnm} (see~\cite{Ivanov:2022mrd} for a review), which consistently accounts for all possible dependencies of the observed galaxy density on the dark matter distribution via a finite set of unknown EFT (Wilson) coefficients. These can be marginalized over when fitting to data, and allow utilization of the full scale-dependence of the galaxy distribution. 

Owing to the aforementioned non-Gaussianity, the information content of galaxy surveys is not fully captured by two-point correlations of the galaxy density field, implying that higher-order correlations should also be analyzed. It is therefore natural to consider the three-point correlation function, which in Fourier space is described by the bispectrum \cite[e.g.,][]{Scoccimarro:1999ed,Scoccimarro:2000sn,Sefusatti:2006pa}. Indeed, many recent works have focused on extracting information from the galaxy bispectrum in a robust fashion, both alone and in joint analyses with the galaxy power spectrum \cite[e.g.,][]{Ivanov:2021kcd,Ivanov:2023qzb,Gil-Marin:2014sta,Gil-Marin:2016wya,DAmico:2019fhj,Oddo:2019run,Oddo:2021iwq,Rizzo:2022lmh,Tsedrik:2022cri,Cabass:2024wob,Cabass:2022wjy,Cabass:2022ymb,Chen:2024vuf,Philcox:2022frc,NovellMasot:2025fju,DAmico:2022osl}. In most cases, these works have restricted themselves to the leading order (tree-level) bispectrum. The computation of the one-loop bispectrum is a formidable task; it requires a fourth-order bias expansion in the density~\cite{Eggemeier:2018qae,Eggemeier:2021cam} and (to account for redshift-space distortions) velocity fields~\cite{Philcox:2022frc,DAmico:2022ukl,DAmico:2022osl}. Crucially, this fourth-order expansion must be supplemented by all possible counterterms and higher-derivative corrections that appear at this order in perturbation theory and are consistent with Galilean invariance,  equivalence principle, and EFT power counting rules.

Moreover, as with the power spectrum, going beyond leading order in perturbative calculations requires computing loop integrals that capture non-linear mode-coupling. This process is computationally demanding; thus, innovative methods are essential to make cosmological inference feasible. In the case of the one-loop galaxy power spectrum, discrete Mellin transforms (also known as FFTLog~\cite{Hamilton:1999uv}) have been established as an efficient way to incorporate loop corrections~\cite{Schmittfull:2016jsw,McEwen:2016fjn,Simonovic:2017mhp}. These techniques are now implemented in most standard codes used to compute the one-loop galaxy power spectrum~\cite{Chudaykin:2020aoj,DAmico:2020kxu,Chen:2020fxs}. In contrast, incorporating next-to-leading order (one-loop) corrections in the bispectrum has proven to be significantly more challenging. 

Specifically, FFTLog-based methods have turned out to be too computationally intensive for efficient cosmological inference with the one-loop bispectrum.
The difficulty arises from the fact that the linear power spectrum enters the one-loop bispectrum via loop integrals, which introduces a steep scaling with the number of $k$-dependent basis elements (\textit{i.e.}\ FFTLog frequencies) used to decompose the power spectrum. As a result, the one-loop bispectrum calculation becomes highly sensitive to the number of basis elements, with computational cost scaling as $N_{\rm basis}^3$, with $N_{\rm basis}=\mathcal{O}(100)$ for FFTLog (required to accurately capture the 
shape of baryon acoustic 
oscillations).
For this reason,~\cite{Philcox:2022frc} limited their analysis to variations in the amplitude parameter $\sigma_8$, which affects the loop integrals through an overall rescaling. 

Alternatively,
\cite{Anastasiou:2022udy} 
considered the approach 
of removing the BAO wiggles 
from the spectrum and 
approximating the 
`non-wiggly" part of the linear power
spectrum over a small set of inverse polynomial
functions denoted in \cite{Anastasiou:2022udy} as `massive propagators.' This approach, just like FFTLog, permits analytical loop evaluation with a reduced number of
required
basis functions. However, because this method is tailored to enable specific analytic solutions, it performs suboptimally when aiming to 
accurately 
represent the linear power spectrum with a small number of basis elements, especially when accuracy on both BAO and broadband is required. 
While this may be satisfactory for datasets with 
relatively large statistical errors, 
the precision targets
of ongoing and upcoming large-scale structure surveys 
demand new powerful methods that can provide 
sub-percent precision. This work presents
one such method to meet these 
precision requirements. 

Recently, \cite{Bakx:2024zgu} suggested the use of an optimal numerically-determined basis for decomposing the linear power spectrum, and demonstrated its use in analyzing the galaxy power spectrum. This approach, dubbed \cobra, offers numerous advantages, namely (i) the numerical basis yields the optimal (\textit{i.e.}\ smallest) number of linear basis functions needed to reach a given accuracy, thus minimizing storage requirements and maximizing computational speed, (ii) integration over the basis functions does not rely on closed-form analytic expressions and is therefore applicable to any $N$-point function at arbitrary order in perturbation theory. 

In this work, we re-assess the potential of the galaxy one-loop bispectrum to enhance constraints on (i) the standard $\Lambda$CDM cosmological parameters; notably the matter density $\Omega_m$, the Hubble parameter $H_0$, and the amplitude of fluctuations $\sigma_8$ and (ii) non-Gaussianity of the equilateral type. Concretely, we improve upon previous works by (i) deriving several new stochastic contributions to the one-loop bispectrum 
required by rigorous 
power counting 
and (ii) consistently varying all relevant cosmological parameters in the one-loop bispectrum using \cobra.
We validate our pipeline extensively using the state-of-the-art high-fidelity 
`PT-Challenge' simulation suite, whose cumulative volume of $V=566h^{-3}\mathrm{Gpc}^{-3}$ far exceeds that of any current or future galaxy survey and thus provides an ideal setting to validate the theory model. 

The structure of this paper is as follows. In Section \ref{sec:cobra}, we summarize the main features of \cobra as relevant to this work. Section \ref{sec:theory} describes the theoretical template used for the one-loop bispectrum including: the fourth-order bias expansion, the impact of large-scale bulk flows on the baryon acoustic oscillation feature in the bispectrum (following \cite{Blas:2015qsi,Blas:2016sfa}); the counterterms and stochastic contributions needed to render our predictions UV-independent. In Section \ref{sec:comp}, we detail the practical implementation of the bispectrum, including binning, coordinate distortions, and the precomputation of tensors required to calculate the loop integrals in the \cobra basis. Section \ref{sec:data} details our results on the baseline $\Lambda$CDM parameters and subsequently equilateral PNG, as well as the impact of including higher multipoles of the bispectrum. We conclude in Section \ref{sec:concl}.

\section{The \texorpdfstring{\cobra}{COBRA} Basis}\label{sec:cobra}
\noindent The efficient computation of loop integrals in perturbative approaches to LSS presents a significant challenge, especially for analyses beyond the one-loop power spectrum.  
The recently proposed \cobra formalism \cite{Bakx:2024zgu} allows us to express the linear power spectrum, $P_{11}(k)$, as a factorized set of cosmology-dependent coefficients, $w_i(\Theta)$, and scale-dependent basis functions $\mathcal{V}_i(k)$: 
\beq\label{eq: cobra-pk}
    P_{11}(k;\Theta) \approx \sum_{i=1}^{N_{\rm COBRA}} w_i(\Theta) \mathcal{V}_i(k),
\eeq
such that loop integrals reduce to small tensor multiplications after integrations over $\mathcal{V}_i$ have been performed in a pre-processing step. Given a specified range of cosmological parameters and wavenumbers, the basis functions $\mathcal{V}_i(k)$ are constructed from a template bank using a singular value decomposition (see \citep[e.g.,][]{Philcox:2020zyp} for related techniques). This only needs to be done once and yields an optimal projection of the linear power spectrum onto a low-dimensional subspace; typically $N_{\rm COBRA} < 10$ suffices to reach $\sim 0.1 \%$ precision for $\Lambda$CDM power spectra, which is the only case we consider here. We also set the neutrino mass to zero in all of what follows, though this is not a limitation of the approach. 

More precisely, a singular value decomposition is applied to the matrix consisting of template power spectra normalized by their mean $\bar{P}$. That is, for a grid of $N_t$ template spectra evaluated at $N_k$ wavenumbers we form 
$\hat{P}_{lm} = P_{11}(k_m;\Theta_l)/\bar{P}(k_m)$ to obtain
\beq\label{eq:svd}
\hat{P} \approx \hat{U} \Sigma \hat{\mathcal{V}}^T 
\eeq
where $\hat{\mathcal{V}}_{mi} = \hat{\mathcal{V}}_i(k_m) = \mathcal{V}_i(k_m) / \bar{P}(k_m)$. Since the normalized scale functions $\hat{\mathcal{V}}_i$ are orthonormal by definition, we can obtain the cosmology-dependent weights for any power spectrum via a simple projection:\footnote{While \citep{Bakx:2024zgu} also emulates the resulting coefficients $w_i(\Theta)$, in this work we instead directly call the Boltzmann solver \texttt{CLASS} to compute $P_{11}(k;\Theta)$ at each iteration during MCMC sampling. This does not lead to an appreciable increase in computation time.} 
\beq\label{eq : proj}
w_i(\Theta) = \sum_{m=1}^{N_k}\hat{\mathcal{V}}_i(k_m)\hat{P}_{11}(k_m;\Theta). 
\eeq
where $\hat{P}_{11} = P_{11}/\bar{P}$.  Notably, while our notation uses the linear power spectrum $P_{11}(k)$ for simplicity, this procedure can be applied to any quantity that enters a loop integral (such as an infrared-resummed power spectrum). To perform the singular value decomposition we use a grid of $25$ evenly-spaced values for $\omega_{\rm cdm}\in[0.095,0.145]$ and $12$ evenly spaced values for $\omega_{\rm b}\in[0.0202,0.238]$, 
$n_s\in[0.91,1.01]$, equal to the `default' parameter range of in \citep{Bakx:2024zgu}. 
he dependence on the parameter $A_s$ is captured via the rescaling $P_{\rm lin}
\to A_s P_{\rm lin}$, which is exact up to IR resummation effects, which  will be discussed in detail later.

In the PTChallenge analysis below, we will fix the baryon density and the spectral index in order to maintain consistency with previous works. Strictly speaking, the decomposition we use here could thus be optimized further. However, by using the more general grid we also illustrate that marginalizing over the spectral index and baryon density, as done in current state-of-the-art full-shape analyses \cite[e.g.,][]{DESI:2024jis} is computationally feasible.

Note that the shape of the matter power spectrum
(expressed in units of comoving Mpc) does not depend  
on the dimensionless Hubble parameter $h$~\cite{Lesgourgues:2006nd}. 
Hence, one can compute the linear power spectrum 
template bank on a grid of 
wavenumbers in units of $h_{*}$/Mpc, where $h_*$
is a fixed constant, compute the 
one-loop bispectrum, 
and simply rescale these computations 
by appropriate powers of $h$
in order to get the answer in units of $h$/Mpc in a given cosmology.
However, an important limitation is 
binning, which is quite important for the 
bispectrum~\cite{Ivanov:2021kcd,Oddo:2019run,Oddo:2021iwq}.
The binned bispectrum predictions are computed 
for a fixed grid in units of $h_{\rm fid}$/Mpc,
where $h_{\rm fid}$ is a fiducial dimensionless Hubble parameter used to convert redshifts and angles to distances. 
When the theory predictions are computed 
in a given cosmology in units of $h$/Mpc, one has to rescale 
the \cobra computations by appropriate powers of $h/h_{*}$ first, then 
interpolate this result onto a grid of $h_{\rm fid}$/Mpc,
and finally use it to compute the binning integrals numerically. This explicit binning integration creates an
additional computational bottleneck. To optimize 
this procedure, 
we include the computation of the 
binning integrals in the full one-loop calculation, \textit{i.e.}\ we apply the \cobra factorization technique 
to the \textit{binned} bispectrum prediction. 
To that end we use $h_*=h_{\rm fid}$ 
to pre-compute the \cobra templates, 
and treat the necessary 
$h/h_{\rm fid}$ rescaling 
as part of the Alcock-Paczynski (AP) distortions.
When combined with the 
usual AP effect, such a rescaling can be fully implemented by 
using the distortion parameters
\beq\label{eq:newAP}
    \alpha^h_\parallel = \frac{H_{\rm fid}(z)}
    {H_{\rm true}(z)}
    , \quad \alpha^h_\perp = 
    \frac{D_{\rm true,A}(z)}{D_{\rm fid,A}(z)}\,,
\eeq
where $H_{\rm true},D_{\rm true,A}$ are the Hubble parameter and the angular diameter distance in the 
cosmology in question (in $\mathrm{Mpc}$ units). 
In what follows we call this rescaling procedure ``AP+$h$'' to distinguish it from the conventional AP distortions.

\section{Bispectrum Theory}\label{sec:theory}
\noindent In this Section, we summarize our EFT theory model for the bispectrum. This builds upon \citep{Philcox:2022frc} with a more comprehensive treatment of stochastic and counterterm contributions (see also \citep{DAmico:2022ukl} and earlier works~\cite{Ivanov:2019pdj,Philcox:2021kcw,Ivanov:2021kcd}). In Section \ref{sec:cobra}, we will discuss how this model can be computed in practice using the \cobra basis outlined above.

\subsection{Bias and Redshift-Space Operators}\label{subsec: bk-det}
\noindent We begin our overview of the theory model with the deterministic 
contribution, \textit{i.e.}\ contributions to the galaxy overdensity, $\delta_g$, that depend only on the initial 
density field $\delta_1$. 
We use the basis of 
\citep{Eggemeier:2021cam} used in
\citep{Philcox:2022frc}, which are 
equivalent to other bases.
We first require the bias expansion up to fourth order,
\beq\label{eq: bias-expan-4th}
    \delta_g &=& \left\{{b_1}\delta\right\} + \left\{\frac{{b_2}}{2}\delta^2+{\gamma_2}\,\G(\Phi_v)\right\}+\,\left\{\frac{{b_3}}{6}\delta^3+{\gamma_2^\times}\,\delta\,\G(\Phi_v)+{\gamma_3}\,\mathcal{G}_3(\Phi_v)+{\gamma_{21}}\,\G(\varphi_2,\varphi_1)\right\}\\\nonumber
    &&\,+\,\left\{{\gamma_{21}^\times}\,\delta\,\G(\varphi_2,\varphi_1)+{\gamma_{211}}\,\mathcal{G}_3(\varphi_2,\varphi_1,\varphi_1)+{\gamma_{22}}\,\mathcal{G}_2(\varphi_2,\varphi_2)+{\gamma_{31}}\,\mathcal{G}_2(\varphi_3,\varphi_1)\right\}+\mathcal{O}(\delta^5) ,
\eeq
where we dropped the terms 
that do not appear in the one-loop
bispectrum directly 
\be
\label{eq:red}
\delta_g\Big|_{\rm redundant}=
\gamma_{2}^{\times \times} \delta^2 \G + \frac{b_4}{4!}\delta^4
+\gamma_{3}^\times 
\delta \mathcal{G}_3 +\gamma_2^{\rm sq}\G^2 \,.
\ee 
Here we introduced
\beq\label{eq:g2g3}
    \G(\Phi_v) &\equiv& \nabla_i\nabla_j \Phi_v\nabla^i\nabla^j\Phi_v-(\nabla^2\Phi_v)^2 , \\\nonumber
    \mathcal{G}_3(\Phi_v) &\equiv& 2\nabla_i\nabla_j\Phi_v\nabla^j\nabla_k\Phi_v\nabla^k\nabla^i\Phi_v-3\nabla_i\nabla_j\Phi_v\nabla^i\nabla^j\Phi_v\nabla^2\Phi_v+(\nabla^2\Phi_v)^3,
\eeq
where the potentials $\varphi_{1},\varphi_2$ and $\Phi_v$ are defined via
($\theta\equiv-\d_iv^i/(f\HH)$):  
\beq
\Phi_v\equiv \nabla^{-2}\theta\,,\quad 
    \nabla^2\varphi_1 = -\delta\,,\qquad \nabla^2\varphi_2 = -\G(\varphi_1) = -\G.
\eeq
These are related to the bias parameters used in the \textsc{class-pt} power spectrum convention \cite{Chudaykin:2020aoj} through 
\be 
\gamma_2=b_{\G}\,,\quad \gamma_{21}=-\frac{4}{7}(b_{\G}+b_{\Gamma_3})\,.
\ee 
The operators in \eqref{eq: bias-expan-4th} which contain multiple instances of $\varphi_1,\varphi_2$ are obtained by replacing instances of $\Phi_v$ in  \eqref{eq:g2g3} by either $\varphi_1$ or $\varphi_2$, \textit{i.e.}\ $\G(\varphi_2,\varphi_1) \equiv \nabla_i\nabla_j \varphi_2\nabla^i\nabla^j\varphi_1-(\nabla^2\varphi_2)(\nabla^2\varphi_1)$ \textit{et cetera}. The exception to this notation is the operator $\mathcal{G}_2(\varphi_3,\varphi_1)$ involving $\varphi_3$, which is separately defined in Eq. (50) of \cite{Eggemeier:2018qae} due to a subtlety in the definition of the third-order LPT potential $\varphi_3$.
Expanding each operator above 
over the linear density field $\delta_1$ in Eulerian 
Standard Perturbation Theory (SPT),
and using the SPT non-linear matter field expansion: 
\be 
\delta=\delta_1+\delta_2+\delta_3+...=\sum_{n=1}\int_{\vq_1\ldots\vq_n}(2\pi)^3\delta_D^{(3)}(\k-\vq_{1...n})F_n(\vq_1,\cdots,\vq_n)\delta_1(\vq_1)\cdots\delta_1(\vq_n),
\ee 
and similarly for the velocity divergence $\theta$ with corresponding kernels $G_n$, we arrive at the following 
SPT perturbative expansion for the 
galaxy density field:
\beq\label{eq: delta-n-exp}
    \delta_{g}(\k) = \sum_{n=1}\int_{\vq_1\ldots\vq_n}(2\pi)^3\delta_D^{(3)}(\k-\vq_{1...n})K_n(\vq_1,\cdots,\vq_n)\delta_1(\vq_1)\cdots\delta_1(\vq_n),
\eeq
where $\vq_{i\cdots j}\equiv \vq_i+\cdots+\vq_j$ and the real-space kernels $K_n$ are given by
\beq\label{eq: real-kernels}
    K_1(\vq_1) &=& {b_1} , \\\nonumber
    K_2(\vq_1,\vq_2) &=&\left\{{b_1}F_2(\vq_1,\vq_2)\right\}+\left\{\frac{{b_2}}{2}+{\gamma_2}\,\kappa(\vq_1,\vq_2)\right\} , \\\nonumber
    K_3(\vq_1,\vq_2,\vq_3)&=& \left\{{b_1}F_3(\vq_1,\vq_2,\vq_3)\right\} + \left\{{b_2}F_2(\vq_1,\vq_2)+2{\gamma_2}\,\kappa(\vq_1,\vq_{23})G_2(\vq_2,\vq_3)\right\}\\\nonumber
    &&\,+\,\left\{\frac{{b_3}}{6}+{\gamma_2^\times}\,\kappa(\vq_1,\vq_2)+{\gamma_3}\,L(\vq_1,\vq_2,\vq_3)+{\gamma_{21}}\kappa(\vq_1,\vq_{23})\kappa(\vq_2,\vq_3)\right\} , \\\nonumber
    K_4(\vq_1,\vq_2,\vq_3,\vq_4)&=&\left\{{b_1}F_4(\vq_1,\vq_2,\vq_3,\vq_4)\right\}\\\nonumber
    &&\,+\,\left\{\frac{{b_2}}{2}\left[F_2(\vq_1,\vq_2)F_2(\vq_3,\vq_4)+2F_3(\vq_1,\vq_2,\vq_3)\right]\right.\\\nonumber
    &&\qquad\left.+{\gamma_2}\left[\kappa(\vq_{12},\vq_{34})G_2(\vq_1,\vq_2)G_2(\vq_3,\vq_4)+2\kappa(\vq_{123},\vq_4)G_3(\vq_1,\vq_2,\vq_3)\right]\right\}\\\nonumber
    &&\,+\,\left\{\frac{{b_3}}{2}F_2(\vq_1,\vq_2)+{\gamma_2^\times}\left[2\kappa(\vq_{12},\vq_3)G_2(\vq_1,\vq_2)+\kappa(\vq_3,\vq_4)F_2(\vq_1,\vq_2)\right]\right.\\\nonumber
    &&\qquad+\,3{\gamma_3}\,L(\vq_1,\vq_2,\vq_{34})G_2(\vq_3,\vq_4)\\\nonumber
    &&\qquad\left.+\,{\gamma_{21}}\left[\kappa(\vq_{12},\vq_{34})\kappa(\vq_1,\vq_2)F_2(\vq_3,\vq_4)+2\kappa(\vq_{123},\vq_4)\kappa(\vq_{12},\vq_3)F_2(\vq_1,\vq_2)\right]\right\}\\\nonumber
    &&\,+\,\left\{{\gamma_{21}^\times}\,\kappa(\vq_1,\vq_{23})\kappa(\vq_2,\vq_3)+{\gamma_{211}}L(\vq_1,\vq_2,\vq_{34})\kappa(\vq_3,\vq_4)\right.\\\nonumber
    &&\qquad+\,{\gamma_{22}}\,\kappa(\vq_{12},\vq_{34})\kappa(\vq_1,\vq_2)\kappa(\vq_3,\vq_4)\\\nonumber
    &&\qquad+\,{\gamma_{31}}\left[\frac{1}{18}\kappa(\vq_1,\vq_{234})\left(\frac{15}{7}\kappa(\vq_{23},\vq_4)\kappa(\vq_2,\vq_3)-L(\vq_2,\vq_3,\vq_4)\right)\right.\\\nonumber
    &&\qquad\qquad\left.\left.+\frac{1}{14}\left(M(\vq_1,\vq_{23},\vq_4,\vq_{234})-M(\vq_1,\vq_{234},\vq_{23},\vq_4)\right)\kappa(\vq_2,\vq_3)\right]\right\},
\eeq
where we have defined
\beq\label{eq: ang-def}
    \kappa(\vq_1,\vq_2) &=& (\hq_1\cdot\hq_2)^2-1 , \\\nonumber
    L(\vq_1,\vq_2,\vq_3) &=& 2(\hq_1\cdot\hq_2)(\hq_2\cdot\hq_3)(\hq_3\cdot\hq_1)-(\hq_1\cdot\hq_2)^2-(\hq_2\cdot\hq_3)^2-(\hq_3\cdot\hq_1)^2+1 , \\\nonumber
    M(\vq_1,\vq_2,\vq_3,\vq_4) &=& (\hq_1\cdot\hq_2)(\hq_2\cdot\hq_3)(\hq_3\cdot\hq_4)(\hq_4\cdot\hq_1).
\eeq
The second important set of ingredients are redshift-space distortions
introduced by the coordinate transformation from the
rest frame of galaxies to the
observer's frame
\cite{Kaiser:1987qv}
\beq\label{eq: rsd-mapping}
    \delta_g^{(s)}(\k) = \delta_g(\k) + \int d\vx\,e^{-i\k\cdot\vx}\left[e^{-ik_z f u_z(\vx)}-1\right]\left(1+\delta_g(\vx)\right),
\eeq
where $v_i = (f\HH)u_i$, $f$ is the logarithmic growth factor, $z^i$
is the line-of-sight unit vector,
$u_z=u_i\hat{z}^i$, $k_z=k_i\hat{z}^i=\mu k$. 
The Taylor 
expansion of the redshift-space mapping
to
fourth order yields
\be
\label{eq:RSDexp}
\delta_\k^{(s)} = \delta_\k -ifk_z[(1+\delta_g)u_z]_\k
+\frac{i^2f^2}{2}k_z^2 [(1+\delta_g)u^2_z]_\k -\frac{i^3f^3}{3!} k_z^3[(1+\delta_g)u^3_z]_\k 
+ \frac{i^4f^4}{4!} k_z^4 [u_z^4(1+\delta_g)]_\k \,.
\ee
Using the perturbative expansion from~\eqref{eq: delta-n-exp}
and the SPT expansion for the
dark matter velocity field 
we obtain the perturbative series 
in redshift-space:
\beq\label{eq:rsd_Zn}
    \delta^{(s)}_{g,(n)}(\k) = 
    \sum_{n}\int_{\vq_1\ldots\vq_n}(2\pi)^3\delta_D^{(3)}(\k-\vq_{1...n})Z_n(\vq_1,\cdots,\vq_n)\delta_1(\vq_1)\cdots\delta_1(\vq_n)\,,
\eeq
with $\mu_{i\cdots j}\equiv \mu_{\vq_i+\cdots+\vq_j}$, defining the redshift-space kernels
\beq\label{eq: RSD-kernels}
    Z_1(\vq_1) &=& K_1 + f\mu_1^2,\\\nonumber
    Z_2(\vq_1,\vq_2) &=& K_2(\vq_1,\vq_2) + f\mu_{12}^2G_2(\vq_1,\vq_2)+\frac{f\mu_{12}q_{12}}{2}K_1\left[\frac{\mu_1}{q_1}+\frac{\mu_2}{q_2}\right]+\frac{(f\mu_{12}q_{12})^2}{2}\frac{\mu_1}{q_1}\frac{\mu_2}{q_2} , \\\nonumber
    Z_3(\vq_1,\vq_2,\vq_3) &=& K_3(\vq_1,\vq_2,\vq_3)+f\mu_{123}^2G_3(\vq_1,\vq_2,\vq_3)\\\nonumber
    &&\,+\,(f\mu_{123}q_{123})\left[\frac{\mu_{12}}{q_{12}}K_1G_2(\vq_1,\vq_2)+\frac{\mu_3}{q_3}K_2(\vq_1,\vq_2)\right]\\\nonumber
    &&\,+\,\frac{(f\mu_{123}q_{123})^2}{2}\left[2\frac{\mu_{12}}{q_{12}}\frac{\mu_3}{q_3}G_2(\vq_1,\vq_2)+\frac{\mu_1}{q_1}\frac{\mu_2}{q_2}K_1\right]+\frac{(f\mu_{123}q_{123})^3}{6}\frac{\mu_1}{q_1}\frac{\mu_2}{q_2}\frac{\mu_3}{q_3} , \\\nonumber
    Z_4(\vq_1,\vq_2,\vq_3,\vq_4) &=&  K_4(\vq_1,\vq_2,\vq_3,\vq_4)+f\mu_{1234}^2G_4(\vq_1,\vq_2,\vq_3,\vq_4)\\\nonumber
    &&\,+\,(f\mu_{1234}q_{1234})\left[\frac{\mu_{123}}{q_{123}}K_1G_3(\vq_1,\vq_2,\vq_3)+\frac{\mu_4}{q_4}K_3(\vq_1,\vq_2,\vq_3)\right.\\\nonumber
    &&\qquad\qquad\qquad\qquad+\,\left.\frac{\mu_{12}}{q_{12}}G_2(\vq_1,\vq_2)K_2(\vq_3,\vq_4)\right]\\\nonumber
    &&\,+\,\frac{(f\mu_{1234}q_{1234})^2}{2}\left[2\frac{\mu_{123}}{q_{123}}\frac{\mu_4}{q_4}G_3(\vq_1,\vq_2,\vq_3)+\frac{\mu_{12}}{q_{12}}\frac{\mu_{34}}{q_{34}}G_2(\vq_1,\vq_2)G_2(\vq_3,\vq_4)\right.\\\nonumber
    &&\qquad\qquad\qquad\qquad\left.+\,2\frac{\mu_{12}}{q_{12}}\frac{\mu_3}{q_3}K_1G_2(\vq_1,\vq_2)+\frac{\mu_1}{q_1}\frac{\mu_2}{q_2}K_2(\vq_3,\vq_4)\right]\\\nonumber
    &&\,+\,\frac{(f\mu_{1234}q_{1234})^3}{6}\left[3\frac{\mu_{12}}{q_{12}}\frac{\mu_3}{q_3}\frac{\mu_4}{q_4}G_2(\vq_1,\vq_2)+\frac{\mu_1}{q_1}\frac{\mu_2}{q_2}\frac{\mu_3}{q_3}K_1\right]\\\nonumber
    &&\,+\,\frac{(f\mu_{1234}q_{1234})^4}{24}\frac{\mu_1}{q_1}\frac{\mu_2}{q_2}\frac{\mu_3}{q_3}\frac{\mu_4}{q_4}\,.
\eeq

In SPT, the above expansion is sufficient to generate the tree and one-loop contributions to the three-point function, inserting 
the Gaussian
distribution 
for the linear density field,
\be 
\langle \delta_1(\k)\delta_1(\k')\rangle =(2\pi)^3\delta_D^{(3)}(\k+\k')P_{11}(k,z)\equiv 
(2\pi)^3\delta_D^{(3)}(\k+\k')\widetilde{P}_{11}(k)D_+^2(z)
\,,
\ee 
where $\widetilde{P}_{11}$ is the linear theory
matter power spectrum at redshift zero, $\widetilde{P}_{11} = P_{11}(z=0)$
and $D_+(z)$ is the scale-independent
$\Lambda$CDM
linear theory growth factor normalized to unity today. In what follows we will suppress the explicit 
redshift dependence and use 
the primed correlators with the 
stripped off Dirac delta function, 
\textit{i.e.}\ 
\be 
\langle \delta_1(\k)\delta_1(\k')\rangle' =P_{11}(k)\,.
\ee 
At the zero loop (\textit{i.e.}\ tree-level) order the deterministic bispectrum reads
\beq
    \langle \delta_{g,2}(\k_3)\delta_{g,1}(\k_2)\delta_{g,1}(\k_1)\rangle'
    +\text{2 cyc.}=B_{211} (\k_1,\k_2,\k_3) = 2\,Z_2(\k_1,\k_2)Z_1(\k_1)Z_1(\k_2)P_{11}(k_1)P_{11}(k_2) + \text{2 cyc.},
\eeq
where ``2 cyc." labels two terms 
obtained by non-identical 
cyclic permutations
of the wavenumbers $\{\k_1,\k_2,\k_3\}$.
At the one-loop order we have 
four distinct contributions:
\cite{Bernardeau:2001qr,Baldauf:2014qfa,Philcox:2022frc}:
\beq\label{eq: Bk-one-loop}
    B_{222}(\k_1,\k_2,\k_3) &=& 8\int_{\vq}Z_2(\k_1+\vq,-\vq)Z_2(\k_1+\vq,\k_2-\vq)Z_2(\k_2-\vq,\vq)
    P_{11}(q)P_{11}(|\k_1+\vq|)
    P_{11}(|\k_2-\vq|), \\\nonumber
    B_{321}^I(\k_1,\k_2,\k_3) &=&  6\,Z_1(\k_1)P_{11}(k_1)\int_{\vq}Z_3(-\vq,\vq-\k_2,-\k_1)Z_2(\vq, \k_2-\vq)
    P_{11}(q)P_{11}(|\k_2- \vq|)+\text{5 perm.}, \\\nonumber
    B_{321}^{II}(\k_1,\k_2,\k_3) &=& 6\,Z_2(\k_1,\k_2)Z_1(\k_2)
    P_{11}(k_1)
    P_{11}(k_2)\int_{\vq}Z_3(\k_1,\vq,-\vq)P_{11}(q)+\text{5 perm.}, \\\nonumber
    B_{411}(\k_1,\k_2,\k_3) &=& 12\,Z_1(\k_1)Z_1(\k_2)
    P_{11}(k_1)P_{11}(k_2)\int_{\vq}Z_4(\k_1,\k_2,\vq,-\vq)P_{11}(q)+\text{2 cyc.}.
\eeq
The contributions written above
represent the leading 
bispectrum corrections whose 
fast 
calculation has represented a challenge 
in the past. Producing a pipeline for their 
efficient computation is a key result of our work. 

Before going forward, it is important to comment on the fate
of the redundant operators~\eqref{eq:red} whose 
contributions are degenerate with lower order
operators at the order we are working with. The one-loop bias parameters (both redundant and non-redundant) lead to unobservable redefinitions of the tree-level bias parameters \citep[e.g.,][]{Assassi:2014fva,Eggemeier:2018qae}:
\be 
\begin{split}
& b_1^{(R)}=b_1+\sigma^2(\Lambda)\left[
\frac{34}{21}b_2+\frac{b_3}{2}
-\frac{4}{3}\gamma_{2}^\times
\right]\,,\\
& b_2^{(R)}=b_2+\sigma^2(\Lambda)\left[
\frac{8126}{2205}b_2+\frac{68}{21}b_3
-\frac{16}{3}\gamma_2^{\times \times}
+\frac{b_4}{2}
+\frac{32}{15}\gamma_{21}^{\times  }
-\frac{764}{10592}\gamma_2^{\times}
+\frac{64}{15}\gamma_2^{\rm sq}
\right]\,,\\
& b_{\G}^{(R)}=b_{\G}+\sigma^2(\Lambda)\left[
\frac{127}{2205}b_2
-\frac{16}{35}\gamma_2^\times
+\gamma_2^{\times \times}
-\frac{2}{5}\gamma_{21}^\times
-\gamma_3^\times
+\frac{8}{15}\gamma_2^{\rm sq}
\right]\,,
\end{split}
\ee 
where $b_1,b_2,b_{\G}$ are the `bare' bias parameters and $b_1^{(R)},b^{(R)}_2,b^{(R)}_{\G}$ are the renormalized bias parameters. We have also defined the filtered mass variance
\be 
\sigma^2(\Lambda)=\int_{\p,p\leq \Lambda}P_{11}(p)\,.
\ee 
If one is using bare operators/parameters and a renormalization scheme with an explicit cutoff, one would have to shift the tree-level 
bias parameter following these expressions once the one-loop
corrections are taken into account.
One can see, however, that such a 
renormalization scheme is 
quite impractical as it leads to an
order $\mathcal{O}(1)$
redefinition of all bias parameters
at every new higher order 
calculation. This makes it complicated to include new calculations
in the analysis and does not allow one to 
use the measurements of bias
parameters from fitting the data with 
the tree-level predictions. 
Alternatively, one may choose
to work with the 
re-normalized bias parameters
directly~\cite{Assassi:2014fva}, but in this case one will have to carry out 
appropriate 
subtractions of divergent loop contributions from the bias operators, see e.g.~\cite{Bakx:2025cvu}
for a recent example of the 
two-loop
galaxy power spectrum
computation.
In our work, however, we work within the 
dimensional regularization scheme automatically implemented by the FFTLog loop computation technique, 
in which all divergent integrals 
are identically set to zero \citep[cf.,][]{Philcox:2022frc,Simonovic:2017mhp}. In this case $b_1^{(R)}=b_1,b_2^{(R)}=b_2,b_{\G}^{(R)}=b_{\G}$, such that 
the ``bare'' parameters
are automatically renormalized, and higher order
computations do not lead to the redefinition 
of bias parameters. 

\subsection{Power Counting in a Power-Law Universe}

\noindent The cornerstone of the EFT philosophy 
is power counting, \textit{i.e.}\ estimates of
various terms that one needs to keep 
in the theoretical description based
on the relevant energy scale
of the experiment. In the context 
of the EFT of LSS, one can obtain reliable estimates
by utilizing the power-law approximation
to the linear matter power spectrum 
(which is the seed for perturbative loop
corrections)~\cite{Pajer:2013jj,Ivanov:2022mrd}. 

The leading order EFT result for the 
bispectrum scales as $B_{211}\sim P^2_{11}$, whilst the SPT one-loop corrections considered above
scale as $P^2_{11}(P_{11}k^3)$. 
In EFT we expect various corrections 
to the above standard perturbation theory result. These corrections can be 
cast into five groups: (i) the higher derivative (`counterterm') contributions $\sim k^2 B_{211}$; (ii)
the `mixed' stochastic-deterministic contributions $\sim P\bar n^{-1}$ where $\bar n$ is the galaxy number
density; (iii) the `pure' stochastic terms 
$\sim \bar n^{-2}$; (iv,v) the $k^2$ corrections 
to (ii) and (iii). In addition, 
the `mixed' stochastic-deterministic terms
obey a loop expansion themselves. 

To understand the importance of the above corrections, we estimate their size in a power-law cosmology with $k_{\rm NL}^3P_{11}\sim (k/k_{\rm NL})^n$. Let's start with the power spectrum. We have the following expressions
for the usual deterministic 
$L$-loop corrections to the position space 
density variance of matter and galaxies (assuming all bias parameters are 
$\mathcal{O}(1)$ numbers):
\be 
\begin{split}
\Delta^2_{L-\text{loop}}(k)\equiv \frac{k^3}{2\pi^2} P_{L-\text{loop}}(k)\sim \left(\frac{k}{k_{\rm NL}}\right)^{(1+L)(3+n)}\,,
\end{split}
\ee 
while the $k^2$ counterterms scale as 
\be 
k^2P_{11}\sim \left(\frac{k}{k_{\rm NL}}\right)^{5+n}\,.
\ee 
Next, the galaxy stochasticity produced by a stochastic field $\epsilon$~\cite{Philcox:2022frc} is described as $(\bar n)^{-1}(1+(k/k_\text{NL})^2+...)$, so that the stochastic part scales as
\be 
\Delta^2_{\rm stoch}=\frac{k^3}{\bar n} + \frac{k^3}{\bar n }\frac{k^2}{k_{\rm NL}^2}+...
\ee 
In combination, we find:
\be 
\begin{split}
\Delta^2\sim  
&\underbrace{\left(\frac{k}{k_{\rm NL}}\right)^{3+n}}_{\rm tree}+
\underbrace{\left(\frac{k}{k_{\rm NL}}\right)^{2(3+n)}}_{\rm 1-loop}+
\underbrace{\left(\frac{k}{k_{\rm NL}}\right)^{3(3+n)}}_{\rm 2-loop}+
\underbrace{\left(\frac{k}{k_{\rm NL}}\right)^{5+n}}_{\rm counterterm}+\underbrace{\frac{k^3}{\bar n}}_{\rm LO~stoch.} + \underbrace{\frac{k^3}{\bar n }\frac{k^2}{k_{\rm NL}^2}
}_{\rm NLO~stoch.}+...\,.
\end{split}
\ee 
In our universe $n\approx -1.5$ for $k\sim 0.2~\hMpc$~\cite{Pajer:2013jj,Cabass:2022wjy}.
Realistic galaxy surveys such 
as BOSS or DESI operate 
in the regime where
$\bar n b_1^2 P_{11}(k_{\rm max}\approx 0.2~\hMpc)\simeq 1$~\cite{Reid:2015gra}, which implies
\be 
\frac{k_\text{max}^3}{\bar n}\approx \left(\frac{\kmax}{k_{\rm NL}}\right)^{1.5}\quad \text{for}\quad \kmax\approx 0.2\,\hMpc\,,
\ee 
where we choose $\kmax=0.2~\hMpc$
consistent with the values
used in EFT-based full-shape analyses~\cite{Ivanov:2019pdj,Chudaykin:2020aoj,Philcox:2021kcw,Chudaykin:2024wlw}.
This produces the following estimate:
\be 
\begin{split}
\Delta^2(\kmax)\sim  
&\underbrace{\left(\frac{\kmax}{k_{\rm NL}}\right)^{1.5}}_{\rm tree}+
\underbrace{\left(\frac{\kmax}{k_{\rm NL}}\right)^{3}}_{\rm 1-loop}+
\underbrace{\left(\frac{\kmax}{k_{\rm NL}}\right)^{4.5}}_{\rm 2-loop}+
\underbrace{\left(\frac{\kmax}{k_{\rm NL}}\right)^{3.5}}_{\rm counterterm}+\underbrace{\left(\frac{\kmax}{k_{\rm NL}}\right)^{1.5}}_{\rm LO~stoch.} + \underbrace{\left(\frac{\kmax}{k_{\rm NL}}\right)^{3.5}
}_{\rm NLO~stoch.}+...\,,
\end{split}
\ee 
\textit{i.e.}\ we get the following hierarchy: 
\be 
\begin{split}
P_{\rm tree}&\sim P_{\rm stoch.~LO} \\
&\gg P_{\rm 1-loop} \\
&> P_{\rm ctr.}\sim P_{\rm stoch.~NLO} \\
&\gg P_{\rm 2-loop}\,.
\end{split}
\ee 
Importantly, the leading order stochasticity is 
as important as the tree-level
deterministic power spectrum
$P_{11}$, while the $k^2$-corrected stochasticity 
is as important as the counterterm. 

Next, we consider the bispectrum.
Here, the relevant quantity
is the dimensionless amplitude of the three-point
fluctuations in position space
$I_B(k)$
generated by Fourier modes of 
wavenumber $k$~\cite{Baldauf:2014qfa} (restricting to equilateral configurations for simplicity). We have the usual deterministic loop corrections
\be 
\begin{split}
I_{B}^{L-\rm loop}(k)\equiv\frac{k^6}{(2\pi^2)^2}B_{L-\rm loop}(k,k,k)\sim \left(\frac{k}{k_{\rm NL}}\right)^{(2+L)(3+n)}\,,
\end{split}
\ee 
the one-loop counterterms $(B\sim k^2 P^2_{11})$,
\be 
\begin{split}
I_B^{\rm ctr.}(k)=\left(\frac{k}{k_{\rm NL}}\right)^{2+2(3+n)}\,,
\end{split}
\ee 
and `pure' stochastic terms:
\be 
\begin{split}
I_B^{\rm stoch.}(k)=\frac{k^6}{\bar n^2}+\frac{k^6}{\bar n^2}\frac{k^2}{k_{\rm NL}^2}+...\,.
\end{split}
\ee 
Furthermore, there are mixed stochastic-deterministic contributions 
that stem from the operators like $\epsilon \delta$ (which generates the leading-order stochasticity term $\sim P(k)/\bar{n}$). Importantly, such couplings involve the non-linear density field, which in perturbation theory generates 
terms like $\epsilon \delta^n_1$,
\be 
k^6\langle \epsilon \cdot\epsilon \delta\cdot \delta \rangle'=
\underbrace{\frac{k^3}{\bar n}\left(\frac{k}{k_{\rm NL}}\right)^{3+n}}_{\text{from}~\langle \epsilon \cdot\epsilon \delta_1\cdot \delta_1 \rangle'}
+\underbrace{\frac{k^3}{\bar n}\left(\frac{k}{k_{\rm NL}}\right)^{2(3+n)}
}_{\text{from}~\langle \epsilon \cdot\epsilon \delta_{1}\cdot \delta_1^3 \rangle',\langle \epsilon \cdot\epsilon \delta^2_{1}\cdot\delta^2_1\rangle'}
+\underbrace{\frac{k^3}{\bar n}\left(\frac{k}{k_{\rm NL}}\right)^{3+n}\frac{k^2}{k_{\rm NL}^2}}_{\text{from}~\langle \epsilon \cdot k^2\epsilon \delta_1\cdot \delta_1 \rangle'}+...
\ee 
Putting these terms all together we get:
\be 
\begin{split}
I_B(\kmax)\sim  
&\underbrace{\left(\frac{\kmax}{k_{\rm NL}}\right)^{3}}_{\rm tree}+
\underbrace{\left(\frac{\kmax}{k_{\rm NL}}\right)^{4.5}}_{\rm 1-loop}+
\underbrace{\left(\frac{\kmax}{k_{\rm NL}}\right)^6}_{\rm 2-loop}+
\underbrace{\left(\frac{\kmax}{k_{\rm NL}}\right)^{5}}_{\rm counterterm}\\
&+\underbrace{\left(\frac{\kmax}{k_{\rm NL}}\right)^{3}}_{\rm LO~stoch.} + \underbrace{\left(\frac{\kmax}{k_{\rm NL}}\right)^{5}
}_{\rm NLO~stoch.}+\underbrace{\left(\frac{\kmax}{k_{\rm NL}}\right)^{3}}_{\rm mixed~tree}
+\underbrace{\left(\frac{\kmax}{k_{\rm NL}}\right)^{4.5}}_{\rm mixed~1-loop}
+\underbrace{\left(\frac{\kmax}{k_{\rm NL}}\right)^{5}}_{\rm mixed~ctr}
...\,.
\end{split}
\ee
We see that we get the following hierarchy: 
\be 
\begin{split}
B_{\rm tree}&\sim B_{\rm stoch.~LO}\sim B_{\rm mixed~tree}\\
&\gg B_{\rm 1-loop}
\sim B_{\rm mixed~1-loop} \\
& > B_{\rm ctr.}\sim B_{\rm stoch.~NLO}\sim
B_{\rm mixed~ctr.} \\
& \gg B_{\rm 2-loop}\,.
\end{split}
\ee 
Importantly, the stochastic contribution
$B_{\rm mixed~tree}$
is as important as the tree-level bispectrum, while 
$B_{\rm mixed~1-loop}$ is as important as the 
deterministic one-loop contributions. This has important implications for the stochasticity structure, which we describe below. With these
power counting results 
in mind, let us proceed
to the discussion of 
the relevant EFT terms.

\subsection{Counterterms}\label{subsec: counterterms}

\noindent The overall bispectrum counterterm contribution 
can be written as
\be\label{eq: Bk-ctr}
 \begin{split}
B^{\rm ctr}(\k_1,\k_2,\k_3) & \equiv B^{\rm ctr.~I}(\k_1,\k_2,\k_3) +B^{\rm ctr.~II}(\k_1,\k_2,\k_3)\\
& =2Z_1(\k_1)Z_1(\k_2)Z_2^{\rm ctr}(\k_1,\k_2)P_{11}(k_1)P_{11}(k_2) +\text{2 cyc.} \,,\\
& + 2Z^{\rm ctr}_1(\k_1)Z_1(\k_2)Z_2(\k_1,\k_2)P_{11}(k_1)P_{11}(k_2)
+\text{5 perms.} \,,
 \end{split}
\ee
where the counterterm kernels are defined as
\be 
\delta^{\rm ctr}(\k)=\sum_{n=1}\left[\prod_{i=1}^n \int_{\q_i}\delta_1(\q_i)\right](2\pi)^3\delta_D^{(3)}(\k-\q_{1...n})Z^{\rm ctr}_n(\q_1,...,\q_n)\,,
\ee 
and we further split
$Z^{\rm ctr}_n=F_n^{\rm (r),ctr}+ F_n^{\rm (s),ctr}$ 
where (r) and (s) stand for 
real and redshift-space,
respectively. Let us discuss these contributions separately. 

\subsubsection{Real-Space Counterterms}\label{subsubsec: ctr-real}

\noindent In real-space, the galaxy counterterms are set by the higher-derivative bias, as well as the counterterms of dark matter in real-space, inherited via the bias relations, 
\be 
\delta_g = b_1\delta +b_2\delta^2+b_{\G}\G+...
\supset b_1c_s\nabla^2\delta+b_2(\delta c_s\nabla^2\delta)+...\,.
\ee 
For the former, we have
the following symmetry-based expression at quadratic order
\be 
\begin{split}
&\delta_g\big|_{\rm h.d.}=(b_{\nabla^2\delta}\nabla^2\delta
+b_{\nabla^2\delta^2}\nabla^2\delta^2 + b_{(\nabla\delta)^2}(\nabla\delta)^2
+b_{\nabla^2\G}\nabla^2
\G
+b_{(\nabla t)^2}\d_kt_{ij}\d_kt_{ij})k_{\rm NL}^{-2}
\,,
\end{split}
\ee 
where we kept only the combinations that are linearly 
independent at this order and we have introduced the tidal tensor $t_{ij}=[\d_i\d_j \hat{\Phi}]/(\frac{3}{2}\Omega_m\mathcal{H}^2)\equiv \d_i\d_j\Phi$, where $\hat \Phi$ is the 
standard
Newtonian potential
with the usual normalization.
Note that $\Phi=-\varphi_1$.
Focusing on the higher derivative terms we obtain
\be
\label{eq:realhd}
\begin{split}
 F^{\rm (r), ctr}_1(\k)=&-b_{\nabla^2\delta}\frac{k^2}{k_{\rm NL}^2}\\
 F^{\rm (r), ctr}_2(\k_1,\k_2) =
 &(-k^2_3 b_{\nabla^2\delta}F_2(\k_1,\k_2)-
   b_{\nabla^2\delta^2} k_3^2
  - b_{\nabla^2\G}k_3^2\G-
   (b_{(\nabla\delta)^2}+ b_{(\nabla t)^2})(\k_1\cdot\k_2)-
   b_{(\nabla t)^2}(\k_1\cdot\k_2)\G)\frac{1}{k_{\rm NL}^2}
\end{split}
\ee
where $\G(\k_1,\k_2)\equiv(\hat \k_1\cdot \hat \k_2)^2-1$ (cf.\,\ref{eq: ang-def}).
This form is equivalent to all other bases used in the literature on the bispectrum counterterms \cite{Eggemeier:2018qae,DAmico:2022ukl}.
The higher-derivative bispectrum thus introduces 
five free parameters: $\{b_{\nabla^2\delta},b_{\nabla^2\delta^2},b_{\nabla^2\mathcal{G}_2},b_{(\nabla\delta)^2},b_{(\nabla t)^2}$\}. Terms with more than two derivatives must contain at least four derivatives, and are hence of higher order than what is needed for the one-loop bispectrum. 


One can check explicitly 
that the above basis~\eqref{eq:realhd}
is sufficient for the renormalization
of the UV sensitivity 
in the real-space
SPT loop diagrams. 
As such, we will find it convenient to express the total matter density field in real-space as
\be \label{eq: delta-g-real}
[\delta_g]_R = \delta_g + \delta_g\big|_{\rm h.d.}\,,
\ee 
including both deterministic and higher-derivative bias
(\textit{i.e.}\ counterterms) contributions.

\subsubsection{Redshift-Space Counterterms}

\noindent Next, we consider the deterministic counterterms arising from the redshift-space expansion of \eqref{eq:RSDexp}. Applying a low-pass filter to this expression, we obtain the following renormalized
expressions for products of the real-space density field and velocities analogous to \cite{Philcox:2022frc}\footnote{We use a slightly different notation to \cite{Philcox:2022frc} to streamline Galilean invariance.}:
\be \label{eq: velocity-renorm}
\begin{split}
& [u^i(1+\delta_g)]_R = (1+[\delta_{g}]_R)u^i +\mathcal{O}^i_u\,, \\
& [u^i u^j(1+\delta_g)]_R = (1+[\delta_{g}]_R{})u^i u^j +u^i \mathcal{O}^j_{u} 
+ u^j \mathcal{O}{}^i_{u} + \mathcal{O}{}^{ij}_{u^2} \,,\\
& [u^i u^j u^k(1+\delta_g) ]_R = 
(1+[\delta_{g}]_R)
u^i 
u^j 
u^k   
+ (u^iu^j \mathcal{O}_{u}^{k} +\text{2 perm.})
+ (u^i \mathcal{O}_{u^2}^{jk} +\text{2 perm.})
+ \mathcal{O}_{u^3}^{ijk} \,,\\
& [u^i u^j u^k u^l (1+\delta_g)]_R = 
(1+[\delta_{g}]_R)
u^i 
u^j 
u^k 
u^l 
+(u^iu^ju^k \mathcal{O}_{u}^{l}+\text{3 perm.}) \\
&\quad \quad \quad \quad \quad \quad
+ (u^iu^j \mathcal{O}_{u^2}^{kl} 
+\text{5 perm.})
+ (u^i \mathcal{O}_{u^3}^{jkl}+\text{3 perm.})
+\mathcal{O}_{u^4}^{ijkl}\,,
\end{split}
\ee
where $[\delta_g]_R$ is the renormalized real-space field defined in \eqref{eq: delta-g-real}. Galilean invariance requires 
that all $\mathcal{O}_{u^n}^{i_1i_2...i_n}$
operators transform as scalars under
the Galilean transformations. The first operator, $\mathcal{O}^i_u$ can thus be expressed in terms of the density and tidal fields as
\be\label{eq:oi_u}
k^2_\text{NL}\mathcal{O}^i_u \supset
{e}_1 \d_i \delta + {e}_2 \d_i(t_{kl}t^{kl})+{e}_3 \d_i \delta^2+e_4 t_{ik}\d^k\delta+{e}_5 \frac{\d_i\d_j\d_k}{\nabla^2}\left(
\frac{\d_j\d_l\delta}{\nabla^2}
\frac{\d_l\d_k\delta}{\nabla^2}
\right)\,,
\ee
keeping only the 
linearly independent terms, and introducing Wilson coefficients $e_{1\cdots 5}$. This neglects stochastic contributions, which are discussed below. 
The last term above stems from
the non-locally contributing velocity, as discussed in \cite{DAmico:2022ukl}. 
Jumping ahead, let us note that 
only $e_1,e_4$ and $e_5$ terms will produce 
non-degenerate contributions 
at the level of the one-loop
bispectrum. 
In addition, the higher derivative term 
in the galaxy density field
$[\delta_g]_R\supset b_{\nabla^2\delta}\nabla^2\delta$, will generate the contribution 
\be 
\begin{split} 
& k_\text{NL}^2(-i f k_z z^i[u^i (1+\delta_g)]_R)\to 
(-b_{\nabla^2\delta}) f k_z k_1^2 \frac{k_{2z} }{k_2^2}
\delta_{1\k_1}
\delta_{1\k_2}=
b_{\nabla^2\delta} f k_{3z} k_1^2 \frac{k_{2z} }{k_2^2}
\delta_{1\k_1}
\delta_{1\k_2}
\,,
\end{split}
\ee
where we used $k_{3z}=-k_{z}=-k_{1z}-k_{2z}$.
Then, writing all 
non-trivial Galilean-symmetric operators with two 
$SO(3)$ indices we arrive at 
\be 
\begin{split}
k^2_\text{NL}\mathcal{O}^{ij}_{u^2} \supset & 
 {c}_0 \delta_{ij}
 +
{c}_1 \delta_{ij}\delta +
{c}_2 t_{ij}   
+ {c}_3 \delta^2 \delta_{ij}
+{c}_{4}t_{il}t_{lj} 
+{c}_{5}\delta t_{ij}
+{c}_{6}\delta_{ij}t^2
+{c}_7\d_i\d_j(\Phi-\Phi_v)\,.
\end{split}
\ee 
The operators
$O_{u^3}^{ijk}$
and 
$O_{u^4}^{ijkl}$,
do not contribute to the bispectrum at this order. 
Combining results and removing 
all the degenerate contributions we arrive at the redshift-space counterterm kernels
\be 
\label{eq:Z1s}
Z^{\rm ctr.~(s)}_1(\k) =\frac{k^2}{k_{\rm NL}^2} \left(-b_{\nabla^2\delta}+\left(e_1 -\frac{1}{2}c_1 f\right)f\mu^2 -\frac{1}{2}c_2 f^2\mu^4\right) \,.
\ee 
which matches Eq.~5.41 in \cite{DAmico:2022osl}. At quadratic order we find\footnote{Note that the contribution of $c_0$ at non-zero momenta generated by the velocity dressings is exactly degenerate with 
$c_2$ and $c_7$, and that its coefficient is fixed by the condition $\langle \delta^{(s)}\rangle=0$.}
\be
\begin{split}
& k_{\rm NL}^2 F^{\rm ctr.~(s)}_2(\k_1,\k_2) = b_{\nabla^2\delta}f k_{3z}\left(\frac{k_1^2k_{2z}}{2k_2^2}+(2\leftrightarrow 1)\right)  +e_1 \left(f k_{3z}^2 F_2(\k_1,\k_2)+\frac{f^2k_{3z}^2k_{1z}k_{2z}}{2}\left(\frac{1}{k_1^2}+\frac{1}{k_2^2}\right) \right) \\
&+ e_5f k_{3z}^2\frac{(\k_1\cdot\k_2)(\k_1\cdot\k_3)(\k_2\cdot\k_3)}{k_1^2 k_2^2 k_3^2} 
- e_4fk_{3z}\left(\frac{k_{1z}(\k_1\cdot \k_2)}{2k_1^2}+(2\leftrightarrow 1)\right)
\\
&
+c_1\left(-\frac{ f^2}{2}k^2_{3z}F_2(\k_1,\k_2)+f^3\frac{k^3_{3z} k_{1z}}{4 k^2_{1}}+(2\leftrightarrow 1)\right)
+c_2\left(-\frac{f^2}{2}
\frac{k^4_{3z}}{k_3^2}F_2(\k_1,\k_2)+f^3 k_{3z}^3\frac{k_{1z}k^2_{2z}}{4k_1^2k_2^2}+(2\leftrightarrow 1)\right) \\
& -\frac{c_3f^2k^2_{3z}}{2} 
 -c_4\frac{f^2}{2}k_{3z}^2\frac{(\k_1\cdot\k_2)k_{1z}k_{2z}}{k_1^2k_2^2}
-\frac{c_6}{2}f^2k_{3z}^2\frac{(\k_1\cdot \k_2)^2}{k_1^2k_2^2}
+\frac{c_7}{2}f^2 \frac{k_{3z}^4}{k^2_3}\frac{2}{7}\G(\k_1,\k_2)~\,.
\end{split}
\ee 
Note that this expression is completely 
equivalent to that of \cite{DAmico:2022osl} (see e.g., Eqs.~5.42, 5.43 and App.~D~2).\footnote{To prove this, a useful identity is 
\be 
\frac{\k_{12}^i \k_{12}^j}{2k_{12}^2}(\k_1\cdot\k_2)\left(
\frac{1}{k_1^2}
+\frac{1}{k_2^2}
\right)=\frac{(\k_1\cdot\k_2)(k_1^ik_1^j+k_2^ik_2^j)}{2k_1^2k_2^2}+
\frac{(\k_1\cdot\k_2)(k_1^ik_2^j+k_2^ik_1^j)}{2k_1^2k_2^2}-\frac{\k_{12}^i \k_{12}^j}{k_{12}^2}\frac{(\k_1\cdot\k_2)^2}{k_1^2k_2^2}
\ee 
where $\k_{12}\equiv \k_1 + \k_2$ (and $k_{12}$ its length) which leads to 
\be 
t_{ij}^{(2)}=\frac{\d_i\d_j\delta_2}{\d^2}=\frac{\d_i\d_j\d_k\delta_1}{\d^2}
\frac{\d_k\delta_1}{\d^2}
+\frac{\d_i\d_l\delta_1}{\d^2}
\frac{\d_j\d_l\delta_1}{\d^2}
+\frac{5}{2}\d_i\d_j(\Phi-\Phi_v)\,.
\ee } 
Combining this with the real-space counterterms we end up with 14 
free counterterm parameters that should be fitted from the data.
The degeneracies in the above formula are eliminated 
with the following identities between the counterterm kernels
\be
\begin{split}
& F^{\rm ctr.~(s)}_{2,c_1}=f\left(F^{\rm ctr.~(s)}_{2,c_5}-\frac{1}{2}F^{\rm ctr.~(s)}_{2,e_1}\right)~\,, \quad F^{\rm ctr.~(s)}_{2,c_3}=-\frac{f}{2}F^{\rm ctr.~(s)}_{2,e_3}\,,\quad F^{\rm ctr.~(s)}_{2,c_6}=-\frac{f}{2}F^{\rm ctr.~(s)}_{2,e_2}\,.
\end{split}
\ee 

Finally, following~\cite{Ivanov:2019pdj,Chudaykin:2020hbf}
(see also
\cite{Lewandowski:2015ziq}),
we include a single higher order
redshift-space 
counterterm $\tilde{c}$
in the power spectrum 
theory prediction. It is defined as a contribution 
to eq.~\eqref{eq:Z1s}: 
\be 
Z_1^{\rm ctr.~(s)}\to Z_1^{\rm ctr.~(s)}-\frac{\tilde{c}}{2}
\frac{k^4}{k_{\rm NL}^4}
\mu^4 f^4 (b_1+f\mu^2)\,.
\ee 
This contribution is necessary in order to account for the large
velocity dispersion
of luminous red galaxies~\cite{Ivanov:2021fbu,Ivanov:2024xgb,Ivanov:2024dgv}. 
We consistently propagate 
this contribution to the relevant 
mixed stochastic bispectrum terms. 

\subsection{Stochasticity}

\subsubsection{Real-Space}
\noindent 
Next, we consider the stochastic contributions to the bispectrum, \textit{i.e.}\ those depending on the galaxy density, $\bar{n}$. In real-space we have the combination of 
pure stochastic and mixed stochastic-deterministic operators,
\be \label{eq: stoch-expan}
\begin{split}
& \delta_g\supset \delta^{\rm stoch}_g +   \delta^{\rm mix~ (r)}_{g}\,,\\
& \delta^{\rm stoch}_g  = \epsilon \,,\\
& 
\delta^{\rm mix~(r)}_{g}  = \frac{\d_i\delta}{\d^2}\d_i\epsilon + \frac{1}{2}\frac{\d_i\delta}{\d^2}\frac{\d_j\delta}{\d^2}\d_i\d_j\epsilon 
+ \frac{1}{6}\frac{\d_i\delta}{\d^2}\frac{\d_j\delta}{\d^2}\frac{\d_k\delta}{\d^2}\d_i\d_j\d_k\epsilon 
\\
&+\left(\epsilon+\d_k\epsilon\frac{\d_k\delta}{\d^2}
+\frac{1}{2}\d_k\d_m \epsilon\frac{\d_k\delta}{\d^2}
\frac{\d_m\delta}{\d^2}
\right)(d_1 \delta + 
\frac{d_2}{2}\delta^2 + 
 d_{\G} \G
+d_{\Gamma_3}   \Gamma_3) \\
& +\epsilon_{ij}t^{ij} 
+  \epsilon^{ijk}\d_k t_{ij} + \epsilon^{ijkl}\d_i\d_j t_{kl} ...\,,
\end{split}
\ee
where $\epsilon,\epsilon_i,\epsilon_{ij}$ \textit{et cetera} are stochastic fields 
that are uncorrelated with the
linear matter density and $d_i$ are Wilson coefficients.
At the level of the one-loop bispectrum, 
$\epsilon \delta$ produces contributions degenerate with those of 
$\epsilon_{ij}t^{ij}$; we keep the 
$\epsilon \delta$ operator explicitly however, because it makes
the discussion of the higher order corrections, such 
as $\epsilon \delta^2$, more transparent. Moreover, $\epsilon_{ij}t^{ij}$ gives a non-degenerate 
contribution only at $\mathcal{O}(k^2)$. 
Beyond these terms, we will focus only on the operators that produce non-degenerate contributions -- for this reason, we have not included operators such as $\epsilon^{ij}\d_i\d_j \delta$, $\epsilon^i \d_i \delta$.
The IR terms in \eqref{eq: stoch-expan} such as $\d_i \epsilon\frac{\d_i\delta}{\d^2}$ are the flow terms
whose origin is best understood
in the context of the Lagrangian
bias expansion, see~\cite{Desjacques:2016bnm}
for an original derivation
and discussions. These do not contain free coefficients, since their structure is dictated by the equivalence 
principle; furthermore, since they are produced by 
the Zel'dovich linear displacements,
we can equivalently 
replace $\delta\to \delta_1$ in their expressions. Note that we have not included the 2LPT displacements 
in the above flow terms because they lead to trivial redefinitions of the
quadratic 
bias parameters $d_{\G},d_2$
at the one-loop bispectrum order.

Let us start with the
the pure stochastic terms. 
In EFT various correlators of the
$\epsilon_{i_1...i_n}$ fields can be expressed
via Taylor series in $k^2$. For $\epsilon$ this gives
\be 
\label{eq:eps_corr0}
\langle \epsilon(\k)\epsilon(\k')\rangle'=\frac{1}{\bar n}\left(
1+P_{\rm shot}
+a_0 \frac{k^2}{k_{\rm NL}^2} + O(k^4)
\right)\,,
\ee 
where $P_{\rm shot},a_0$ are Wilson coefficients, whose amplitudes are zero for pure Poissonian stochasticity. Likewise, for the pure stochastic 
bispectrum 
component, we require only a symmetry-based
expansion in $k^2$:
\be \label{eq: Bk-stoch-pure}
B_{\rm stoch}(\k_1,\k_2,\k_3)=\frac{1}{\bar n^2}\left(A_{\rm shot}+a_1\frac{1}{k_{\rm NL}^2}(k_1^2+k_2^2+k_3^2)+O(k^4)\right)\,,
\ee 
for free $A_{\rm shot},a_1$ with $A_{\rm shot}=1, a_1=0$ in the Poisson limit. 

{To derive the expressions for the stochastic correlators in general, we use the following symmetry-based EFT expansions which generalize eq.~\eqref{eq:eps_corr0}:
\be 
\begin{split}
&\langle \epsilon(\k)\epsilon_{ij}(\k)\rangle'=\delta_{ij} a'_0 +
\frac{k_ik_j}{k_{\rm NL}^2} a'_1 + \frac{k^2}{k_{\rm NL}^2}\delta_{ij} a'_2+... \\
& \langle \epsilon_{ij}(\k')\epsilon_{kl}(\k) \rangle'= \delta_{ij}\delta_{kl}\, a''_0 
+ (\delta_{ik}\delta_{jl}+\delta_{il}\delta_{jk})\, a''_1 
+ \frac{1}{k_{\rm NL}^2}
\Big[
(k_ik_j\delta_{kl}+k_kk_l\delta_{ij})\,a''_2 \\
&\quad \quad \quad 
+(k_ik_k\delta_{jl}+k_ik_l\delta_{jk}+k_jk_k\delta_{il}+k_jk_l\delta_{ik})\,a''_3 
+ k^2\delta_{ij}\delta_{kl}\,a''_4
+ k^2(\delta_{ik}\delta_{jl}+\delta_{il}\delta_{jk})\,a''_5
\Big]+\ldots \\
&\langle \epsilon(\k')\epsilon_{ijk}(\k)\rangle'=i\frac{1}{k_{\rm NL}}(\delta_{ij}k_k+\delta_{jk}k_i+\delta_{ik}k_j )b'_0 +... \\
& \langle \epsilon(\k)\epsilon_{ijkl}(\k)\rangle'=(\delta_{ij}\delta_{kl}+\delta_{ik}\delta_{jl}+\delta_{il}\delta_{jk}) a'''_0 +
\frac{1}{k_{\rm NL}^2}(k_ik_j\delta_{kl}+
k_ik_l\delta_{jk}+k_ik_k\delta_{jl}+k_{j}k_k\delta_{il}+k_jk_l\delta_{ik}+k_kk_l\delta_{ij}) a'''_1 \\
&+ \frac{k^2}{k_{\rm NL}^2}(\delta_{ij}\delta_{kl}+\delta_{ik}\delta_{jl}+\delta_{il}\delta_{jk}) a'''_2+...\, \\
\end{split}
\ee
In redshift space, there are additional 
terms involving the stochastic vector field
$\epsilon_i$. For completeness, we present their correlators relevant for the one-loop calculation below:
\be 
\begin{split}
&\langle \epsilon(\k')\epsilon_j(\k) \rangle'=i \frac{k^j}{k_{\rm NL}} b''_0+...\,,\\
&\langle \epsilon_i(\k')\epsilon_j(\k) \rangle'=\delta_{ij}b'''_0+\frac{k^2}{k_{\rm NL}^2}\delta_{ij}b'''_1+\frac{k_ik_j}{k_{\rm NL}^2}b'''_2+...\,,\\
&\langle \epsilon_i(\k')\epsilon_{jk}(\k) \rangle'= i \frac{b''''_0}{k_{\rm NL}} k_i \delta_{jk}+i \frac{b''''_1}{k_{\rm NL}} (k_j \delta_{ik}+k_k \delta_{ij})+...\,,
\end{split}
\ee 
where $a'_n, b'_n, a''_n, b''_n$
are $k$-independent constants. }

Next, consider the mixed terms, \textit{i.e.}\ those involving $\epsilon\delta$ and beyond. For the one-loop power spectrum, the mixed terms produce 
contributions that can be absorbed into 
the tree-level
EFT parameters. For the bispectrum 
at the tree-level we have~\cite{Ivanov:2021kcd}
\be 
\begin{split}\label{eq: Bk-stoch-mixed-tree}
B^{\rm tree}_{\rm mixed}(\k_1,\k_2,\k_3)
&=
\langle K_1\delta_1(\k_1)\cdot\epsilon(\k_2)\cdot d_1[\epsilon\delta]_{\k_3} \rangle'+\langle K_1\delta_1(\k_1)\cdot d_1[\epsilon\delta]_{\k_2}\cdot \epsilon(\k_3) \rangle' + \text{2 cyc.}\\
& =2d_1b_1\langle |\epsilon|^2\rangle' (P_{11}(k_1)+
P_{11}(k_2)+
P_{11}(k_3))\\
&=\frac{B_{\rm shot}}{\bar n}(P^{\rm tree}_{gg}(k_1)+ \text{2 cyc.})\,,
\end{split}
\ee 
where $b_1 B_{\rm shot}\equiv 2d_1(1+P_{\rm shot})$ and $K_1 = b_1$. 

The situation becomes
more interesting 
at one-loop
order. 
First, let us first ignore the flow terms. At one-loop order, one can have mixed 13- and 22-type stochasticity contributions stemming from correlators such as 
\be 
\label{eq:1l_mix}
\langle[\delta^2]_{\k_1}\cdot \epsilon(\k_2)\cdot [\epsilon \delta^2]_{\k_3} \rangle\,,\quad 
\langle 
 \delta(\k_1)\cdot\epsilon(\k_2)\cdot[\epsilon \delta^3]_{\k_3}
\rangle\,,\quad 
\langle 
[\delta^3]_{\k_1}\cdot\epsilon(\k_2)\cdot [\epsilon \delta]_{\k_3} 
\rangle~.
\ee
Taken together they give
\be
\begin{split}\label{eq: Bk-stoch-mixed-loop}
& B^{\rm 1-loop~(r)}_{\rm mixed}(\k_1,\k_2,\k_3)=\frac{2(1+P_{\rm shot})}{\bar n}\Bigg(d_1b_1 P_{\rm 1-loop}^{\rm mm}(k_1)
+ \frac{d_2b_2}{4} I_{\delta^2\delta^2}(k_1)
+(d_1 b_2+d_2 b_1)\frac{1}{2}I_{\delta^2}(k_1)\\
& + (d_1 b_{\G}+d_{\G} b_1) I_{\G}(k_1) + (d_{\G}b_2  + b_{\G}d_2)\frac{1}{2} I_{\delta^2\G}(k_1) + d_{\G} b_{\G} I_{\G\G}(k_1) \\
& 
+(\frac{2}{5} d_{\Gamma_3}b_1 + 
\frac{2}{5} b_{\Gamma_3}d_1 +
d_1b_{\G}
+b_1 d_{\G}
)\mathcal{F}_{\G}(k_1)+
\text{2 cyc.}\Bigg)\,,
\end{split}
\ee 
utilizing the one-loop tracer power spectrum notation of~\cite{Assassi:2014fva}. This term takes the form of the Poissonian contribution 
involving 
a one-loop
cross-spectrum
of two tracers
with two bias parameter sets:
$\{b_1,b_2,b_{\G},b_{\Gamma_3}\}$
and 
$\{d_1,d_2,d_{\G},d_{\Gamma_3}\}$.

To streamline the calculation
of the mixed one-loop components including the flow terms, 
it is convenient to rewrite the $\delta^{\rm mix,(r)}_{g}$ terms involving operators 
scaling like $k^0\epsilon\delta^n$ (for $n=1,2,3$) in a form 
similar to perturbative expressions in standard perturbation theory:
\be \label{eq: delta-g-mix}
\delta_{g}^{\text{mix (r)}}(\k)\Big|_{O(\epsilon\delta^n)}
=\sum_{n=1}\left[\prod_{i=1}^n \int_{\q_i}\delta_1(\q_i)\right]\int_{\q_{n+1}}\epsilon(\q_{n+1})(2\pi)^3\delta_D^{(3)}(\k-\q_{1...n+1})\mathcal{K}_n(\q_1,...,\q_{n+1})~\,.
\ee 
The new kernels $\mathcal{K}_n(\q_1,...,\q_{n+1})$
are very similar to the usual 
bias kernels $K_n(\q_1,...,\q_{n})$ from \eqref{eq: delta-n-exp}.
In particular, they satisfy 
$\mathcal{K}_n(\q_1,...,\q_n,\q_{n+1}=0)=K_n(\q_1,...,\q_n)$.
The mixed
one-loop
deterministic-stochastic
contribution
can be written as 
\be\label{eq:mixed1loopstoch_r}
\begin{split}
& B^{\rm 1-loop~(r)}_{\rm mixed}(\k_1,\k_2,\k_3)=\frac{1+P_{\rm shot}}{\bar n}(\mathcal{P}_{22}^{\text{(r)}}(k_1,k_2) + \mathcal{P}^{I~\text{(r)}}_{13}(k_1,k_2) + \mathcal{P}^{II~\text{(r)}}_{13}(k_1,k_2) +\text{5 perms} )\,,\\
& \mathcal{P}_{22}^{\text{(r)}}(k_1,k_2) = 2\int_{\q} \mathcal{K}_2(\q,\k_1-\q,\k_2)K_2(-\q,-\k_1+\q)P_{11}(q)P_{11}(|\k_1-\q|)\,,\\
& \mathcal{P}^{I~\text{(r)}}_{13}(k_1,k_2) = 3K_1(\k_1) P_{11}(\k_1)\int_{\q} \mathcal{K}_3(\k_1,\q,-\q,\k_2)P_{11}(q) \,,\\
& \mathcal{P}^{II~\text{(r)}}_{13}(k_1,k_2) = 3\mathcal{K}_1(\k_1,\k_2) P_{11}(\k_1)\int_{\q} K_3(\k_1,\q,-\q)P_{11}(q)\,.
\end{split}
\ee 
The inclusion of the flow terms simply leads to unobservable 
shifts of the mixed bias operators, e.g., $d_1\to d_1-1/2$, \textit{i.e.}\ the \eqref{eq:mixed1loopstoch_r} terms can be absorbed by redefining $d_{\mathcal{O}}$. In what follows $d_{\mathcal{O}}$ will refer to the shifted parameters.

Finally, let us discuss the $\mathcal{O}(k^2P_{11}/\bar n)$ terms.
One correction of this type stems from 
the shift term $\frac{\d_i\delta}{\d^2}\d_i\epsilon$, with $\langle \epsilon^2 \rangle$ expanded to order $k^2$. This generates 
\be 
\begin{split}
& B_{\rm mixed}^{\rm ctr.~(r)~I}(\k_1,\k_2,\k_3)\supset -\frac{a_0 K_1(\k_1)P_{11}(k_1)}{\bar n}\left[
\frac{k_1^2(k_2^2+k_3^2)-(k^2_2-k_3^2)^2
}{2k_1^2k_{\rm NL}^2}
\right]+\text{2 cyc.}\,..
\end{split}
\ee 
Two more contributions can be obtained
from the $\epsilon^{ij}t_{ij}$ term in \eqref{eq: stoch-expan} expanded to $\mathcal{O}(k^2)$
as described in~\cite{Philcox:2022frc}.
Combining these with contributions from the $\epsilon^{ijkl}\d_i\d_jt_{kl}$ term produces the following 
cumulative contribution
\be \label{eq: Bk-stoch-derivI}
B_{\rm mixed}^{\rm ctr.~(r)~I}(\k_1,\k_2,\k_3)=\sum_{n=1}^4 s_n F^{(n)}_{k^2\bar n^{-1}P}(\k_2,\k_3) \frac{K_1(\k_1)P_{11}(k_1)}{\bar n}+\text{2 cyc.}\,,
\ee 
where $s_n=\{a_0,a_3,a_4,a_5\}$ are Wilson coefficients. This defines the kernels
\be 
\label{eq:real_space_mix_kern}
\begin{split}
& F^{(1)}_{k^2\bar n^{-1}P}(\k_2,\k_3)=-\frac{k_1^2(k_2^2+k_3^2)-(k^2_2-k_3^2)^2
}{2k_1^2k_{\rm NL}^2}\,, \quad 
 F^{(2)}_{k^2\bar n^{-1}P}(\k_2,\k_3)=-\frac{k_1^4 + (k_2^2-k_3^2)^2}{2k_1^2k_{\rm NL}^2} \\
& F^{(3)}_{k^2\bar n^{-1}P}(\k_2,\k_3)=-\frac{k_1^2}{k_{\rm NL}^2}\,, \quad \quad \quad 
 F^{(4)}_{k^2\bar n^{-1}P}(\k_2,\k_3)=-\frac{(k_2^2+k_3^2)}{k_{\rm NL}^2}\,.
\end{split}
\ee 
Note that the counterterm 
$F^{(1)}_{k^2\bar n^{-1}P}$
is not independent: it is a linear 
combination of $F^{(2)}_{k^2\bar n^{-1}P}$, $F^{(3)}_{k^2\bar n^{-1}P}$, and $F^{(4)}_{k^2\bar n^{-1}P}$,
\be 
F^{(1)}_{k^2\bar n^{-1}P}=\frac{1}{2}F^{(4)}_{k^2\bar n^{-1}P}+
\frac{1}{2}F^{(3)}_{k^2\bar n^{-1}P}-F^{(2)}_{k^2\bar n^{-1}P}\,.
\ee 
This degenerancy is broken once redshift-space distortions are taken into account, as discussed in the next section. Finally, there will also be a contribution generated by contracting $\delta_{\rm mix}, \epsilon$
and the power spectrum counterterms,
\be \label{eq: Bk-stoch-derivII}
\begin{split}
B_{\rm mixed}^{\rm ctr.~(r)~II}(\k_1,\k_2,\k_3)&=\langle K_1^{\rm ctr~(r)}\delta_1(\k_1)\cdot\epsilon(\k_2)\cdot\delta_{\rm mix}(\k_3)\rangle'+ 
\langle K_1^{\rm ctr~(r)}\delta_1(\k_1)\cdot\delta_{\rm mix}(\k_2)\cdot\epsilon(\k_3)
\rangle'\\
&=\frac{1}{\bar{n}}\frac{k_1^2}{k_{\rm NL}^2}(b_1 B_{\rm shot})\left(-b_{\nabla^2\delta}\right)P_{11}(k_1)+\text{2 cyc.}\,,
\end{split}
\ee 
which does not introduce any new parameters.

In summary, we find that the real-space bispectrum has four pure stochastic counterterms with free parameters
$\{P_{\rm shot},a_0,A_{\rm shot},a_1\}$ \eqref{eq: Bk-stoch-pure}, one mixed tree-level stochastic counterterm $B_{\rm shot}$ \eqref{eq: Bk-stoch-mixed-tree}, three mixed one-loop `bias'
parameters $d_2,d_{\G},d_{\Gamma_3}$ \eqref{eq: Bk-stoch-mixed-loop}, and three mixed counterterms with free parameters
$\{a_3,a_4,a_5\}$ scaling as $B\sim k^2 P^2_{11}$ (\ref{eq: Bk-stoch-derivI} and \ref{eq: Bk-stoch-derivII}). Noting that $d_1$ is determined by $B_{\rm shot}$ and $P_{\rm shot}$, this yields 11 free coefficients.

\subsubsection{Redshift-Space}
\noindent The redshift-space stochastic contributions arise 
from the renormalization of the local operators appearing in \eqref{eq: velocity-renorm}. At leading order,
\be 
\label{eq:stoch_cont}
\begin{split}
& k_{\rm NL}^2\mathcal{O}^i_u \supset \epsilon^i + \d_k \epsilon^i \frac{\d_k\delta}{\d^2} 
+\tilde{a}_1\epsilon^{ijk}\d_i\d_j\Phi 
+ \tilde{a}_2\epsilon^{ijkl}\d_i\d_j\d_l \Phi
+ \tilde{a}_3 \frac{\d_i\d_j\d_k}{\d^2}
\left(\epsilon^{jl} \frac{\d_l\d_k\delta}{\d^2}\right)\\
&k_{\rm NL}^2\mathcal{O}^{ij}_{u^2} \supset\epsilon^{ij} + \d_k \epsilon^{ij} \frac{\d_k\delta}{\d^2}+\tilde{a}_4 \epsilon^{ijkl}\d_k\d_l\Phi~\,,
\end{split}
\ee 
keeping only the non-degenerate operators relevant at one-loop order. For the purely stochastic components we find
\be 
\begin{split}
& P_{\rm stoch}(k,\mu) = \frac{1}{\bar n}\left(1+P_{\rm shot}+a_0\frac{k^2}{k^2_{\rm NL}} +a_2\mu^2 \frac{k^2}{k^2_{\rm NL}} \right)~\,,\\
& B_{\rm stoch}(\k_1,\k_2,\k_3) = \frac{1}{\bar n^2}\left(
A_{\rm shot} + \frac{1}{k_{\rm NL}^2}(a_1(k_1^2+k_2^2+k_3^2)+a_6(k_{1z}^2+k_{2z}^2+k_{3z}^2))
\right)\,,
\end{split}
\ee 
where $a_2$ and $a_6$ are new parameters encoding the anisotropic stochasticity. The latter term is equivalent to $k_{1z}k_{2z} + \text{cyc.}$ 
by virtue of 
momentum conservation.

As before, we also find mixed terms, which take the form
\be 
\label{eq:Bmixs}
\begin{split} 
& B_{\rm mix}(\k_1,\k_2,\k_3) = \langle \delta_{\rm det}(\k_1)\cdot\delta_\epsilon(\k_2)\cdot\delta_{\rm mix}(\k_3) \rangle'+ 
\langle \delta_{\rm det}(\k_1)\cdot\delta_{\rm mix}(\k_2)\cdot\delta_\epsilon(\k_3)
\rangle'+\text{2 cyc.}~\,,
\end{split}
\ee 
where $\delta_{\rm det}\equiv\sum_n Z_n \delta_1^n$
is the deterministic contribution, 
while $\delta_\epsilon
\equiv\{\epsilon,-if k_{\rm NL}^{-2} k_zz^i\epsilon_{i},-\tfrac{1}{2}f^2 
k_{\rm NL}^{-2} 
k_z^2z^iz^j\epsilon_{ij}\}$ is the purely stochastic field. 
For the tree-level mixed terms $\delta_{\rm det}=\delta^{(s)}_{g,(1)}$, yielding 
\be\label{eq:mixedtree_s}
B^{\rm tree~(s)}_{\rm mixed}(\k_1,\k_2,\k_3) = \frac{(b_1 B_{\rm shot}+(1+P_{\rm shot})f\mu^2_1)}{\bar n}Z_1(\k_1)P_{11}(k_1)+\text{2 cyc.}\,,
\ee 
which matches \cite{Ivanov:2021kcd}.
To compute the mixed one-loop corrections, we insert
$\delta^{\rm mix}_g$ \eqref{eq: delta-g-mix} into the
redshift-space mapping, which gives 
\be
\label{eq:RSDexp_mix}
\begin{split}
\delta_\k^{\text{mix (s)}} = \delta_\k^{\text{mix (r)}}  -ifk_z[ \delta^{\text{mix (r)}}u_z]_\k
+\frac{i^2f^2}{2}k_z^2 [ \delta^{\text{mix (r)}}u^2_z]_\k -\frac{i^3f^3}{3!} k_z^3[ \delta^{\text{mix (r)}}u^3_z]_\k 
+ \frac{i^4f^4}{4!} k_z^4 [ \delta^{\text{mix (r)}}u_z^4]_\k \,.
 \end{split} 
\ee
Keeping terms of order 
$k^0\epsilon\,\delta^n$
up to $n=3$
we can rewrite the above expression as
(noting that only $\epsilon$ contributions are relevant at the one-loop order)
\be 
\delta_{g}^{\text{mix (s)}}(\k)\Big|_{\mathcal{O}(\epsilon\delta^n)}
=\sum_{n=1}\left[\prod_{i=1}^n \int_{\q_i}\delta_1(\q_i)\right]\int_{\q_{n+1}}\epsilon(\q_{n+1})(2\pi)^3\delta_D^{(3)}(\k-\q_{1...n+1})\mathcal{Z}_n(\q_1,...,\q_{n+1})~\,,
\ee 
where the new SPT-like kernels
are
\be 
\begin{split}
\mathcal{Z}_1(\k_1,\k_2) &  =d_1 + f\frac{k_{12z}k_{1z}}{k_1^2}\\
 \mathcal{Z}_2(\k_1,\k_2,\k_3) &=d_1 F^{\rm SPT}_2(\k_1,\k_2)+f\frac{k_{123z}k_{12z}}{k_{12}^2}G_2^{\rm SPT}(\k_1,\k_2)+\frac{fk_{123z}d_1}{2}\left(\frac{k_{1z}}{k_1^2}+\frac{k_{2z}}{k_2^2}\right)\\
 & ~~~~~ + \frac{f^2 k_{123z}^2 }{2}\frac{k_{1z}}{k_1^2}\frac{k_{2z}}{k_2^2}+\frac{d_2}{2}+d_{\G} \G(\k_1,\k_2) \,,
\end{split}
\ee 
where $F_n^{\rm SPT}$
and $G_n^{\rm SPT}$
are density and velocity 
kernels in standard perturbation theory~\cite{Bernardeau:2001qr}, while 
for the $\mathcal{Z}_3$
kernel we have the following expression
that has to be symmetrized over
$\{\k_1,\k_2,\k_3\}$:
\be 
\begin{split}
& \mathcal{Z}_3(\k_1,\k_2,\k_3,\k_4) = d_1F^{\rm SPT}_3(\k_1,\k_2,\k_3)
+f \frac{k_{1234z}k_{123z}}{k_{123}^2}G^{\rm SPT}_3(\k_1,\k_2,\k_3)
+f k_{1234z}\frac{k_{1z}}{k_1^2}d_1 F_2^{\rm SPT}(\k_2,\k_3)
\\
&+ f^2\frac{k_{1234z}^2k_{1z}}{k_1^2}G_2^{\rm SPT}(\k_2,\k_3)\frac{k_{23z}}{k_{23}^2}+fk_{1234z}d_1\frac{k_{23z}}{k_{23}^2}G_2^{\rm SPT}(k_2,\k_3)
+d_1\frac{f^2k_{1234z}^2}{2}
\frac{k_{1z}k_{2z}}{k_1^2k_2^2}\\
&+\frac{f^3k_{1234z}^3}{6}
\frac{k_{1z}}{k_1^2}
\frac{k_{2z}}{k_2^2}
\frac{k_{3z}}{k_3^2}+d_2F_2^{\rm SPT}(\k_2,\k_3) +2 d_{\G}\G(\k_2+\k_3,\k_1)F_2^{\rm SPT}(\k_2,\k_3) +d_2 \frac{fk_{1234z}k_{1z}}{2k_1^2}\\
& + d_{\G} fk_{1234z}
\G(\k_2,\k_3)\frac{k_{1z}}{k_1^2}+2d_{\Gamma_3}\G(\k_2+\k_3,\k_1)(F^{\rm SPT}_2(\k_2,\k_3)-G_2^{\rm SPT}(\k_2,\k_3))~\,.
\end{split}
\ee 
These kernels
satisfy 
$\mathcal{Z}_n(\q_1,...,\q_n,\q_{n+1}=0)=Z_n(\q_1,...,\q_n)$. 
Note that $\mathcal{Z}_n(f=0) = \mathcal{K}_n$.
The mixed
one-loop
deterministic-stochastic
contribution
can be written as 
\be\label{eq:mixed1loopstoch_s}
\begin{split}
& B_{\rm mixed}^{\rm 1-loop~(s)}(\k_1,\k_2,\k_3)=\frac{1+P_{\rm shot}}{\bar n}(\mathcal{P}_{22}^{\text{(s)}}(k_1,k_2) + \mathcal{P}^{I~\text{(s)}}_{13}(k_1,k_2) + 
\mathcal{P}^{II~\text{(s)}}_{13}(k_1,k_2) +\text{5 perms} )\,,\\
& \mathcal{P}^{\text{(s)}}_{22}(k_1,k_2) = 2\int_{\q} \mathcal{Z}_2(\q,\k_1-\q,\k_2)
Z_2(-\q,-\k_1+\q)P_{11}(q)P_{11}(|\k_1-\q|)\,,\\
& \mathcal{P}^{I~\text{(s)}}_{13}(k_1,k_2) = 3Z_1(\k_1) P_{11}(\k_1)\int_{\q} \mathcal{Z}_3(\k_1,\q,-\q,\k_2)P_{11}(q) \,,\\
& \mathcal{P}^{II~\text{(s)}}_{13}(k_1,k_2) = 3\mathcal{Z}_1(\k_1,\k_2) P_{11}(\k_1)\int_{\q} Z_3(\k_1,\q,-\q)P_{11}(q)\,.
\end{split}
\ee 
The above loop corrections are computed using FFTLog along the lines of~\cite{Chudaykin:2020aoj}. 
Note that the contribution above is 
\textit{different} from the naive
Poissonian expectation~\cite{1980lssu.book.....P} 
\be 
\label{eq:poiss_mix}
\frac{1}{\bar n}\left(P^{\rm 1-loop}_{gg}(k_1,\mu_1)+P^{\rm 1-loop}_{gg}(k_2,\mu_2)+P^{\rm 1-loop}_{gg}(k_3,\mu_3)\right)
\ee 
where $P^{\rm 1-loop~(s)}_{gg}$
is the one-loop power spectrum.
Eq.~\eqref{eq:poiss_mix} can be obtained 
from eq.~\eqref{eq:mixed1loopstoch_s} if we 
formally substitute $k_{2z}\to -k_{1z}$
in $\mathcal{P}_{22},\mathcal{P}^{I~(s)}_{13},\mathcal{P}^{II~(s)}_{13}$.
However, for actual kinematic configurations
that appear in the bispectrum this 
substitution does not take place.
First, the loops 
with the linear dependence on the external momentum $k_{2z}$
add up to $k_{2z}+k_{3z}=-k_{1z}$
resulting in corrections 
twice smaller than those appearing in $P^{\rm 1-loop~(s)}$.
On top of that, the terms quadratic in $ k_{2z}$
generate contributions $\sim k^2_{2z}+k^2_{3z}\neq k^2_{1z}$.
However, for the 
bispectrum monopole, 
in the limit 
$P_{\rm shot}\to 0$,
$B_{\rm shot}\to 1$,
$2d_{\mathcal{O}}\to b_{\mathcal{O}}$
all the new terms combined are numerically similar  
to the naive Poissonian result 
to an accuracy better
than one percent. 
This is unsurprising given that the result is exact in real-space, 
and implies that one could 
expand $B_{\rm mixed}^{\rm 1-loop~(s)}$ around
the Poisson limit in practice.

Finally, we require the terms at order $k^2P\bar n^{-1}$. The first type are
obtained by computing various 
contractions of the terms from~\eqref{eq:stoch_cont} in~\eqref{eq:Bmixs}
and expending the stochastic correlators to the appropriate orders in $k$, yielding: 
\be 
B_{\rm mixed}^{\rm ctr.~(s)~I}(\k_1,\k_2,\k_3)=\sum_{n=1}^{10} s_n Z^{(n)}_{k^2\bar n^{-1}P}(\k_2,\k_3) \frac{1}{\bar{n}}Z_1(\k_1)P_{11}(k_1)+\text{2 cyc.}\,,
\ee 
where $s_n=\{a_0,a_3,a_4,a_5,a_7,...,a_{12}\}$ (note that $a_1,a_2,a_6$ are absent in the list) are 10 Wilson coefficients and we define the kernels
\be 
\begin{split}
&  Z^{(1)}_{k^2\bar n^{-1}P}(\k_2,\k_3)=-\frac{k_1^2(k_2^2+k_3^2)-(k^2_2-k_3^2)^2
+2fk_{1z}(k_{3z}k_2^2+k_{2z}k_3^2)
}{2k_1^2k_{\rm NL}^2}\,, \\
& 
 Z^{(2)}_{k^2\bar n^{-1}P}(\k_2,\k_3)=-\frac{k_1^4 + (k_2^2-k_3^2)^2}{2k_1^2k_{\rm NL}^2}\,, \quad Z^{(3)}_{k^2\bar n^{-1}P}(\k_2,\k_3)=-\frac{k_1^2}{k_{\rm NL}^2}\,,\quad 
 Z^{(4)}_{k^2\bar n^{-1}P}(\k_2,\k_3)=-\frac{(k_2^2+k_3^2)}{k_{\rm NL}^2}\,,\\
 & Z^{(5)}_{k^2\bar n^{-1}P}(\k_2,\k_3)=f\frac{\left[(k_3^2-k_1^2-k_2^2)k_{2z}^2+(k_2^2-k_3^2-k_1^2)k^2_{3z}+2k_{3z}k_{2z}k_1^2-2f k^2_{1z}k_{2z}k_{3z}\right]}{2k_1^2k_{\rm NL}^2}\,,\\
 &  
 Z^{(6)}_{k^2\bar n^{-1}P}(\k_2,\k_3)=\frac{f^2}{2k_{\rm NL}^2}k^2_{1z}(1-f\mu_1^2)\,,\quad Z^{(7)}_{k^2\bar n^{-1}P}(\k_2,\k_3)=f\frac{k_{1z}^2}{k_{\rm NL}^2}\,,\\
 & Z^{(8)}_{k^2\bar n^{-1}P}(\k_2,\k_3)=f\frac{k_{2z}k_{3z}}{k_{\rm NL}^2}\,,\quad 
Z^{(9)}_{k^2\bar n^{-1}P}(\k_2,\k_3)=
\frac{f}{4k_1^2k_2^2k_3^2k_{\rm NL}^2}
\left(
(k_2^2-k_3^2-k_1^2)^2 k_2^2 k^2_{3z}
+(k_3^2-k_2^2-k_1^2)^2 k_3^2 k^2_{2z}
\right) \,,\\
& Z^{(10)}_{k^2\bar n^{-1}P}(\k_2,\k_3)=\frac{f}{k_{\rm NL}^2}\left(\frac{k_{1z}}{k_1^2}\left(
k_{1z}(k_1^2+k_2^2-k_3^2)
+2k_{2z}(k_2^2-k_3^2)
\right)\right)\,.
\end{split}
\ee 
Note that naively there should be one more term, $\propto k_{2z}k_{3z} f^2(1-f\mu_1^2)$, but 
it appears degenerate with the other operators by virtue of
\be 
\frac{k_{2z}k_{3z}}{k_{\rm NL}^2}f^2(1-f\mu_1^2)=f\left(Z^{(5)}_{k^2\bar n^{-1}P}+Z^{(7)}_{k^2\bar n^{-1}P}-Z^{(8)}_{k^2\bar n^{-1}P}-\frac{1}{2}Z^{(10)}_{k^2\bar n^{-1}P}\right)\,.
\ee 

Note that the combination of our $B_{\rm mixed}^{\rm ctr.~(s)~I}$
and $B_{\rm mixed}^{\rm tree~(s)}$ from \eqref{eq:mixedtree_s}
(which amounts to 12 free coefficients in total) is equivalent to 
$B_{321}^{r,h,(I),\epsilon}$
from~\cite{DAmico:2022ukl} (12 free coefficients). 

The second type of mixed $O(k^2P\bar n^{-1})$ terms are given by contracting $\delta_{\rm mix}$ with $\epsilon$ as before: 
\be 
\begin{split}
B_{\rm mixed}^{\rm ctr.~(s)~II}(\k_1,\k_2,\k_3)&=\langle Z_1^{\rm ctr~(s)}(\k_1)\delta(\k_1)\cdot\delta_\epsilon(\k_2)\cdot\delta_{\rm mix}(\k_3)\rangle'+ 
\langle Z_1^{\rm ctr~(s)}(\k_1)\delta(\k_1)\cdot\delta_{\rm mix}(\k_2)\cdot\delta_\epsilon(\k_3)
\rangle'+\text{2 cyc.}\\
&=\frac{1}{\bar{n}}\frac{k_1^2}{k_{\rm NL}^2}(b_1 B_{\rm shot}+f\mu_1^2(1+P_{\rm shot}))\\
&\quad \quad \times \left(-b_{\nabla^2\delta}+\left(e_1-\frac{1}{2}c_1 f\right)f\mu_1^2 -\frac{1}{2}c_2 f^2 \mu_1^4-\frac{\tilde{c}}{2}\frac{k^2_1}{k_{\rm NL}^2}f^4\mu^4_1(b_1+f\mu_1^2)\right)P_{11}(k_1)+\text{2 cyc.}\,.
\end{split}
\ee 
These terms involve only on the free parameters that we have introduced before, 
$P_{\rm shot},B_{\rm shot}$ and $\{b_{\nabla^2\delta},e_1,c_1,c_2,\tilde{c}\}$.
Note that this term was included in the analysis of~\cite{Philcox:2022frc}, but missed by
\cite{DAmico:2022ukl}.\footnote{The $c_2$-dependent terms in this expression
cannot be absorbed by $c_{i}^{\rm St}$ of \cite{DAmico:2022ukl}. } 

It is important to note that
there is another contribution into 
$B_{\rm mixed}^{\rm ctr.~(s)~II}$
that stems from the velocity 
dressing in the RSD map when the velocity
field gets renormalized via $u_z\to u_z+u_z^{\rm ctr}$. 
However, this entire contribution
except the higher-order $\tilde{c}$ term 
can be absorbed into the $a_i$ counterterms
at the one loop bispectrum order, making it
completely redundant. Since the $\tilde{c}$ term
is beyond our power counting, it is consistent
to ignore its contribution there.

Before closing this section, let us briefly discuss
the differences between the above model and that of previous works.
Most of the operators in our model 
have were introduced in \cite{Philcox:2022frc,DAmico:2022ukl}. 
The main difference with respect to \cite{Philcox:2022frc} is that we propagate all constraints stemming from 
Galilean symmetry and the universality of EFT counterterms $b_{\nabla^2\delta},e_1,c_1,c_2$ for both the power spectrum and 
bispectrum (modulo the above discussion of $b_{\nabla^2\delta}$.) We have imposed all the relevant consistency constraints which link the coefficients
treated as independent in \cite{Philcox:2022frc}.
In addition, we include the 
non-locally contributing 
velocity counterterms derived 
in~\cite{DAmico:2022ukl}, and self-consistently include 
the mixed one-loop stochastic terms
$B_{\rm mixed}^{\rm 1-loop~(s)}$
for the first time. 
These terms were omitted 
in the previous bispectrum literature
\cite{Eggemeier:2018qae,Philcox:2022frc,DAmico:2022ukl}, though~\cite{Philcox:2022frc}
did include the $B_{\rm mixed}^{\rm ctr.~(s)~II}$ piece discussed above which 
is required by power-counting at order $O(k^2\bar n^{-1}P)$,
and which
has the same origin
as $B_{\rm mixed}^{\rm 1-loop~(s)}$.
Neither $B_{\rm mixed}^{\rm ctr.~(s)~II}$
nor $B_{\rm mixed}^{\rm 1-loop~(s)}$ (which we find to be important on power-counting grounds as well as in practical fitting the data) were included in the analysis of~\cite{DAmico:2022ukl}.
We stress that some 
of the terms entering these two
contributions 
do not require 
any free parameters,
\textit{i.e.}\ they are fully predictable 
and should be non-zero 
for physically
expected nearly-Poissonian
stochasticity. 
In fact, these terms 
are required 
in order to 
reproduce 
classic large-scale
structure result~\cite{1980lssu.book.....P} on the mixed 
deterministic-stochastic bispectrum contributions in~\eqref{eq:poiss_mix}.

Finally, a comment on the 
renormalization
is in order. 
It was shown before 
that the counterterms 
appearing in $B^{\rm ctr.(s)~I-II}$ and $B^{\rm ctr.(s)~I}_{\rm mixed}$
are sufficient for the renormalization
of the SPT one-loop diagrams~\cite{DAmico:2022ukl}.
This discussion, however, 
did not include the mixed 
one-loop diagrams. However, 
one can check that the UV dependence of these diagrams 
is canceled by the same 
counterterms that have already appeared in $B^{\rm ctr.(s)~I}_{\rm mixed}$, though require
renormalization conditions
different from those 
of~\cite{DAmico:2022ukl}.
For instance, the leading order
scale-dependent 
($\sim k^2P/\bar{n}$)
UV sensitivity 
of  $B_{\rm mixed}^{\rm 1-loop~(r)}$ is renormalized by the $a_4$ counterterm from
eq.~\eqref{eq:real_space_mix_kern}.

\subsection{Summary of the theory model}

\noindent All in all, our complete theory model for the power spectrum and bispectrum reads
\be 
\begin{split}
P&=P^{\rm tree~(s)}_{\rm SPT}[b_1]+
P^{\rm 1-loop~(s)}_{\rm SPT}[b_1,b_2,b_{\G},b_{\Gamma_3}]+P^{\rm ctr.~(s)}[b_{\nabla^2\delta},e_1,c_1,c_2,\tilde{c}]+P_{\rm stoch}[P_{\rm shot},a_0,a_2]\,,\\
B&= B^{\rm tree~(s)}_{\rm SPT}[b_1,b_2,b_{\G}]+
B^{\rm 1-loop~(s)}_{\rm SPT}[b_1,b_2,b_{\G},b_{\Gamma_3},b_3,\gamma_2^\times,\gamma_3,\gamma_{21}^\times,\gamma_{211},\gamma_{22},\gamma_{31}] 
\\
&+ 
B^{\rm ctr.~(s)~I}[b_1,b_{\nabla^2\delta},b_{\nabla^2\delta^2},b_{\nabla^2\G},b_{(\nabla \delta)^2},b_{(\nabla t)^2},e_1,e_4,e_5,c_1,c_2,c_3,c_4,c_6,c_7]  \\
&+B^{\rm ctr.~(s)~II}[b_1,b_2,b_{\G},b_{\nabla^2\delta},e_1,c_1,c_2] 
+B_{\rm mixed}^{\rm tree~(s)}[b_1,P_{\rm shot},B_{\rm shot}]\\
&+B_{\rm mixed}^{\rm 1-loop~(s)}[b_1,b_2,b_{\G},b_{\Gamma_3},P_{\rm shot},B_{\rm shot},d_2,d_{\G},d_{\Gamma_3}]
+B_{\rm mixed}^{\rm ctr.~(s)~I}[b_1,a_0,a_3,...,a_5,a_7,...,a_{12}]
\\
&+B_{\rm mixed}^{\rm ctr.~(s)~II}[b_1,P_{\rm shot},B_{\rm shot},b_{\nabla^2\delta},e_1,c_1,c_2,\tilde{c}]+B_{\rm stoch}[A_{\rm shot},a_1,a_6]\,.
\end{split} 
\ee 
At one-loop order, this depends on 11 bias parameters, 14+1 deterministic counterterms, and 19 stochastic counterterms, \textit{i.e.}\ 45 parameters in total. 
This can be contrasted with 
11 bias parameters,
18 deterministic counterterms, 
and 13 stochastic 
parameters used in~\cite{Philcox:2022frc},
and 11 bias parameters, 
14 deterministic counterterms, 
and 16 stochastic counterterms
used in~\cite{DAmico:2022ukl}. The differences with respect to previous works
are discussed 
at the end of the previous section.
Notably, our model does account for infrared resummation -- this procedure is outlined in Section~\ref{subsec: ir-resummation}.

Notably, many parameters enter both the power spectrum
and bispectrum simultaneously, which helps to break some of the degeneracies. For instance,
$e_1$ and $c_1$ are completely degenerate at the power spectrum level,
but this degeneracy is lifted by the one-loop bispectrum.
The only power spectrum parameter that does not appear
in the bispectrum is the redshift-space
stochastic counterterm $a_2$: the bispectrum contribution associated with this parameter 
is exactly degenerate 
with other stochastic 
one-loop bispectrum 
counterterms.

\section{Efficient Computation using \texorpdfstring{\cobra}{COBRA}}\label{sec:comp}
\noindent Next, we discuss how to practically implement the bispectrum model discussed above. This modifies the treatment of \citep{Philcox:2022frc} to incorporate \cobra basis functions \citep{Bakx:2024zgu} and bispectrum multipoles, with the latter following the tree-level treatment of \citep{Ivanov:2023qzb}. 

\subsection{COBRA Factorization}\label{subsec: loops}
\noindent As discussed in Section~\ref{subsec: bk-det}, the deterministic contributions to the one-loop bispectrum can be written
\beq
    B(\k_1,\k_2,\k_3;\beta,\Theta) = \big(B_{211}\big) + \big(B_{222}+B_{321}^{(I)}+B_{321}^{(II)}+B_{411}\big) \equiv B_{\rm tree} + B_{\rm 1-loop},
\eeq
where each term is a function of scale, cosmology parameters ($\Theta$), and bias parameters ($\beta$, including the growth factor, $f$). As shown in \eqref{eq: Bk-one-loop}, each term is a three-dimensional integral over the linear power spectrum and redshift-space kernels, $Z_n$, e.g., for $B_{222}$:
\beq\label{eq: B-loop-def}
    B_{222}(\k_1,\k_2,\k_3;\Theta,\beta) &=& 8\int_{\vq}Z_2(\k_1+\vq,-\vq;\beta)Z_2(\k_1+\vq,\k_2-\vq;\beta)Z_2(\k_2-\vq,\vq;\beta)\\\nonumber
    &&\quad\,\times\,P_{11}(q;\Theta)P_{11}(|\k_1+\vq|;\Theta)P_{11}(|\k_2-\vq|;\Theta),
\eeq
where we include the dependence on $\Theta$ and $\beta$ explicitly. Importantly, the $Z_n$ kernels depend only polynomially on the set of bias parameters and $f(z)$, thus we can separate the dependence on $\beta$:
\beq
    B_{222}(\k_1,\k_2,\k_3;\Theta,\beta) = \sum_{n=1}^{N_{\rm bias}}f_n(\beta)B^{(n)}_{222}(\k_1,\k_2,\k_3;\Theta),
\eeq
where $\{f_n(\beta)\}$ are a set of $N_{\rm bias}$ combinations of bias parameters, e.g., $b_1b_2f^2$. We require $N_{\rm bias}\approx 100$ to compute the full one-loop bispectra. Inserting the \cobra decomposition \eqref{eq: cobra-pk} into \eqref{eq: B-loop-def}, we obtain a factorized bispectrum template
\beq\label{eq: cobra-bispec}
    B_{222}(\k_1,\k_2,\k_3;\Theta,\beta) \approx \sum_{i_1,i_2,i_3=1}^{N_{\rm COBRA}}\sum_{n=1}^{N_{\rm bias}}w_{i_1}(\Theta) w_{i_2}(\Theta) w_{i_3}(\Theta)f_n(\beta) \mathsf{M}^{i_1i_2i_3,n}_{222}(\k_1,\k_2,\k_3),
\eeq
where the rank-3 tensor $\mathsf{M}_{222}$ is given explicitly by
\beq\label{eq: M222}
    \mathsf{M}^{i_1i_2i_3,n}_{222}(\k_1,\k_2,\k_3) &=& 8\int_{\vq}\frac{\partial}{\partial f_n(\beta)}\big[Z_2(\k_1+\vq,-\vq;\beta)Z_2(\k_1+\vq,\k_2-\vq;\beta)Z_2(\k_2-\vq,\vq;\beta)\big]_{\beta=0}\\\nonumber
    &&\quad\,\times\,\mathcal{V}_{i_1}(q)\mathcal{V}_{i_2}(|\k_1+\vq|)\mathcal{V}_{i_3}(|\k_2-\vq|).
\eeq
This is simply the integral of the bispectrum kernel corresponding to $f_n(\beta)$ multiplied by the \cobra basis functions.
Similar expressions can be derived for $B^{(I,II)}_{321}$ and $B_{411}$; these are somewhat simpler since the integrals factorize, e.g., $\mathsf{M}_{411}^{i_1i_2i_3,(n)}\sim \mathcal{V}_{i_1}\mathcal{V}_{i_2}F_{i_3}^{(n)}+\text{2 cyc.}$ for some $F$. 

Thus, \eqref{eq: cobra-bispec} is the desired result: we have factorized the bias and cosmology dependence from the one-loop bispectrum, allowing the full result to be computed as a matrix multiplication, following precomputation of the relevant loop integrals. In practice, this is much more expensive than for the power spectrum case considered in \citep{Bakx:2024zgu}, since (i) the loop integrals involve (at most) three power spectra instead of two and (ii) the number of bias coefficients is larger than in the power spectrum case. As discussed above, setting $N_{\rm COBRA}=\mathcal{O}(10)$ results in highly accurate bispectra: to test our pipeline, we compute the $\mathsf{M}^{i_1i_2i_3}$ matrices using $N_{\rm COBRA}=\{8,10,12\}$ for the power spectra entering the $\{B_{222},B_{321}^{(I)},B_{321}^{(II)},B_{411}\}$ loops, and $N_{\rm COBRA}=12$ for any external power spectra (e.g., in $B_{411}$), noting that the $B_{222}$ term is both more expensive to evaluate (since all basis functions appear inside the integral) and easier to approximate (since it is smoother). In practice, we find excellent performance when setting $N_{\rm COBRA}=8$ everywhere; this will be assumed in the below (see Figure \ref{fig: cobra-bispectrum-comparison} for validation).

\subsection{Practical Computation with FFTLog}
\noindent Next, we discuss how to assemble the high-dimensional $\mathsf{M}$ matrices that underlie our decomposition. We first note that the matrix integrands (e.g., \ref{eq: M222}) are polynomial in the redshift-space angles $\mu_i\equiv \hk_i\cdot\hn$.\footnote{The integrand also depends on the orientation of $\vq$. This can be accounted for by rewriting the integrals as functions of $\vq\cdot\k_i$ and $\mu_i$ invoking isotropic tensor algebra \citep[e.g.,][]{Philcox:2022frc}. An alternative approach is to retain the angular dependence and modify the (analytic) method used to compute the integrals \citep{Lee:2024nqu}. While the latter approach is more efficient and less sensitive to numerics, we adopt the former in this work for consistency with previous pipelines.} These can be rewritten these in terms of the standard angular coordinates $\mu,\phi$ defined by 
\beq\label{eq: mu-i-angles}
    \mu_1 = \mu, \qquad \mu_2 = \mu\cos\zeta-\chi\sin\zeta, \qquad \mu_3 = -\frac{k_1}{k_3}\mu_1-\frac{k_2}{k_3}\mu_2
\eeq
for $\chi \equiv \sqrt{1-\mu^2}\cos\phi$, $\cos\zeta\equiv\hk_1\cdot\hk_2$ \citep[e.g.,][]{Scoccimarro:2015bla,Philcox:2022frc}. This leads to
\beq\label{eq: M-matrices}
    \mathsf{M}^{i_1i_2i_3,n}(\k_1,\k_2,\k_3) &=& \sum_{a=1}^{N_{\rm ang}}g_a(\mu,\chi)\mathsf{M}^{i_1i_2i_3,n,a}(k_1,k_2,k_3)
\eeq
where $g_a(\mu,\chi)$ is a polynomial in $\mu$ and $\chi$ and $N_{\rm ang}\approx 50$. This fully accounts for the angular dependence of the integrand (up to infrared-resummation effects, as discussed below), and requires $N_{\rm ang}$ scalar integrals for each triplet of wavenumbers. 

Given the (somewhat oscillatory) \cobra basis functions, we require a scheme to evaluate the loop integrals. While one could use brute-force numerical integration (perhaps expedited using Monte Carlo methods), an alternative approach is to further decompose the $\mathcal{V}_i(k)$ functions onto a basis for which the loop integrals can be analytically computed, such as the FFTLog \citep{Simonovic:2017mhp} or the massive propagator \citep{Anastasiou:2022udy} form. We stress however that such methods are not strictly necessary in order for \cobra to be applied. Following \citep{Philcox:2022frc}, we adopt the FFTLog method in this work, approximating the \cobra functions as a sum of $N_{\rm FFT}$ complex power laws:
\beq
    \mathcal{V}_i(k) \approx \sum_{m=-N_{\rm FFT}/2}^{N_{\rm FFT}/2}\mathsf{C}^m_i k^{\nu+i\eta_m}
\eeq
\citep{Simonovic:2017mhp}, where $\nu$ is the FFTLog bias, $\eta_m = 2\pi m/\log(k_{\rm max}/k_{\rm min})$, $\mathsf{C}_i^m$ is the $N_{\rm COBRA}\times N_{\rm FFT}$ matrix of coefficients, and we set 
$k_{\rm min}=10^{-5}~h_{\rm fid}\text{Mpc}^{-1}, k_{\rm max}=10~h_{\rm fid}\text{Mpc}^{-1}$
in this work, where $h_{\rm fid}$
is the true $h$ of the PTChallenge 
cosmology.
Switching basis to $k = k_1,x = k_3^2/k_1^2,y=k_2^2/k_1^2$, the loop integrals can be expressed as
\beq\label{eq: cobra-vs-fftlog}
    \mathsf{M}^{i_1i_2i_3,n,a}(k_1,k_2,k_3) &=& \sum_{m_1,m_2,m_3=-N_{\rm FFT}/2}^{N_{\rm FFT}/2}\mathsf{C}^{m_1}_{i_1}\mathsf{C}^{m_2}_{i_2}\mathsf{C}^{m_3}_{i_3}\mathsf{N}_{m_1m_2m_3}^{n,a}(k,x,y).
\eeq
involving a total of $N_{\rm FFT}^3N_{\rm ang}N_{\rm bias}$ loop integrals, as a function of $k,x,y$. As discussed in \citep{Simonovic:2017mhp,Philcox:2022frc}, the $\mathsf{N}$ matrices can be computed analytically and expressed in terms of gamma functions and hypergeometric functions (with $k$ factoring out). Given that the $\mathsf{C}_i^m$ matrices are easy to compute from the tabulated \cobra basis functions, $\mathcal{V}_i(k)$, this allows the one-loop bispectrum components to be computed efficiently.

In the above approach, we first decompose the input power spectra onto the \cobra basis (to separate cosmology and scale-dependence) and then onto FFTLog basis (to efficiently compute loop integrals). An alternative option would be to drop the first decomposition, instead working directly with the FFTLog power laws, whose coefficients would encode cosmology. We do not adopt this scheme here, since the FFTLog basis is far less efficient than the \cobra scheme. As shown in \citep{Bakx:2024zgu}, accurate power spectrum approximations can be obtained with $N_{\rm COBRA}\sim 10$ compared to $N_{\rm FFT}\sim 100$, thus using FFTLog directly would reduce the speed of the bispectrum pipeline by $\sim 1000\times$, and increase its memory consumption by a similar factor (noting that the $\mathsf{N}$ matrices can be deleted after preprocessing in our two-step approach).

Overall, our pipeline for computing the one-loop bispectrum is as follows:
\begin{enumerate}
    \item Using \textsc{mathematica}, compute the FFTLog integration kernels. This requires computing products of the kernels, $Z_1$ to $Z_4$, symmetrizing, and expressing the result (exactly) as a polynomial series in $k, x, y, |\k_1+\vq|, |\k_2-\vq|,q$. Here, we additionally decompose all angular dependence into functions of $\mu_i$ (see \citep{Lee:2024nqu} for an alternative approach, which avoids the need to rewrite products of $\k_i\cdot\vq$ in terms of $|\k_i\pm\vq|$).
    \item Compute the analytic derivatives of the kernels with respect to the bias parameter functions, $f_n(\beta)$, and angular parameters, $g_a(\mu,\chi)$. We output a set of tables containing the relevant analytic coefficients.
    \item Compute the $\mathsf{C}_i^m$ matrices from the \cobra basis functions. We adopt $N_{\rm FFT} = 64, 96, 128, 128$ for the $B_{222}, B_{321}^{(I)}, B_{321}^{(II)}, B_{411}$ terms, noting that the latter terms are easier to compute due to the factorization properties.    
    \item Using parallelized \textsc{cython} code, compute the $\mathsf{N}$ matrices for each bias and angular function. This is described in \citep{Philcox:2022frc} and requires careful choice of the FFTLog bias $\nu$.\footnote{Strictly, this is only true if we require that each individual term, e.g., $B_{222}$ and $B_{321}^{(I)}$, is accurate -- the sum converges for a wide range of $\nu$ values. We have validated that our combined results are stable when we change $\nu$.} We compute results for a grid of $64$ points in $N_k$ and $N_{x,y}=40$ points in each of $x$ and $y$, assuming $x\leq y\leq 1.$
    \item Transform from the FFTLog basis to the \cobra basis, using the $\mathsf{C}$ matrices \eqref{eq: cobra-vs-fftlog}. This is performed in \textsc{cython} for speed.
    \item Perform angular integration and integrate over $k$-bins, as described below. This is computed using \textsc{python} and is specific to each analysis set-up.
\end{enumerate}
The result is a set of bin-integrated $\mathsf{M}$ matrices, which, when combined with the relevant bias parameters and \cobra $w_i(\Theta)$ parameters, can be used to obtain the full one-loop bispectrum. We caution that the $\mathsf{M}$ matrices are very large: the dimensionality scales as $N_{\rm COBRA}^3N_{\rm bias}N_{\rm ang}N_xN_yN_k$ before bin-integration. After bin integration, the dimensionality scales as $N_{\rm COBRA}^3N_{\rm bias}N_{\rm ang}N_\text{bin}$ where $N_\text{bin}$ is the number of triangle bins employed.  This is the quantity that needs to be stored and re-used. Note that $N_\text{bin} = \mathcal{O}(300) \ll N_x N_y N_k \approx 10^5$ (see Section \ref{subsec: ang-bin}). 

\subsection{Infrared Resummation}\label{subsec: ir-resummation}
\noindent The loop integrals discussed above do not correctly account for the damping effects induced by long-wavelength displacements. As discussed in previous works \citep[e.g.,][]{Philcox:2022frc,Blas:2016sfa,Blas:2015qsi,Ivanov:2018gjr,Chen:2024pyp,Baldauf:2015xfa,Crocce:2007dt,Senatore:2014via}, this requires infrared resummation of the theoretical model. At the precision required for the one-loop contributions, this can be implemented by replacing the linear power spectra in the loop integrals (e.g., \ref{eq: B-loop-def}) with the infrared (IR) resummed equivalent:
\be 
\begin{split}\label{eq: IR-pk}
    & P_{11}(k)\to P_{\rm IR}(k) \equiv P_{11}^{\rm nw}(k) + e^{-k^2\Sigma^2_{\rm tot}}P_{11}^{\rm w}(k),\\
    & \Sigma^2_{\rm tot}=\Sigma^2(1+f\mu^2(2+f))+\delta \Sigma f^2\mu^2(\mu^2-1)\,,\\
    &\Sigma^2 = \frac{1}{6\pi^2}\int^{\Lambda_{\rm IR}}_0 {\rm d}p~P_{11}(p)[1-j_0(r_s p)+2j_2(r_sp)]\,,
    \quad \delta\Sigma^2 = \frac{1}{2\pi^2}\int^{\Lambda_{\rm IR}}_0 {\rm d} p~P_{11}(p)j_2(r_sp)\,,
\end{split}
\ee 
where $\Lambda_{\rm IR}=0.1~\hMpc$,
$r_s\approx 110~\Mpch$ is the comoving sound horizon, 
$j_\ell(x)$ are spherical Bessel functions, and 
$P^{\rm nw,w}$ refer to no-wiggle and wiggle contributions to the linear power spectrum, obtained as in \citep{Chudaykin:2020aoj}. Strictly, the damping factor $\Sigma^2_{\rm tot}$ contains angular dependence; as in \citep{Philcox:2022frc}, we neglect the angular-dependence of the damping factor in the loop integrals, though include it at tree-level, alongside the necessary $\mathcal{O}(k^2)$ corrections to \eqref{eq: IR-pk}. This higher angular dependence in the loops could also be included by expanding the exponent in some polynomial Pbasis in angle (e.g. Legendre polynomials), although we expect the corrections to be suppressed relative to the leading term.

\begin{figure}
    \centering
    \includegraphics[width=0.5\linewidth]{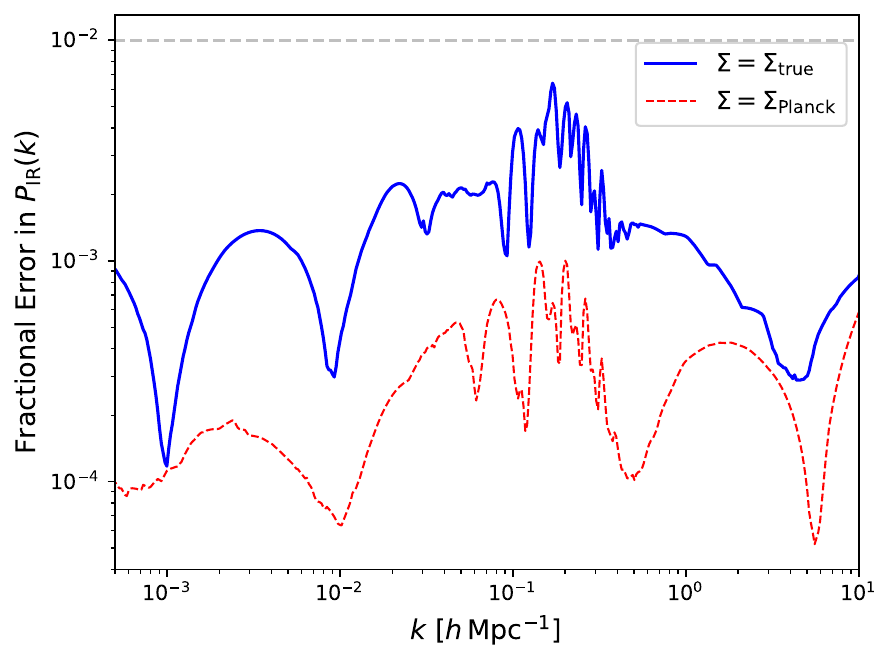}
    \caption{Accuracy of the \cobra decomposition applied to infrared-resummed power spectra at $z=0.61$, using $N_{\rm COBRA}=8$. For each k, we plot the $95$th percentile of the errors across $120$ test cosmologies, drawn from the default $\Lambda$CDM range discussed in \citep{Bakx:2024zgu}. The basis functions are computed using an SVD applied to a grid of infrared-resummed power spectra, computing using a fixed damping exponent, 
    $\Sigma_{\rm Planck}$. The dashed red curve shows the error under the same approximation (analyzing power spectra computed with fixed damping, which isolates the pure compression error), while the solid blue curve indicates the {total} true error, projecting the full resummed spectra (with cosmology-dependent $\Sigma=\Sigma_{\rm true}$) onto the  basis {computed with the fixed-cosmology BAO damping}. We find subpercent precision in all cases, justifying our choice of basis and $N_{\rm COBRA}$.}
    \label{fig: IR-resummation-check}
\end{figure}

To implement IR resummation in our efficient bispectrum framework, we have two options: (1) use linear theory to compute the \cobra basis functions $\mathcal{V}_i(k)$, then damp these via \eqref{eq: IR-pk}; (2) create a new set of basis functions $\mathcal{V}_i^{\rm IR}(k)$ using the set of resummed linear power spectra (\textit{i.e.}\ applying \eqref{eq: IR-pk} to the template bank). Here, we assume (2) since it results in a more efficient decomposition. When computing the set of template spectra underlying the \cobra decomposition, we typically sample only `shape' parameters, with all `amplitude' parameters such as $\sigma_8$ acting only multiplicatively \citep{Bakx:2024zgu,sanchez_evol,neutrino_evol,egge_comet}. Since $\Sigma_{\rm tot}$ depends on $\sigma_8$, IR resummation breaks this simplification, and na\"ively requires varying both shape and amplitude parameters upon template construction, and thus result in a less efficient decomposition. Here, we instead fix $\Sigma_{\rm tot}$ to the \textit{Planck} value when computing the \cobra basis; this retains factorizability at the expense of a slight loss of accuracy. In Figure~\,\ref{fig: IR-resummation-check}, we demonstrate that the resulting basis can capture the IR-resummed power spectra to sub-percent precision with just $N_{\rm COBRA}=8$ basis elements (the value used throughout this work).

\subsection{Angular Integration \& Binning}\label{subsec: ang-bin}
\noindent The final step is to transform the unbinned bispectra to the observational quantities of interest: binned bispectrum multipoles. For a general bispectrum $B(\k_1,\k_2,\k_3)$ we define the multipoles as
\beq\label{eq: B-mult-no-AP}
    B_\ell^{\rm ideal}(k_1,k_2,k_3) = (2\ell+1)\int_0^{2\pi}\frac{\md\phi}{2\pi}\int_{-1}^{1}\frac{\md\mu}{2}B(k_1,k_2,k_3,\mu_1,\mu_2,\mu_3)\mathcal{L}_\ell(\mu)
\eeq
\citep{Ivanov:2023qzb} where $\mu_i \equiv \hk_i\cdot\hn$ as before, and we perform angular integration with respect to $\k_1$, using the angular definitions in \eqref{eq: mu-i-angles}. While this quantity is straightforward to compare to observations, it does not encode the full angular dependence -- this could be obtained by introducing additional Legendre polynomials, decomposing in spherical harmonic moments, $\mathcal{L}_\ell(\mu)\to Y_{\ell m}(\cos^{-1}\mu,\phi)$, or inserting tripolar spherical harmonics \citep{Gualdi:2020ymf,Sugiyama:2018yzo,Scoccimarro:2015bla,Bakx:2025siw}, though some of these can be difficult to estimate from the data.

In practice, we must supplement \eqref{eq: B-mult-no-AP} with a prescription encoding the fiducial cosmology and bin integration. The first effect is specified by the distortion parameters
\beq
    \alpha_\parallel = \frac{H_{\rm fid}(z)H_{0,\rm true}}{H_{\rm true}(z)H_{0,\rm fid}}, \quad \alpha_\perp = \frac{D_{\rm true,A}(z)H_{0,\rm true}}{D_{\rm fid,A}(z)H_{0,\rm fid}}
\eeq
{for the tree-level bispectrum,
and $\alpha_\parallel^h$,
$\alpha_\perp^h$ (cf.\,\ref{eq:newAP})
for the one-loop bispectrum. The latter include the distortions due to the mismatch 
between the \cobra basis cosmology
and the true cosmology.}
All together, these transform both magnitudes and angles: 
\beq
    k_i\to k_i'(k_i,\mu) \equiv k_i\left[\frac{\mu_i^2}{\alpha_\parallel^2}+\frac{1-\mu_i^2}{\alpha_\perp^2}\right]^{1/2}, \qquad \mu_i\to \mu_i'(\mu_i) \equiv \frac{\mu_i}{\alpha_\parallel}\left[\frac{\mu_i^2}{\alpha_\parallel^2}+\frac{1-\mu_i^2}{\alpha_\perp^2}\right]^{-1/2},
\eeq
\textit{i.e.} rescaling parallel and perpendicular components of $\k_i$ by $1/\alpha_\parallel$ and $1/\alpha_\perp$ respectively. 
{The coordinate distortions 
in the PTChallenge data we use in this study are computed for the fiducial $\Omega_m=0.3$, and $h_{\rm fid}$ equal to the true 
Hubble parameter of the simulation (which is masked for the purpose of the PTChallenge).}
For the second effect, we assume a set of linear bins with $k_i\in[\bar{k}_i-\delta k,\bar{k}_i+\delta k)$, whose centers satisfy the triangle conditions, \textit{i.e.}\ $|\bar{k}_1-\bar{k}_2|\leq \bar{k}_3\leq \bar{k}_1+\bar{k}_2$, and further restrict to $\bar{k}_1\geq \bar{k}_2\geq \bar{k}_3$ to avoid degeneracies. 

The full binned and distorted bispectrum is defined as
\beq\label{eq: B-mult-bin-w-AP}
    B_\ell(\bar{k}_1,\bar{k}_2,\bar{k}_3)&=&\frac{1}{\mathcal{N}(\bar{k}_1,\bar{k}_2,\bar{k}_3)}\frac{2\ell+1}{\alpha_\parallel^2\alpha_\perp^4}\prod_{i=1}^3\left[\int_{\bar{k}_i-\delta k}^{\bar{k}_i+\delta k}k_i \md k_i\right]\int_0^{2\pi}\frac{\md \phi}{2\pi}\int_{-1}^{1}\frac{\md \mu}{2}\mathcal{I}(k_1,k_2,k_3,\mu,\phi)\\\nonumber
    &&\qquad\qquad\qquad\qquad\qquad\,\times\,B(k_1',k_2',k_3',\mu_1',\mu_2',\mu_3')\mathcal{L}_\ell(\mu),
\eeq
where $\mathcal{I}(k_1,k_2,k_3,\mu,\phi)$ is one if $\k_1,\k_2,\k_3$ obey the triangle conditions and zero else.\footnote{Note that this is equivalent to stating that $\vec k'_1,\vec k'_2,\vec k'_3$ must obey the triangle conditions.} 
This integrates over the true bispectra, weighting by the observed line-of-sight angle $\mu$. The normalization is given by
\beq
    \mathcal{N}(\bar{k}_1,\bar{k}_2,\bar{k}_3) = \prod_{i=1}^3\left[\int_{\bar{k}_i-\delta k}^{\bar{k}_i+\delta k}k_i \md k_i\right]\int_0^{2\pi}\frac{\md\phi}{2\pi}\int_{-1}^{1}\frac{\md\mu}{2}\mathcal{I}(k_1,k_2,k_3,\mu,\phi).
\eeq
Notably, the restriction $\bar{k}_1\geq \bar{k}_2\geq \bar{k}_3$ does not imply that $k_1'\geq k_2'\geq k_3'$ in \eqref{eq: B-mult-bin-w-AP}: other orderings can occur if two or more bins are equal, due to the finite bin widths and coordinate distortions.
For computational efficiency, it is useful to express the bispectrum entirely in the ordered form with $k_1\geq k_2\geq k_3$. This can be obtained by expanding the angular dependence of $B$ using a permuted
version of \eqref{eq: M-matrices}:
\beq
    B_{\rm perm}(k_1',k_2',k_3',\mu_1',\mu_2',\mu_3') = \sum_{a=1}^{N_{\rm ang}} g_a(\mu'_{123},\chi'_{123})B^a(K_1,K_3^2/K_1^2,K_2^2/K_1^2),
\eeq
where $K_1,K_2,K_3$ are the longest/mediumest/shortest of $k_1',k_2',k_3'$, with angular components $\mu_{123}',\chi_{123}'$ corresponding to $\k_1$. 
This can be inserted into \eqref{eq: B-mult-bin-w-AP}, allowing numerical integration over angles and sides.\footnote{Even in the absence of coordinate distortions, performing the angular integrals analytically is non-trivial, since finite-bin effects mix-up the $\mu_i$ angles, and the tree-level IR resummation is anisotropic.}
In practice, we compute the integral via Gauss-Legendre quadrature, defining a grid of $3^3\times 8^2$ values of $\{k_1,k_2,k_3,\mu,\phi\}$ for each bin, which are used to compute $\{k_1',k_2',k_3',\mu_1',\mu_2',\mu_3'\}$, and, following ordering, the bispectrum components $B^a$.\footnote{An alternative approach would be to switch the order of angular averaging and loop integration, \textit{i.e.}\ integrating over the raw bispectrum kernels (e.g., the integrand of \eqref{eq: M222}). This would reduce the size of both $\mathsf{M}$ and $\mathsf{N}$ matrices, and simplify the preprocessing step. While useful for the bispectrum monopole, this is more difficult for higher multipoles due to the ordering ambiguities induced by finite-bin and coordinate distortion effects. While this could be ameliorated by performing angular integration with respect to all three lines-of-sight, the resulting matrix can be numerically unstable.}
Note that our discreteness
treatment here
is an improvement over
~\cite{Ivanov:2021kcd}
because the original method used there assumed a continuous distribution
of the triangles inside the bin, \textit{i.e.}\ used the approximation
$\mathcal{I}\to 1$.

As discussed in~\cite{Ivanov:2021kcd}, the above integration schemes in general do not fully account
for binning effects in the bispectrum, due to the discrete nature of the Fourier-space grid and the finite fundamental frequency. To account for these effects, we correct the bin-integrated bispectra by `discreteness weights', defined
as the ratio between a discretely computed bispectrum and the continuous approximation discussed above for a given fiducial bispectrum:
\be 
w_\ell(\bar{k}_1,\bar{k}_2,\bar{k}_3) = \frac{B^{\rm discrete}_\ell(\bar{k}_1,\bar{k}_2,\bar{k}_3)}{B^{\rm integral}_\ell(\bar{k}_1,\bar{k}_2,\bar{k}_3)}\,.
\ee 
These weights are 
approximately 
cosmology independent, at least 
for the bispectrum monopole 
and for the tree-level bispectrum
multipoles \cite{Ivanov:2021kcd}. 
Incorporating the $\mathcal{I}$
kernel outlined above, 
we have found that the discreteness weights
for the bispectrum monopole 
amount
to sub-percent 
corrections 
that do not have any practical 
effect. As such, we do not apply the 
discreteness weights to the monopole in this work.

\begin{figure}[t]
    \centering
    \includegraphics[width=0.49\linewidth]{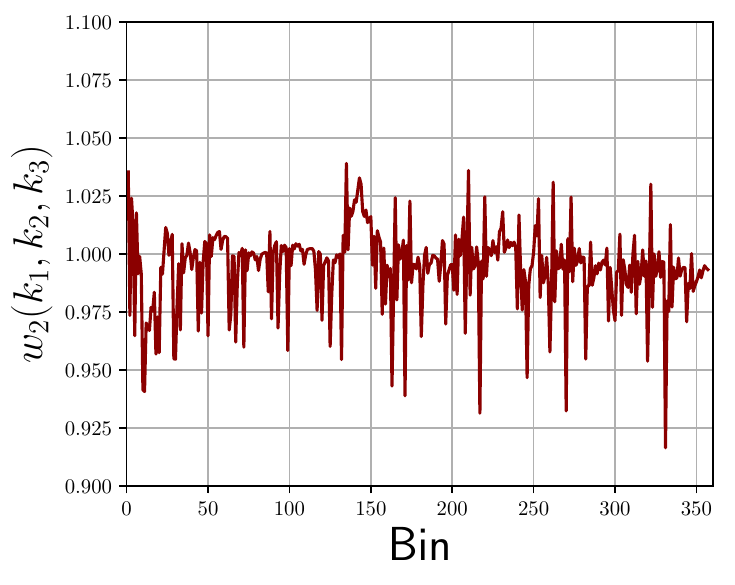}
    \includegraphics[width=0.49\linewidth]{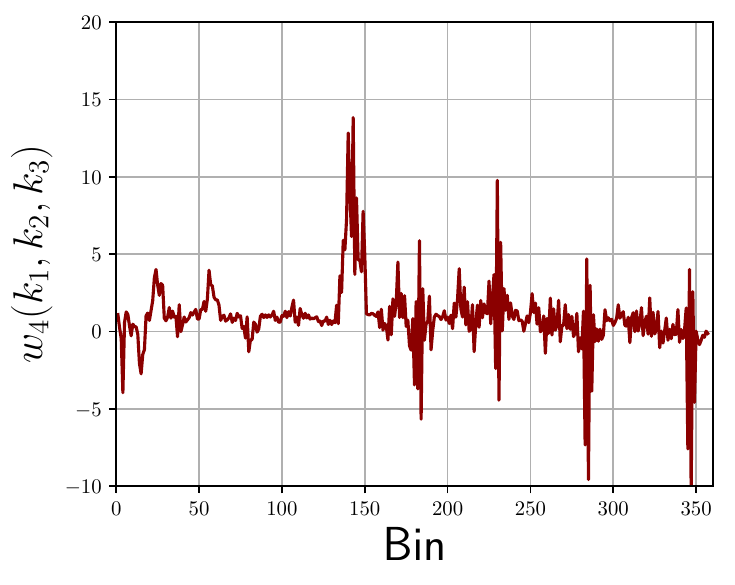}
    \caption{Discreteness weights for the bispectrum quadrupole 
    and hexadecapole moments. These are equal to the ratio of the bispectrum computed on a finite grid of $\k$-modes to that binned assuming a continuous distribution. Here, we show results obtained from a box of side-length $L=3800~\Mpch$, matching the PTChallenge simulations, for all $k$ modes in range $0.01~\hMpc\leq k\leq 0.08~\hMpc$, the site length $L=2000~\Mpch$ for bins with the largest wavenumber $0.08~\hMpc< k \lesssim 0.11 ~\hMpc$, 
    and $L=1000~\Mpch$ for bins with the largest wavenumber $0.11~\hMpc \lesssim k$.
    The smaller box sizes are chosen 
    to reduce computational cost. The bins 
    correspond to triangles
    with $0.01~\hMpc\leq k\leq 0.15~\hMpc$
    used the bispectrum multipole analysis.
    The discreteness weights for the monopole 
    are close to 1 for the entire range of scales used in the analysis ($0.01~\hMpc\leq k\leq 0.20~\hMpc$),
    and therefore 
    are thus omitted both here and in the analysis.
    \label{fig:disc_weights}}
\end{figure}

For the multipoles,
the discreteness weights
are somewhat more 
sizeable. 
We show the discreteness
weights for the quadrupole
and hexadecapole moments
for the bins in the range $0.01~\hMpc\leq k\leq 0.15~\hMpc$
used in our analysis
computed for a best-fit 
PTChallenge cosmology and 
bias parameters in 
Figure~\ref{fig:disc_weights}. 
For the bispectrum quadrupole,
given the maximal deviation of $9\%$, it suffices to compute the weights using the tree-level bispectrum. Since the one-loop bispectrum quadrupole is itself only $\sim 10\%$ of the tree-level piece (on the scales relevant to this work), the impact of the one-loop-induced discreteness corrections are at most $1\%$, which is of the same 
order of magnitude as the two-loop contributions, and can thus be neglected. That said, we will still apply the (tree-level-derived) weights to the one-loop quadrupole, which leads to a small improvement in the $\chi^2$ statistic. 

We caution that the above argument does not apply to the hexadecapole, since both the discreteness effects and loop contributions can be large (see Figures~\ref{fig:disc_weights}\,\&\,\ref{fig: bispectrum-model}). 
That said, the errorbars on this
multipole are also larger, which suggests that using the tree-level
discreteness weights for the full one-loop bispectrum  hexadecapole may still be 
accurate. In Section \ref{subsec: bisp-mults}, we explicitly test this assumption, which is found to be accurate at least up to $\kmax=0.15~\hMpc$.

In the \cobra formalism, we apply the above angular integration and binning directly to the $\mathsf{M}^{i_1i_2i_3,n,a}$ matrices.
Given that the coordinate distortion effects are typically small, we compute a Taylor expansion around {$\alpha^h_\parallel = \alpha^h_\perp=1$},\footnote{Strictly, we expand around {$\alpha^h_\parallel=1+\epsilon,\,\alpha^h_\perp=1+\epsilon$ for $|\epsilon|\ll1$} to avoid ordering ambiguities in equilateral configurations.} requiring three matrices for each Legendre moment. At run-time, the bispectrum is assembled as
\beq
    B^{\rm 1-loop}_\ell(\bar{k}_1,\bar{k}_2,\bar{k}_3;\alpha^h_\parallel,\alpha^h_\perp,\Theta,\beta) &=& \sum_{i_1i_2i_3=1}^{N_{\rm COBRA}}\sum_{n=1}^{N_{\rm bias}}w_{i_1}(\Theta)w_{i_2}(\Theta)w_{i_3}(\Theta)f_n(\beta)\\\nonumber
    &&\times\,\bigg[\mathsf{M}^{i_1i_2i_3,n}_\ell(\bar{k}_1,\bar{k}_2,\bar{k}_3)+(\alpha^h_\parallel-1)\left[\frac{\partial\mathsf{M}}{\partial\alpha_\parallel}\right]^{i_1i_2i_3,n}_\ell(\bar{k}_1,\bar{k}_2,\bar{k}_3)\\\nonumber
    &&\qquad\qquad\qquad\qquad\qquad+(\alpha^h_\perp-1)\left[\frac{\partial\mathsf{M}}{\partial\alpha_\perp}\right]^{i_1i_2i_3,n}_\ell(\bar{k}_1,\bar{k}_2,\bar{k}_3) \bigg];
\eeq
in total, this requires $3N_\ell N_{\rm bin}N_{\rm bias}$ matrices $\mathsf{M}$ of size $N_{\rm COBRA}^3$. Via tensor multiplication, the full bispectrum can be computed in $\mathcal{O}(1)$ second.

Finally, to test the accuracy our 
treatment of the $h$-dependence of the 
matter power spectrum via the Alcock-Paczynski
rescaling, we compare it with 
the alternative method based 
on the \cobra template bank computed 
with an explicit sampling of $h$.
We find negligible difference with respect to our fiducial AP+$h$ treatment, see Appendix~\ref{app:APtest} for more detail.

\subsection{Validation}
\noindent To validate our pipeline, we compare the binned bispectra obtained using the \cobra-factorization to those computed directly from an input power spectrum. To obtain the latter, we adopt the same procedure as above (and \citep{Philcox:2022frc}), simply replacing the \cobra basis functions by the true power spectrum of interest (thus dropping the $i$ indices). This is $\sim N_{\rm COBRA}^3$ times cheaper to compute than the $\mathsf{M}$ matrices above: however, it is far too expensive to be run on-the-fly in real cosmological analyses. This process is repeated across ten widely varying cosmologies, which facilitates stringent tests of our pipeline.

\begin{figure}[t]
    \centering
    \includegraphics[width=0.6\linewidth]{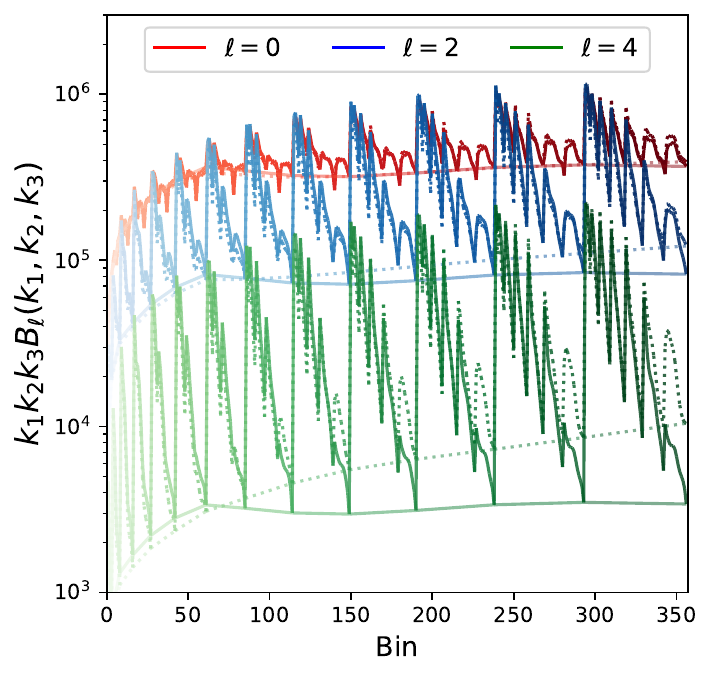}
    \caption{Example of the bispectrum multipoles computed using our pipeline. The solid red, blue, and green lines show (respectively) the tree-level bispectrum monopole, quadrupole, and hexadecapole, with the dotted lines including also the deterministic one-loop contributions. The color-scale ranges indicates the maximum $k$ of each bispectrum bin ranging from $k=0.01~\hMpc$ (lightest) to $k = 0.15~\hMpc$ (darkest). For reference, we highlight the equilateral configurations, which are connected by the almost horizontal lines. To form this plot, we assume a \textit{Planck}-like cosmology at $z=0.61$ and fix the bias parameters to the PTChallenge best-fit results.}
    \label{fig: bispectrum-model}
\end{figure}

In Figure~\ref{fig: bispectrum-model}, we show an example of the computed bispectrum multipoles (focusing on the deterministic contributions, which are the most difficult to estimate). As in previous works \citep[e.g.,][]{Philcox:2022frc,Ivanov:2023qzb}, the (dimensionless) bispectrum amplitude varies considerably with configuration, and is most prominent for large $k$, where the field becomes more non-linear. Moreover, we find that $B_2$ is suppressed compared to $B_0$ (and $B_4$ even more so), indicating that the higher multipoles are difficult to detect \citep[cf.,][]{Ivanov:2023qzb}. While the one-loop contributions are negligible on large scales, they become significant by $k\sim 0.1~\hMpc$, particularly for the higher multipoles. Indeed, for equilateral configurations with $\ell=4$, the contributions can be $\mathcal{O}(1)$ (though we caution that the signal-to-noise of this multipole is small).

\begin{figure}[t]
    \centering
    \includegraphics[width=0.9\linewidth]{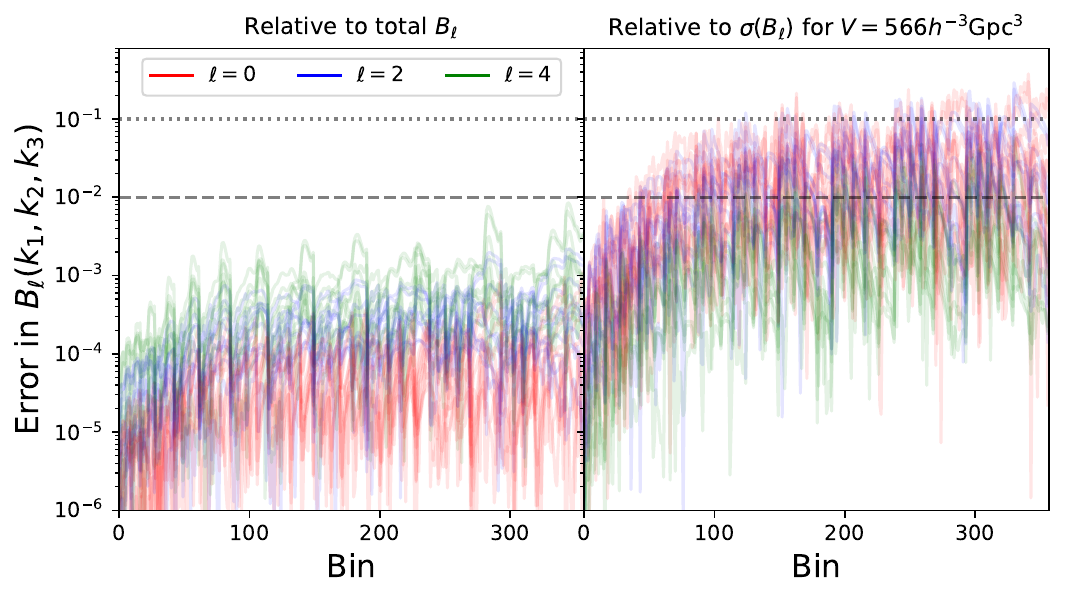}
    \caption{Accuracy of the \cobra decomposition applied to the one-loop galaxy bispectrum multipoles. We compare the \cobra bispectra (which are the main novelty of this work) to directly computed bispectra for 10 choices of primordial power spectrum, which are overplotted on the figure. In all cases, use the same bias parameters as Figure~\ref{fig: bispectrum-model} and assume $N_{\rm COBRA}=8$. The left panel compares the error to the total (tree-plus-one-loop) bispectrum, finding subpercent agreement in all cases, with largest differences for the hexadecapole. The right panel compares the error to the expected bispectrum errorbar, obtained by rescaling the PTChallenge covariance to the cosmology of interest by a factor $\prod_{i=1}^3\sqrt{P(k_i)/P_{\rm fid}(k_i)}$. Despite the huge volume of PTChallenge ($V=566~h^{-3}\mathrm{Gpc}^3$, much larger than any current or upcoming survey), we find $\lesssim 0.2\sigma$ agreement for all triangles with just $N_{\rm COBRA}=8$.}
    \label{fig: cobra-bispectrum-comparison}
\end{figure}

Figure~\ref{fig: cobra-bispectrum-comparison} compares the `exact' (non-factorized) and \cobra-factorized bispectra across the ten test cosmologies. 
Relative to the total bispectrum (which, for $\ell=0$, is dominated by the tree-level contribution, for which we do not require \cobra), we find subpercent errors, with $<0.001\%$ consistency for almost all monopole bins. This is an important result: with just $N_{\rm COBRA}=8$, we reconstruct the full bispectra at very high precision. In practice, it is useful to compare to the error-bar on $B_\ell$ rather than the value itself: this is shown in the right panel of Figure~\ref{fig: cobra-bispectrum-comparison}. For the full volume of the PTChallenge simulations (discussed below), we find $\lesssim 0.2\sigma$ agreement in each monopole bin (with largest errors on small-scales, where the signal-to-noise is greatest), which improves to $\mathcal{O}( 10^{-2})\sigma$ for the quadrupole and $\mathcal{O}(10^{-3})\sigma$ for the hexadecapole. Notably, this is much larger than the (effective) volume expected from any upcoming survey: reducing to $V_{\rm survey}=20~h^{-3}\mathrm{Gpc}^3$ would imply $<0.05\sigma$ errors everywhere (and moreover, small differences can be absorbed by nuisance-parameter marginalization). In conclusion, we obtain highly accurate predictions from \cobra which are easily sufficient for both current and next-generation galaxy surveys.

\section{Comparison with Simulation Data}\label{sec:data}

\noindent In this section, we use the one-loop
bispectrum calculations described above to model galaxy clustering data from the PTChallenge
simulation~\cite{Nishimichi:2020tvu}. Our analysis will be similar to that of \citep{Philcox:2022frc,Ivanov:2023qzb}, except that: (i) we study the impact of the one-loop bispectrum on all the major cosmological parameters, not just  $\sigma_8$ \citep{Philcox:2022frc}; (ii) we will extend the tree-level bispectrum
multipole analysis of \cite{Ivanov:2023qzb} to one-loop order. 

\subsection{Data and Likelihood}

\noindent The PTChallenge simulation
suite comprises 10 $N$-body simulations
with box-size $L=3840$ $\Mpch$,
resulting in a total cumulative volume  
of $V=566h^{-3}\mathrm{Gpc}^{3}$. This is $\mathcal{O}(100)$ times
larger than the total volume of the BOSS survey and $\mathcal{O}(10)$ times larger than that of the full DESI survey, allowing the theoretical error of the EFT calculations to be clearly assessed. 
Each simulated box of the PTChallenge suite was populated with mock galaxies using a custom HOD prescription described 
in~\cite{Nishimichi:2020tvu} and designed to mimic the luminous red galaxy sample of BOSS DR12 LOWZ, CMASS1, and CMASS2. Here, we use the mock galaxies at $z=0.61$ corresponding to the CMASS2 BOSS DR12 
sample, and average the output statistics over the 10 independent simulations. 

Our datavector consists of the power spectrum
multipoles $P_\ell(k)$ ($\ell=0,2,4$), 
the real-space power spectrum proxy $Q_0$~\cite{Ivanov:2021fbu},
and the bispectrum multipoles $\ell=0,2,4$:
\be
\label{eq:datav}
\{P_0,P_2,P_4,Q_0,B_0,B_2,B_4\}\,.
\ee
Our baseline power spectrum data cuts are $k^{P_\ell}_{\rm min}=0$, $k^{P_\ell}_{\rm max}=0.16~\hMpc$,
$k^{Q_0}_{\rm min}=0.16~\hMpc$, $k^{Q_0}_{\rm max}=0.20~\hMpc$, as validated in~\cite{Ivanov:2021fbu,Philcox:2022frc}.
For the bispectra, we use 
$k^{B_{\ell}}_{\rm min}=0.01~\hMpc$ to emulate realistic survey settings whence large-scale modes
are contaminated by systematics and are typically
excluded from the analysis \citep[e.g.,][]{DESI:2024jis}. 
We will vary the maximum $k$-mode of the bispectrum
in our analyses to determine the range up to which the one-loop bispectrum provides an unbiased fit. The bispectrum quadrupole and hexadecapole 
measurements available to us have $\kmax=0.15~\hMpc$, which provides an upper limit on the scale cut for our analyses of these moments.

As in previous works, we adopt a Gaussian likelihood, with a diagonal covariance tuned to the total cumulative volume of the 
PTChallenge suite. The analytic forms for the covariances of the power spectrum multipoles and $Q_0$ statistic can be found in \citep{Chudaykin:2019ock} and \citep{Ivanov:2021fbu} respectively, and utilize the measured power spectrum multipoles. For the bispectrum multipoles, we use the expressions given in \cite{Ivanov:2021kcd,Ivanov:2023qzb}, which are computed using the best-fit cosmological model and bias parameters extracted from the 
combined fit of the galaxy power spectrum and tree-level bispectrum multipoles from~\cite{Ivanov:2023qzb}.

In full, our likelihood takes the form: 
\be\label{eq: loglik}
\begin{split} 
& -2\ln L_{\text{tot}}=-2\ln L_{P_\ell}-2\ln L_{Q_0}-2\ln L_{B_\ell}\,,\\
&-2\ln L_{P_\ell}=\sum_{\ell,\ell'=0,2,4}\sum_{i,j:k_i,k_j< k^{P_\ell}_{\rm max}}[C^P_{(\ell \ell')}]^{-1}_{ij}(P^{\rm EFT}_\ell(k_i)-P^{\rm data}_\ell(k_i))
(P^{\rm EFT}_{\ell'}(k_j)-P^{\rm data}_{\ell'}(k_j))\,,\\
&-2\ln L_{Q_0}=\sum_{i,j:~k^{P_\ell}_{\rm max}\leq k_i,k_j\leq k^{P_\ell}_{\rm max}}[C^{Q_0}]^{-1}_{ij}(Q^{\rm EFT}_0(k_i)-Q^{\rm data}_0(k_i))
(Q^{\rm EFT}_{0}(k_j)-Q^{\rm data}_{0}(k_j))\,,\\
&-2\ln L_{B_\ell}=\sum_{\ell,\ell'=0,2,4}\sum_{i,j:~k_i<k^{B_\ell}_{\rm max},~
k_j\leq k^{B_{\ell'}}_{\rm max}}[C^B_{(\ell \ell')}]^{-1}_{ij}(B^{\rm EFT}_\ell(k_i)-B^{\rm data}_\ell(k_i))
(B^{\rm EFT}_{\ell'}(k_j)-B^{\rm data}_{\ell'}(k_j))\,,
\end{split}
\ee 
Due to the differing importance of non-linear effects and discreteness corrections, we use a different $\kmax$ for each bispectrum multipoles. Notably, we omit the cross-covariance 
between the power spectrum and the real-space proxy $Q_0$ -- this is possible since we include $Q_0$ only for modes with $k\geq k^{P_\ell}_{\rm max}$, which nulls the Gaussian correlation. 

In \eqref{eq: loglik}, we additionally ignore the cross-covariance between the power spectrum and bispectrum multipoles. Previously, \citep{Ivanov:2021kcd} demonstrated that the monopole cross-covariance is negligible for $k<0.08~\hMpc$; furthermore, \citep{Salvalaggio:2024vmx} showed that both the multipole cross-covariance and the non-Gaussian contributions to the bispectrum covariance are negligible for non-squeezed configurations up to $\kmax=0.12~\hMpc$. Given that our analysis restricts to $\kmax=0.15~\hMpc$ and excludes the most squeezed configurations by setting $k^{B_\ell}_{\rm min}=0.01~\hMpc$, the above approximation is expected to be valid. For the remaining squeezed 
triangles, we follow~\cite{Ivanov:2021kcd} and rescale the analytic Gaussian covariance estimates by $N^T_{\rm theory}/N^T_{\rm data}\sim 1.2$, where $N^T_{\rm theory},N^T_{\rm data}$ are the number of fundamental triangles in the bin as predicted theoretically assuming a continuous distribution of modes, and as measured from the data, 
respectively. Finally, we note that the cosmological information is
dominated by the triangle
configurations with similar
wavenumbers, for which our Gaussian approximation is adequate. 

We sample the EFT parameters using the priors similar to the ones described in \cite{Philcox:2022frc}.
Specifically, we use 
\be
\label{eq:priors}
\begin{split}
& B_{\rm shot}\sim \mathcal{N}(1,1^2)\,, \,\quad A_{\rm shot}\sim \mathcal{N}(1,1^2)\,,\,
P_{\rm shot}\sim \mathcal{N}(0,1^2)\,,\,
a_{0,2}\sim \mathcal{N}(0,1^2)\,,b_1\in[0,4]\,, \\
& \{b_{\nabla^2\delta},b_{\nabla^2\delta^2},b_{\nabla^2\G},b_{(\nabla \delta)^2},b_{(\nabla t)^2},e_1,e_5,c_1,...,c_7\}\sim \mathcal{N}(0,8^2)\,,\, \{b_{2},b_{\mathcal{G}_2}\}\sim \mathcal{N}(0,1^2)\,,\\
& 
b_{\Gamma_3}\sim \mathcal{N}(\frac{23}{42}(b_1-1),1^2)\,,\,
\{a_1,a_3,a_4,...,a_{12}\}\sim \mathcal{N}(0,8^2)\,,\, \gamma_{\mathcal{O}}\sim \mathcal{N}(0,10^2)\,,\,\\
&
\{
d'_{2},
d'_{\G},
d'_{\Gamma_3}
\}=
\{
d_{2},
d_{\G},
d_{\Gamma_3}
\}\times 2(1+P_{\rm shot})\sim \mathcal{N}(0,10^2)\,,\,\tilde{c}\sim \mathcal{N}(0,20^2)\,,
\end{split}
\ee 
where we choose the fiducial non-linear scale 
$k_{\rm NL}=0.45~\hMpc$ following~\cite{Ivanov:2021fbu,Philcox:2021kcw}.
The priors on $A_{\rm shot},B_{\rm shot},P_{\rm shot}$ are motivated by halo exclusion
arguments~\cite{Baldauf:2013hka}, whilst those on $b_1,b_2,b_{\Gamma_3},a_{0,2},\tilde{c}$
are motivated by earlier analyses 
of the BOSS data~\cite{Philcox:2021kcw,Ivanov:2023qzb}
and by the field-level measurements
of these parameters for dark matter halos
and simulated galaxies~\cite{Schmittfull:2018yuk,Schmittfull:2020trd,Ivanov:2024hgq,Ivanov:2024xgb,Ivanov:2024dgv,Ivanov:2025qie}.
For the new bispectrum parameters
we choose wide enough priors
in order to be conservative but respect the perturbativity
of the EFT expansion. 

All of EFT parameters except $\{b_1,b_2,b_{\mathcal{G}_2},b_{\Gamma_3}\}$ are marginalized over analytically (and exactly) after imposing Gaussian priors \eqref{eq:priors}, implying that the cosmological parameter posterior can be efficiently sampled. For the mixed term $B_{\rm mixed}^{\rm ctr.~II~(s)}$, direct analytic marginalization is not possible, since it involves both the counterterms 
$\{b_{\nabla^2\delta},e_1,c_1,c_2\}$
and $P_{\rm shot},B_{\rm shot}$. 
To rectify this, we Taylor expand
these terms around the best-fit values
of $P_{\rm shot}$ and $B_{\rm shot}$ (noting that these parameters are 
constrained quite well by the data). 
In cosmological analyses, we sample the 
physical density of dark matter $\omega_{\rm cdm}$, the Hubble parameter
$H_0$, and the redshift-zero matter fluctuation amplitude $\sigma_8$. As in previous works \citep{Nishimichi:2020tvu}, we fix the baryon density $\omega_b$, and the power spectrum tilt, $n_s$, to 
their true values used in the simulations. We note however that sampling these as done in state-of-the-art full-shape analyses \cite{DESI:2024jis} with \cobra does not pose any problems. When performing primordial non-Gaussianity analyses, we sample
the amplitude of the equilateral
non-Gaussianity $f_{\rm NL}$, fixing all other cosmological parameters, as in \citep{Cabass:2022wjy,Cabass:2022epm}. We assume flat uninformative priors on all varied cosmological parameters.\footnote{Given the precision of the data we consider here, the prior ranges used to construct the \cobra template bank in Section \ref{sec:cobra} are completely uninformative.}
The posterior is estimated using the \texttt{MontePython} Markov Chain Monte Carlo sampler \cite{Audren:2012wb,Brinckmann:2018cvx} and post-processing is done with \texttt{GetDist} \cite{Lewis:2019xzd}. 
We consider the chains to be converged
when the Gelman-Rubin metric \cite{gelman-rubin} satisfies
$|R-1|<0.05$ for all sampled parameters. 

\begin{figure*}[htb!]
\centering
\includegraphics[width=0.99\textwidth]{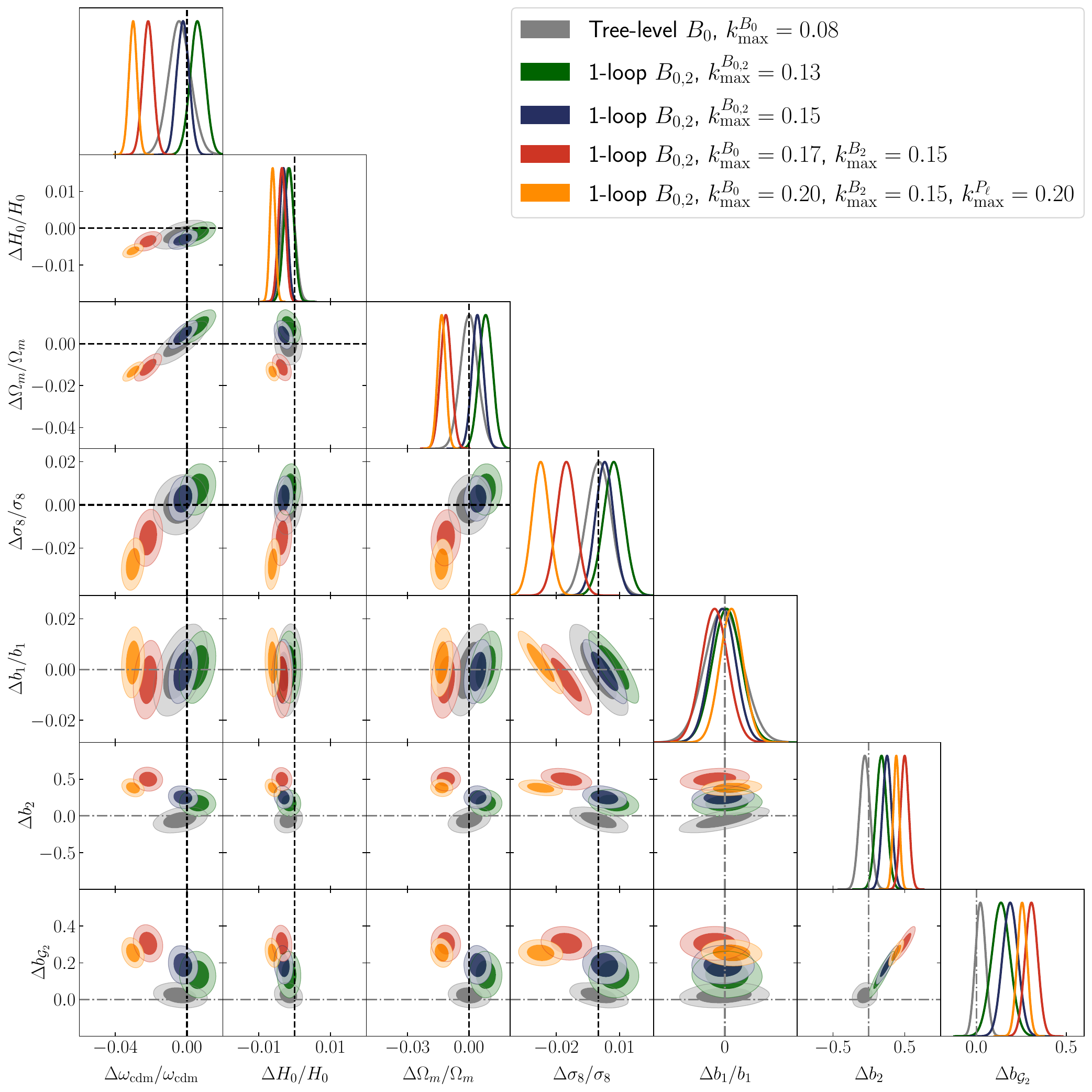}   \caption{PTChallenge constraints on cosmological parameters 
    obtained from analyzing the redshift-space power spectrum and the bispectrum multipoles $B_0,B_2$. As in Table~\ref{tab:tab1}, we report deviations of the cosmological parameters from their true values as well as bias parameters normalized to the values obtained from real-space bispectrum analyses (or the low-$k$ limit of the galaxy-matter cross-spectrum for $b_1$). 
    All quoted $k_{\rm max}$ values in the legend are in the $~\hMpc$ units.
    While the power spectrum likelihood is the same in all analyses, we consider five different variants of the bispectrum multipoles' likelihood: tree-level $B_0$ with $k_{\rm max}^{B_0}=0.08~\hMpc$
   (gray),
   one-loop with $k_{\rm max}^{B_{0,2}}=0.13~\hMpc$ (green),
   $k_{\rm max}^{B_{0,2}}=0.15~\hMpc$ (blue),
   $k_{\rm max}^{B_0}=0.17~\hMpc$ \& 
   $k_{\rm max}^{B_2}=0.15~\hMpc$
   (red) and finally $k_{\rm max}^{B_0}=0.20~\hMpc$ \& $k_{\rm max}^{B_2}=0.15~\hMpc$ where also the power spectrum cutoff is increased to $k_{\rm max}^{P_\ell} = 0.20~\hMpc$ (yellow). (Note that in the first four analyses we also use $Q_0$ with $0.16\leq k/(\hMpc)< 0.2$.) For the bias parameters, the light-dashed lines indicate best-fit values obtained from the PTChallenge 
   real-space
   power spectrum and 
   bispectrum analysis which we use 
   as proxies for the true values
   of the bias parameters. We find consistent results on cosmological parameters for $k_{\rm max}^{B_{0,2}}\lesssim 0.15~\hMpc$ (which are significantly tighter than the tree-level results) but find significant deviations for larger $k_{\rm max}^{B_0}$.
   The $B_2$ analysis is carried out up to $k_{\rm max}^{B_2}=0.15~\hMpc$ due to the unavailability of data at smaller scales.} \label{fig:cosmo}
\end{figure*}

\begin{figure*}[htb!]
\centering
\includegraphics[width=0.99\textwidth]{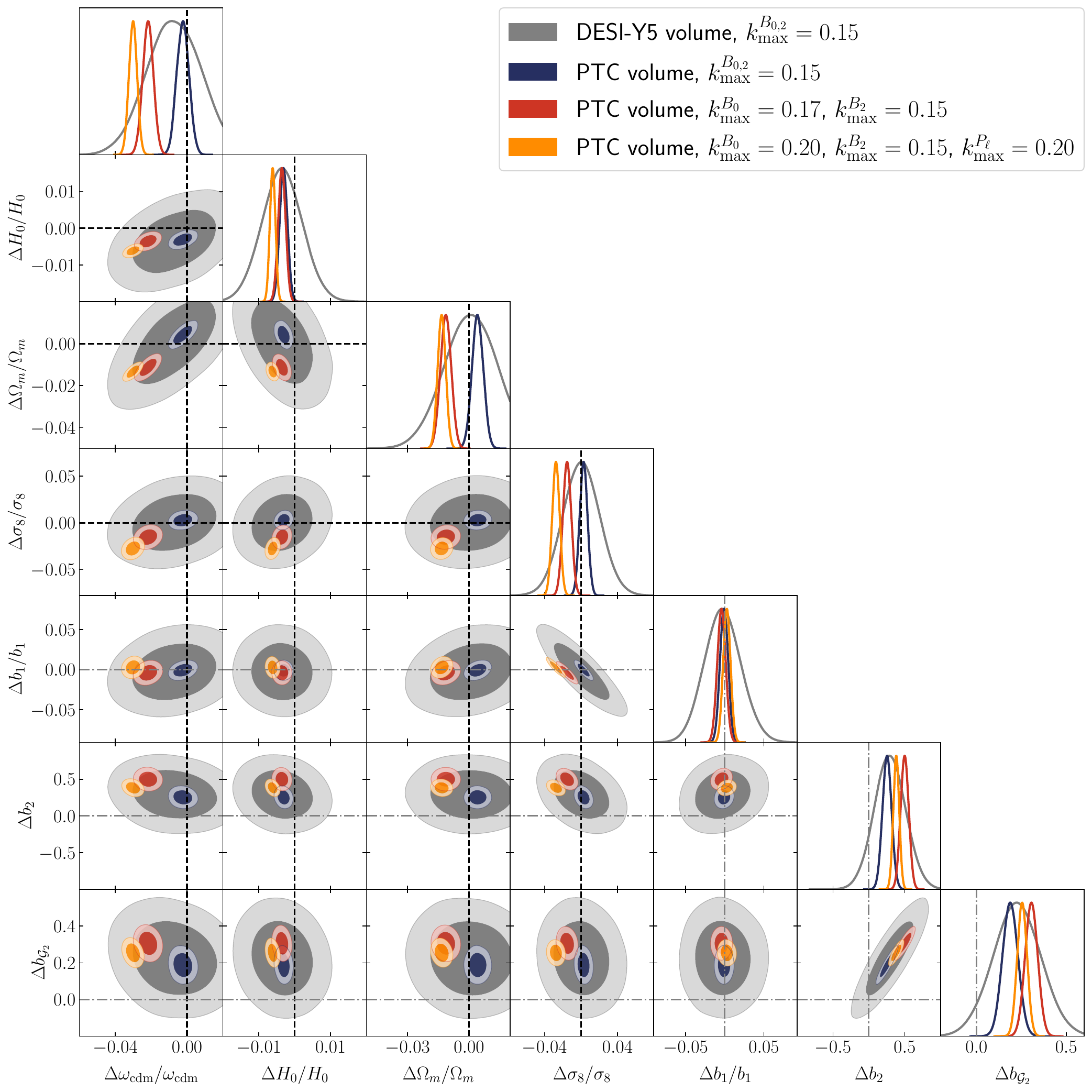}   \caption{Same as fig.~\ref{fig:cosmo}
(but without the tree-level results),
but with the $k_\text{max}^{B_0}=0.15~\hMpc$
data analyzed with the 
covariance rescaled 
to match the DESI-Y5 volume.
    } \label{fig:desi}
\end{figure*}

\subsection{Cosmological Parameters}\label{subsec: monopole}

\noindent 
With the \cobra pipeline for 
the fast evaluation
of the one-loop
bispectrum integrals, 
we can now carefully examine the 
cosmological information
in the redshift-space 
bispectrum 
at one-loop
order.
First, we study the impact of the one-loop galaxy bispectrum monopole and quadrupole by analyzing the PTChallenge simulation. The best-fits and marginalized posteriors from this analysis
are shown in Figure~\ref{fig:cosmo}
and in Table~\ref{tab:tab1}, encompassing both cosmological parameters and the main bias parameters 
$b_1,b_2,b_{\G}$. Results for the bispectrum multipoles will be discussed in Section~\ref{subsec: bisp-mults}.

In Figure~\ref{fig:cosmo}, we show results using a variety of experimental configurations: (i) the one-loop bispectrum monopole and quadrupole moments with various choices of $k_{\rm max}^{B_0}/(\hMpc)\in 0.13,0.15,0.17$ ($k_{\rm max}^{B_2}/(\hMpc)\in 0.13,0.15,0.15$, respectively, as the multipoles' data only up to $\kmax=0.15~\hMpc$ is available to us), in concert with the real-space $Q_0$ statistic (using the fiducial $k$-range); (ii) the tree-level bispectrum at $\kmax=0.08~\hMpc$ combined with $Q_0$; (iii) the one-loop bispectrum at $\kmax=0.2~\hMpc$ without $Q_0$. We include the power spectrum multipoles in all analyses, using the fiducial scale-cuts discussed above, except for case (iii) which uses $k^{P_\ell}_{\rm max}=0.2~\hMpc$. To avoid unblinding the PTChallenge simulations, we present the cosmological parameter and linear bias results in the form $\Delta X/X$, where $X$ is the true value (or that extracted from the low-$k$ limit of the real-space galaxy-matter cross-spectrum for the linear bias: $b_1=\lim_{k\to 0}P_{gm}(k)/P_{mm}(k)$). For the quadratic bias parameters, we study the deviation with respect to the best-fit values extracted from the tree-level real-space bispectrum analysis of \citep{Ivanov:2021kcd}. Whilst these values are subject to measurement errors and noise, they provide a fair baseline for our study, particularly given that they are measured from real-space data, which is less affected by non-linear physics. Finally, we directly quote the values of the cubic non-local bias parameter $b_{\Gamma_3}$
as is because its measurements 
do not jeopardize 
the PTChallenge due to large errors.
\begin{table*}[t!]
\begin{tabular}{l|c|c|cc}
 \multicolumn{5}{c}{PTC $P_\ell+B^{\rm tree}_0$, $k_{\rm max}^{B_0}=0.08~\hMpc$} \\
\hline
 & best-fit & mean$\pm\sigma$ & \multicolumn{2}{c}{95\% lower / upper} \\ \hline
$10^2\Delta \omega_{\rm cdm}/\omega_{\rm cdm}$ &$-0.35$ & $-0.40\pm 0.61$ & $-1.57$ & $0.83$ \\
$10^2\Delta H_0/H_0$ &$-0.17$ & $-0.17\pm 0.16$ & $-0.48$ & $0.14$ \\
$10^2\Delta \sigma_8/\sigma_8$ &$0.075$ & $0.01\pm 0.57 $ & $-1.1$ & $1.1$ \\
$10^2\Delta \Omega_m/\Omega_m$ &$0.051$ & $0.01\pm 0.43   $ & $-0.80$ & $0.86$ \\
$10^2\Delta b_1/b_1$  &$-0.058$ & $-0.04\pm 0.74 $ & $-1.5$ & $1.4$ \\
$ 10^2\Delta b_{2 }$ &$-13$ & $-11.4\pm 7.3 $ & $-25$ & $3$ \\
$10^2\Delta b_{\mathcal{G}_2}$ &$-1.7$ & $-1.2\pm 2.8$ & $-6.7$ & $4.4$ \\
$ b_{\Gamma_3}$ &$0.66$ & $0.64\pm 0.23$ & $0.16$ & $1.11$ \\
 \end{tabular} 
 \vspace{0.19em}
\begin{tabular}{l|c|c|cc}
\multicolumn{5}{c}{PTC $P_\ell+B^{\rm 1-loop}_{\ell}$, $k_{\rm max}^{B_{0,2}}=0.15~\hMpc$} \\
  \hline
& best-fit & mean$\pm\sigma$ & \multicolumn{2}{c}{95\% lower / upper} \\ \hline
$10^2\Delta \omega_{\rm cdm}/\omega_{\rm cdm}$ &$-0.30$ & $-0.23\pm 0.34$ & $-0.88$ & $0.44$ \\
$10^2\Delta H_0/H_0$ &$-0.32$ & $-0.31\pm 0.11$ & $-0.52$ & $-0.10$ \\
$10^2\Delta \sigma_8/\sigma_8$ &$0.26$ & $0.29\pm 0.43$ & $-0.55$ & $1.1$ \\
$10^2\Delta \Omega_m/\Omega_m$ &$0.39$ & $0.43\pm 0.28$ & $-0.12$ & $0.97$ \\
$10^2\Delta b_1/b_1$ &$-0.09$ & $-0.08\pm 0.52$ & $-1.1$ & $0.94$ \\
$ 10^2\Delta b_{2 }$ &$26$ & $25.3\pm 6.3$ & $13$ & $38$ \\
$10^2\Delta b_{\mathcal{G}_2}$ &$19$ & $18.8\pm 4.3$ & $10$ & $27$ \\
$ b_{\Gamma_3}$ &$0.36$ & $0.35\pm 0.16$ & $0.04$ & $0.65$ \\
\end{tabular}
 \caption{Cosmological and EFT parameters obtained from analyzing the PTChallenge simulation data using two models: the tree-level bispectrum monopole at $\kmax=0.08~\hMpc$ (left) and the one-loop bispectrum monopole and quadrupole at $\kmax=0.15~\hMpc$ (right). In all cases, we show the best-fit results, the marginalized posterior means and widths, and the $95\%$ confidence interval. Results are shown with respect to the true parameter values (or the real-space bias fits), and we rescale by $100$ such that the cosmological results are in percentage units. Two-dimensional posteriors are shown in Figure~\ref{fig:ptc_bestfit}.}
 \label{tab:tab1}
 \end{table*}

Inspecting Figure~\ref{fig:cosmo},
we note that the 
$k_\text{max}^{B_{0,2}}=0.13~\hMpc$
and $k_\text{max}^{B_{0,2}}=0.15~\hMpc$ analyses
are roughly consistent with the tree-level results, and, most importantly, the input cosmological parameters. 
We find only a slight $\sim 3\sigma$ bias 
on $h$, which is however
consistent 
with being a statistical fluctuation. 
While we see some
bias at the level of $b_{\G}$
at $k_\text{max}^{B_{0,2}}=0.15~\hMpc$ with respect to the fiducial value, this bias is still quite small in absolute terms and is negligible for current cosmological analyses, given the large errorbars of BOSS DR12 and DESI, including even the full five-year dataset (see below). Moreover, this does not induce significant bias in cosmological parameters, with the results of Table~\ref{tab:tab1} consistent with the true values within 95\% CL. Increasing $k_\text{max}^{B_0}$ to $0.17~\hMpc$ ({at fixed $k_{\rm max}^{B_2}=0.15~\hMpc$}) leads to a strong bias
in cosmological parameters, reaching $2\%$ for $\Omega_m$ and $3\%$ for $\sigma_8$. This is even more apparent in the non-linear bias parameters, with the quadratic biases shifting by $\approx 0.5$, which is around the $1\sigma$ statistical error of DESI DR1~\cite{Chudaykin:2025aux}. This indicates a breakdown of the one-loop analysis at $k_\text{max}^{B_0}=0.17~\hMpc$.  
The shifts of $b_2$
and $b_{\G}$ are clear indicators of bias in the one-loop model since these parameters are primarily measured from the bispectrum itself -- for other parameters, such as $\Omega_m$, much of the constraining power comes from the unbiased
power spectrum likelihood, which yields smaller shifts. 

\begin{figure}[t]
    \centering
    \includegraphics[width=0.8\linewidth]{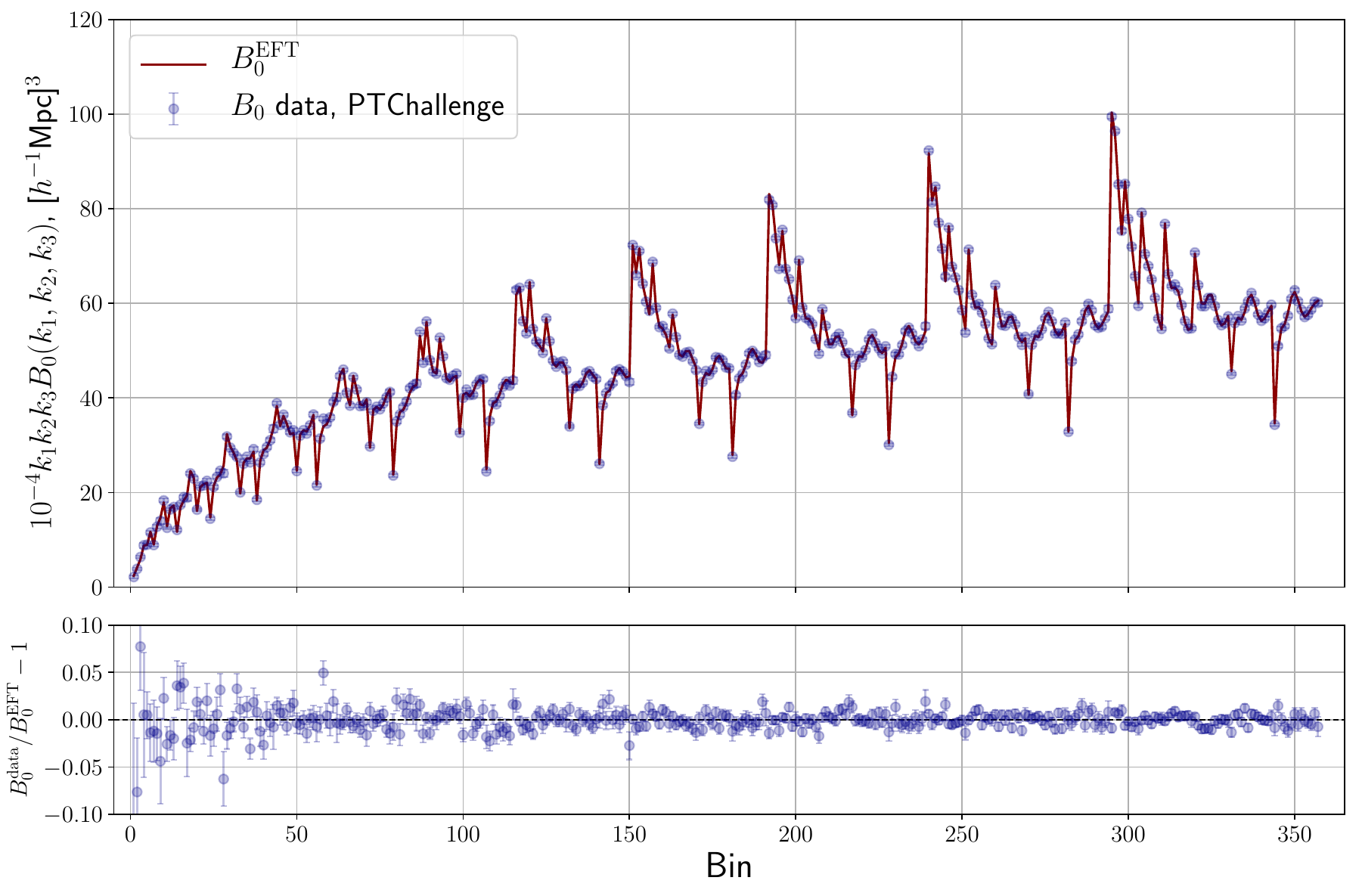}
      \caption{Comparison of the PTChallenge bispectrum monopole data (points) and the best-fit one-loop EFT model (lines), whose parameters are given in Table~\ref{tab:tab1}. The bins contain momenta ranging from $k=0.01\,\hMpc$ to $k = 0.15\,\hMpc$. We find exquisite agreement on these scales, despite the huge volume of the PTChallenge dataset.}
    \label{fig:ptc_bestfit}
\end{figure}

To further examine the breakdown of the one-loop
EFT bispectrum model, we consider simultaneously increasing both the power spectrum and bispectrum monopole scale-cut to $\kmax=0.2~\hMpc$ (similar to \cite{DAmico:2022ukl}). We find that this leads to a strong bias in all cosmological parameters, including a $2\%$ downwards shift in $\Omega_m$ (which could propagate to other parameters including standard-model extensions such as evolving dark energy) and a 4\% downward shift on $\sigma_8$.
Without the mixed one-loop corrections (which were not included in \cite{DAmico:2022ukl}), the bias is even larger, with the downwards shift in $\Omega_m$ reaching $5\%$, for example. This indicates that performing one-loop analyses beyond $k_\text{max}^{B_0}=0.15~\hMpc$ in general induces strong parameter biases. 
This result is somewhat anticipated given the two-loop corrections to the redshift-space power spectrum discussed in \cite{Nishimichi:2020tvu}, which lead to a breakdown of the one-loop power spectrum model at $\kmax\approx 0.15~\hMpc$ in a high-volume analysis~{appropriate for an estimation of the theory-systematic errors due to
the two-loop corrections~\cite{Chudaykin:2024wlw}}.
This is further supported by field-level analyses \citep[e.g.,][]{Schmittfull:2020trd,Ivanov:2024xgb}, which demonstrated that 
the one-loop model, even if extended
with free transfer functions, fails for PTChallenge-like HOD 
galaxies around $\kmax\approx 0.2~\hMpc$.
This motivates the choice of $\kmax= 0.15~\hMpc$
as a baseline cut for cosmological analyses. 

\begin{figure}[t]
    \centering
    \includegraphics[width=0.6\linewidth]{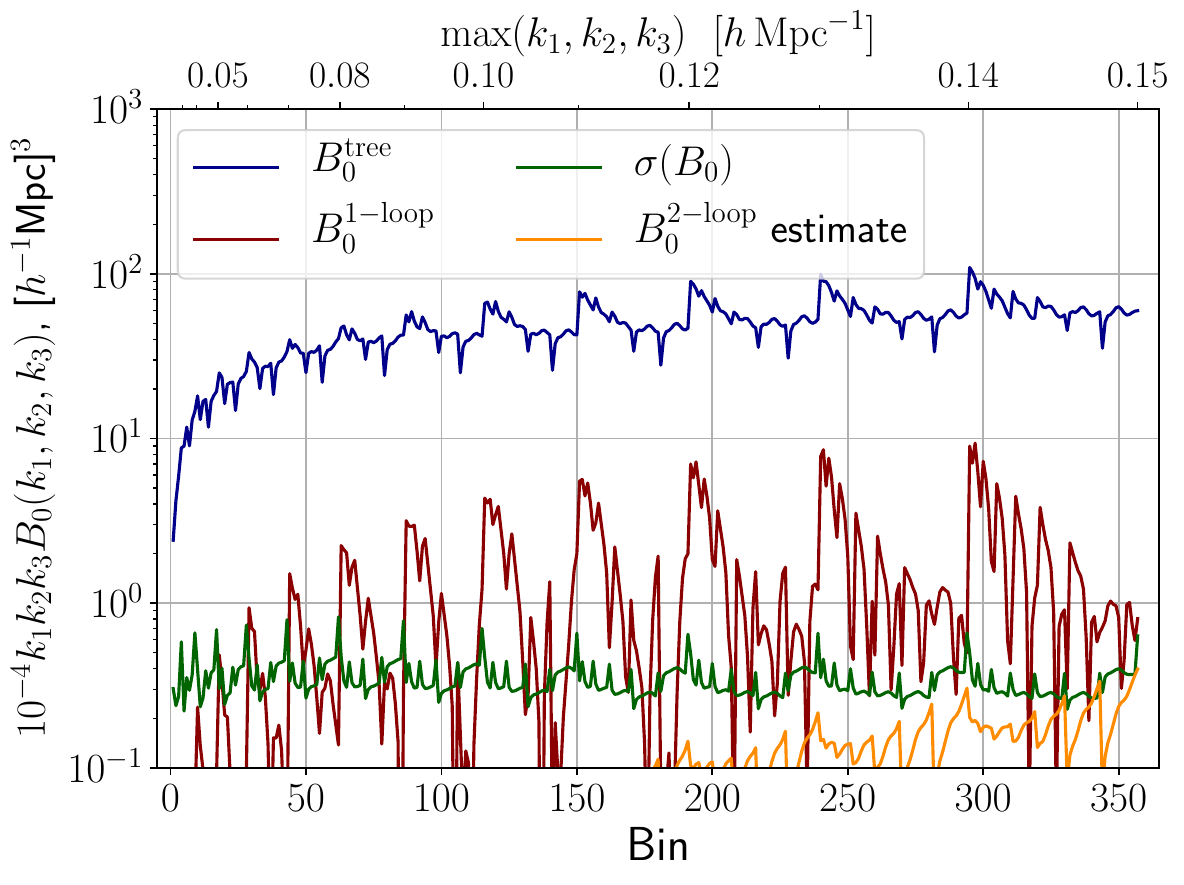}
     \caption{Comparison of the tree-level (blue)
    and one-loop (red) contributions to the 
    best-fit PTChallenge bispectrum monopole. For comparison, we show the statistical errors $\sigma(B_0)$ are shown in green. As in Figure~\ref{fig:ptc_bestfit}, the bins contain momenta ranging from $k=0.01\,\hMpc$ to $k = 0.15\,\hMpc$. {The upper $x$-axis label gives the largest of the three triangle sides in units of $\hMpc$. } Notably, the one-loop contributions are significantly larger than the errorbars on almost all scales, and, at larger $k$, the two-loop bispectra are expected to become important.
    }
    \label{fig:ptc_break}
\end{figure}

Figure~\ref{fig:ptc_bestfit} displays the best-fit EFT
bispectrum monopole 
at $k_\text{max}^{B_0}=0.15~\hMpc$, which is broken down into tree-level and 
and one-loop contributions in Figure~\ref{fig:ptc_break}. In addition, we also show the PTChallenge errorbars and an estimate for the two-loop bispectrum based
on matter two-loop 
contributions from \citep{Baldauf:2016sjb}: 
\be 
B^{\rm 2-loop}_{0, \rm estimate}(k_1,k_2,k_3)=B_0^{\rm tree} (k_1,k_2,k_3)D^4_+(z)\left(\frac{k_1+k_2+k_3}{3 k_\text{NL}}\right)^{3.3}\,,
\ee 
where $D_+(z)$ is the growth factor and $k_\text{NL} = 0.45\,\hMpc$ as before.
We see that the one-loop EFT calculation fits the data to $\lesssim 1\%$ up to $k_\text{max}^{B_0}=0.15~\hMpc$. 
As in Figure~\ref{fig: bispectrum-model}, the monopole model is dominated by the tree-level piece, with the one-loop components  bispectrum contributing at most 10\%
of the total signal. As such, 
our best-fit is consistent 
with the perturbative size 
of the one-loop correction. 
The one-loop terms, however, are much greater than the statistical errors, making their inclusion vital. 
At the same time, the two-loop corrections are estimated to be subdominant for $k\lesssim 0.15~\hMpc$, but become of order the statistical errors on smaller scales, reaffirming the above conclusions.

Comparing the baseline one-loop
$B_{0,2}$ results with those of the tree-level
at $\kmax=0.08~\hMpc$ in Table~\ref{tab:tab1},
we note that the one-loop
bispectrum improves
constraints on $\omega_{\rm cdm}$ 
and 
$\Omega_m$ {by $\approx 44\%$
and $\approx 35\%$ }respectively.\footnote{We measure the improvement 
on an error of parameter 
$p$ after including the new statistic 
as $1-\sigma^{\rm after}_{p}/\sigma^{\rm before}_{p}$ as done in \cite{Philcox:2022frc}.
}
The constraint on $H_0$ is improved 
{by $\approx 32\%$},
while the constraint on $\sigma_8$
is tightened by
{$\approx 25\%$.} 
{Interestingly, 
in the bispectrum monopole-only analysis we found somewhat 
smaller improvements: $\Omega_m$, $H_0$, and $\sigma_8$ improve by $\approx 23\%$,
$\approx 18\%$, and $\approx 2\%$,
respectively. Thus, the bispectrum quadrupole produces noticeable improvements over the monopole analysis only, especially for $\sigma_8$. While $B_2$ has lower signal-to-noise than the monopole, 
it has a different dependence on cosmological parameters, leading to 
additional 
degeneracy breaking (see e.g~\cite{Ivanov:2023qzb,Eggemeier:2025xwi}). }
{Our improvements} can be compared with the results of \cite{Philcox:2022frc}, which found that the one-loop bispectrum monopole sourced a $\approx 10\%$ improvement on the amplitude of scalar fluctuations 
$A_s^{1/2}$ in a fixed-cosmology analysis. Whilst we find a similar improvement in $A_s^{1/2}$ when restricted to $B_0$ only, this does not directly translate to an improved constraint on $\sigma_8$ due to the correlation with  cosmological parameters $h$ and $\omega_{\rm cdm}$.
In the case of $B_0$ only, it is interesting to note
that the constraint on $\sigma_8$
does not noticeably improve
even if we formally push the analysis
to $\kmax=0.2~\hMpc$, which implies that
even the high-$k$ bispectrum 
monopole data cannot 
fully break the degeneracies between
$\sigma_8$ and the many necessary bias parameters. 
The degeneracy is, however,
somewhat lifted upon including 
the bispectrum quadrupole
data. 

To put our results 
in context of the 
ongoing galaxy surveys,
we have repeated 
our
analysis 
with the covariance
matrix 
rescaled to a volume of $V_{\rm DESI}=20$
[$h^{-1}$Gpc]$^3$, similar to that of DESI-Y5.
(The number density 
of the PTChallenge
roughly matches the 
DESI LRG and ELG number density.) 
All other analysis settings
remain unchanged. 
Our main results 
are shown 
in Fig.~\ref{fig:desi}.
First, we see
that the aforementioned shift of $b_2$
and $b_{\mathcal{G}_2}$
from the $V=566~[h^{-1}\text{Gpc}]^3$
analysis,
even if we interpret 
it as an actual bias
in the model, 
is significantly smaller
than $1\sigma$ error-bar
from DESI-Y5. 
Furthermore, the constraints on bias parameters from such a survey will be worse than those found above, since the data will consist 
of multiple chunks at different redshifts with different associated bias parameters (in contrast to the single-sample analysis performed herein). 
This implies that 
the shift in 
$b_2$
and $b_{\mathcal{G}_2}$
of our model at $\kmax=0.15~\hMpc$
can be completely 
ignored even
for the DESI-Y5 volume.
Next, we consider the cosmological parameters. While the 
$\kmax=0.15~\hMpc$ likelihood recovers
all the relevant cosmological
parameters without bias, the mentioned
above theory systematic
bias at 
$\kmax=0.17~\hMpc$ 
and 
$\kmax=0.20~\hMpc$
is quite significant. 
For $\omega_{\rm cdm}$,
$\Omega_m$
and $\sigma_8$
it is more than 
1$\sigma$ of the statistical error 
of DESI-Y5; this implies that improved models will be required to faithfully model these scales.
To aid interpretation of the one-loop posteriors from future multi-chunk surveys, it will additionally be important to quantify posterior projection effects (see both early works \cite{Ivanov:2019pdj,Chudaykin:2020ghx}
and recent discussions~\cite{Maus:2024dzi,Chudaykin:2024wlw,Paradiso:2024yqh,Chudaykin:2025aux}).

Finally, let us comment
on the results of~\cite{DAmico:2022osl}
who found somewhat larger improvement on $\sigma_8$, utilizing an incomplete one-loop EFT model up to $\kmax=0.23~\hMpc$. Even when dropping the mixed one-loop stochastic terms from our theory model (so as to match \citep{DAmico:2022osl}), we do not find agreement with the former work, though our results are in accordance with  \cite{Philcox:2022frc}. 
Moreover, we find that the non-local velocity counterterms ($a_{12}$ and $e_5$ contributions in our model, which were omitted in \citep{Philcox:2022frc}), neither improve the reach of the model nor reduce the bias in parameter estimation, in contrast to the conjecture of \cite{DAmico:2022ukl}.\footnote{We caution that this holds only the HOD-based luminous red galaxies used in this work. These counterterms might be important for other galaxy samples.}
We believe that the difference between our results here and those of~\cite{DAmico:2022osl}
may stem from the following four factors:
the use of less conservative power spectrum scale
cuts in \cite{DAmico:2022osl} 
(see~\cite{Chudaykin:2024wlw} for
detailed discussions), inaccuracies in the modeling of the survey window function (already noted in \cite{DAmico:2022osl}),
numerical differences in the one-loop
calculations due to the use of a different computation technique, 
and the omission of the mixed one-loop stochastic terms, 
which have a significant impact on our analysis.

\subsection{Non-Local Non-Gaussianity}

\begin{figure*}
\centering
\includegraphics[width=0.6\textwidth]{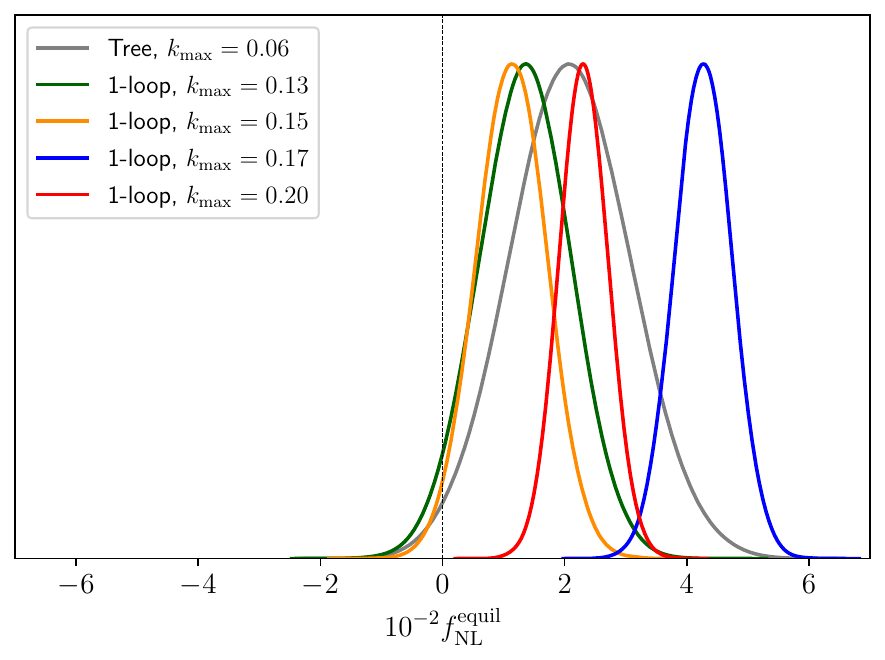}
   \caption{Constraints
   on the primordial non-Gaussianity
   parameter $f_{\rm NL}^{\rm equil}$ from the redshift-space PTChallenge galaxy power 
   spectrum and bispectrum monopole. We show results for several values of the bispectrum momentum cut $\kmax$ (quoted in units of $\hMpc$). A vertical dashed line marks 
   the true value
   of the simulation
   $f_{\rm NL}^{\rm equil}=0$.
   We observe that the constraints are significantly tightened by the inclusion of one-loop corrections, and are broadly consistent with zero up to $k_{\rm max}=0.15~\hMpc$, given the huge simulation volume.
    } \label{fig:fnl}
\end{figure*}
\begin{table}[t!]
\begin{tabular}{c|c|c|cc}
$k_{\rm max}^{B_0}/(\hMpc)$ & best-fit & mean$\pm\sigma$ & \multicolumn{2}{c}{95\% lower / upper} \\ \hline
$0.06$ &$2.12$ & $2.12_{-1}^{+1}$ & $0.09$ & $4.12$ \\
$0.08$  &$3.09$ & $3.16_{-0.73}^{+0.62}$ & $1.86$ & $4.46$ \\
\hline
$0.13$ &$1.29$ & $1.34_{-0.77}^{+0.78}$ & $-0.18$ & $2.8$ \\
$0.15$   &$1.11$ & $1.13_{-0.58}^{+0.6}$ & $-0.03$ & $2.30$ \\
$0.17$   &$4.29$ & $4.27_{-0.48}^{+0.48}$ & $3.33$ & $5.21$ \\
$0.20$    &$2.39$ & $2.31_{-0.42}^{+0.41}$ & $1.51$ & $3.13$ \\
 \end{tabular} 
 \caption{\label{tab:tab2} Constraints on the equilateral primordial non-Gaussianity parameter $10^{-2}f_{\rm NL}^{\rm equil}$ as a function of $k_{\rm max}^{B_0}$. The first two lines give the tree-level constraints whilst the remainder include one-loop contributions. Our one-loop analysis yields significantly tighter constraints, but can suffer from biases if $k_{\rm max}^{B_0}$ is increased beyond $\approx 0.15~\hMpc$.}
\end{table}

\noindent Next, we discuss the impact of the one-loop
bispectrum monopole on primordial non-Gaussianity measurements (hereafter PNG). In this paper we focus
on the non-local equilateral-type PNG, whose
constraints are dominated 
by the bispectrum \citep[e.g.,][]{Cabass:2022wjy,Cabass:2022epm,DAmico:2022gki}. In this case we expect the one-loop
calculation to have the 
largest impact. 
We leave the detailed study
of other types of non-Gaussianity
to future work. Here we 
include only the 
leading PNG contributions 
in our theory model, 
corresponding to the 
$B_{111}$ diagrams, but
ignore the one-loop
PNG bispectrum diagrams
considered in~\cite{Assassi:2015jqa}, which is appropriate for small $f_{\rm NL}$.
{We consider only the $B_0$ data in our analysis in order 
to compare more directly with~\cite{Philcox:2022frc}, noting 
that the higher-order moments 
do not bring significant improvements in the PNG analyses.}

Since the PNG signature is
relatively weak, it becomes sensitive
to higher-order (two-loop) corrections
on larger scales than the cosmological 
parameters. To mitigate possible biases, we reduce
the power spectrum scale cuts to 
$k^{P_\ell}_{\rm max}=0.14~\hMpc$ and $k^{Q_0}_{\rm max}=0.18~\hMpc$. As discussed above, we fix all cosmological parameters to their
true values and vary only the PNG
amplitude $f^{\rm equil}_{\rm NL}$
and EFT parameters in our MCMC chains.
Our results for the one-loop bispectrum likelihood with four choices of momentum cut $\kmax/(\hMpc)=0.13,0.15,0.17,0.20$
are displayed in Table~\ref{tab:tab2}
and Figure~\ref{fig:fnl}. For comparison, we additionally give the tree-level bispectrum monopole result at $\kmax=0.06~\hMpc$, using a low $\kmax$ to ensure that $f_{\rm NL}^{\rm equil}$
is consistent with the fiducial value (zero), within $\approx 2\sigma$,  ensuring that the theory systematic error is small.
In practice, given the finite 
errors of real observational data, one can use a slightly higher $\kmax$ at the expense of 
a relatively small theory systematic error.
For instance, fitting the tree-level monopole data of 
PTC at $k_{\rm max}^{B_0}=0.08~\hMpc$ yields 
$f_{\rm NL}^{\rm equil}=316\pm 67$. (This results is consistent with ~\cite{Cabass:2022wjy}, but their estimate was based on Nseries simulations with smaller volume
than PTChallenge.)
Contrasting this with the 
BOSS
data constraint 
$f_{\rm NL}^{\rm equil}=940\pm 600$~\cite{Cabass:2022wjy},
we can see 
that a $0.5\sigma$
theory systematic 
error on $f_{\rm NL}^{\rm equil}$ is tolerable
in the tree-level analysis
of the BOSS data. The theoretical
error, however, 
may become important
once the simulation-based
priors
are applied~\cite{Ivanov:2024hgq} (also see~\cite{Ivanov:2024xgb,Ivanov:2024dgv,Akitsu:2024lyt,Cabass:2024wob}).

Our results from 
Table~\ref{tab:tab2}
and Figure~\ref{fig:fnl}
show that the one-loop bispectrum EFT
calculation starts to become biased 
after $\kmax\approx 0.15~\hMpc$ which manifests itself in shifts of the posterior distribution
of $f_{\rm NL}^{\rm equil}$ beyond 2$\sigma$ 
for $\kmax>0.15~\hMpc$, with the theory systematic error
at $\kmax=0.17~\hMpc$ reaching $\Delta f_{\rm NL}^{\rm equil}\approx 400$,
which greatly exceeds the current CMB errorbars \citep{Planck:2019kim,Jung:2025nss}.
If one aims to improve over the CMB constraints (which itself will be hard \citep[cf.][]{Cabass:2022epm}), we will likely need to restrict to very large scales.
Also note that a slight reduction
in the bias on $f_{\rm NL}^{\rm equil}$ at $\kmax=0.20~\hMpc$ to about 200
is not a sign of the reduction
of the systematic error. The 
two-loop contributions at different 
scales may be accidentally
fitted with different values of $f_{\rm NL}^{\rm equil}$,
but the theory model is already in the regime of inconsistency after $\kmax=0.15~\hMpc$.
The $\approx 2\sigma$
tension in the recovered
values of $f_{\rm NL}^{\rm equil}$ at $\kmax=0.20~\hMpc$
and $\kmax=0.17~\hMpc$
is a clear sign of this inconsistency. At face value, 
the shift of optimal 
$f_{\rm NL}^{\rm equil}$
at $\kmax=0.20~\hMpc$
implies that the 
bispectrum likelihoods at 
$\kmax=0.20~\hMpc$
and $\kmax=0.17~\hMpc$
are in tension with one another
and hence one is not supposed to combine these likelihoods 
(i.e. increase $k_{\rm max}$)
in the 
strict statistical sense. 
We caution against the practice of 
such inconsistent combinations
of data sets because even
if one can ``tune'' the 
recovered $f_{\rm NL}^{\rm equil}$
values to be consistent with 
zero (or input simulation values) by adjusting $k_{\rm max}$, the corresponding errorbars are severely underestimated and hence cannot 
be trusted, see~\cite{Chudaykin:2024wlw}
for related discussions.

The inclusion of the one-loop bispectrum leads to noticeably tighter $f_{\rm NL}^{\rm equil}$ constraints than possible in a tree-level analysis, with the $\kmax=0.15~\hMpc$ bounds by $\approx 10
\%$
stronger than those without loop corrections if we compare with the $\kmax=0.08~\hMpc$
tree-level results ignoring the systematic error.
If we compare the one-loop 
constraints with  the 
tree-level results
at $\kmax=0.06~\hMpc$
instead,
our improvement becomes
$\approx 40\%$.
This matches the conclusion of \cite{Philcox:2022frc}, and motivates the use of the one-loop bispectrum in future data analyses (see also \cite{DAmico:2022gki}, with the aforementioned caveats).

We now comment on the 
importance of the 
mixed one-loop stochastic terms
that we have included in our 
analysis for the first time. These have a significant 
impact both in terms of the
PNG recovery
and the 
goodness of fit estimated 
by the $\chi^2$ statistic. 
Setting $B_{\rm mixed}^{\rm 1-loop~(s)}=0$ at $\kmax=0.15~\hMpc$, 
the best-fit $\chi^2$  
worsens by $\approx$ 9 units. 
This difference is not due to the 
free parameters alone, as adding $B_{\rm mixed}^{\rm 1-loop~(s)}$ with
all the new parameters fixed as $2d_2=b_2$,
$2d_{\mathcal{G}_2}=b_{\mathcal{G}_2}$,
and $2d_{\Gamma_3}=b_{\Gamma_3}$ (to 
match the Poissonian one-loop prediction)
improves the fit by $\approx 7.5$ units. 
Importantly, without 
$B_{\rm mixed}^{\rm 1-loop~(s)}$ the measured
$f^{\rm equil}_{\rm NL}$
is biased by more than
$2\sigma$ already at 
$\kmax=0.15~\hMpc$
if the mixed one-loop
stochastic terms are omitted. 

\subsection{Bispectrum Hexadecapole}\label{subsec: bisp-mults}

\noindent 
{Finally, we present results from the bispectrum multipole analyses including the hexadecapole $B_4$. 
The main reason we 
separate $B_4$
from the lower order 
multipoles 
is that $B_4$ 
requires substantial 
discreteness corrections (see Fig.\,\ref{fig:disc_weights}), which makes its analysis
more involved 
than that of $B_{0,2}$. Here we explore if
adding $B_4$ produces any significant 
improvements in constraining power.}

\begin{figure*}[htb!]
\centering
\includegraphics[width=0.99\textwidth]{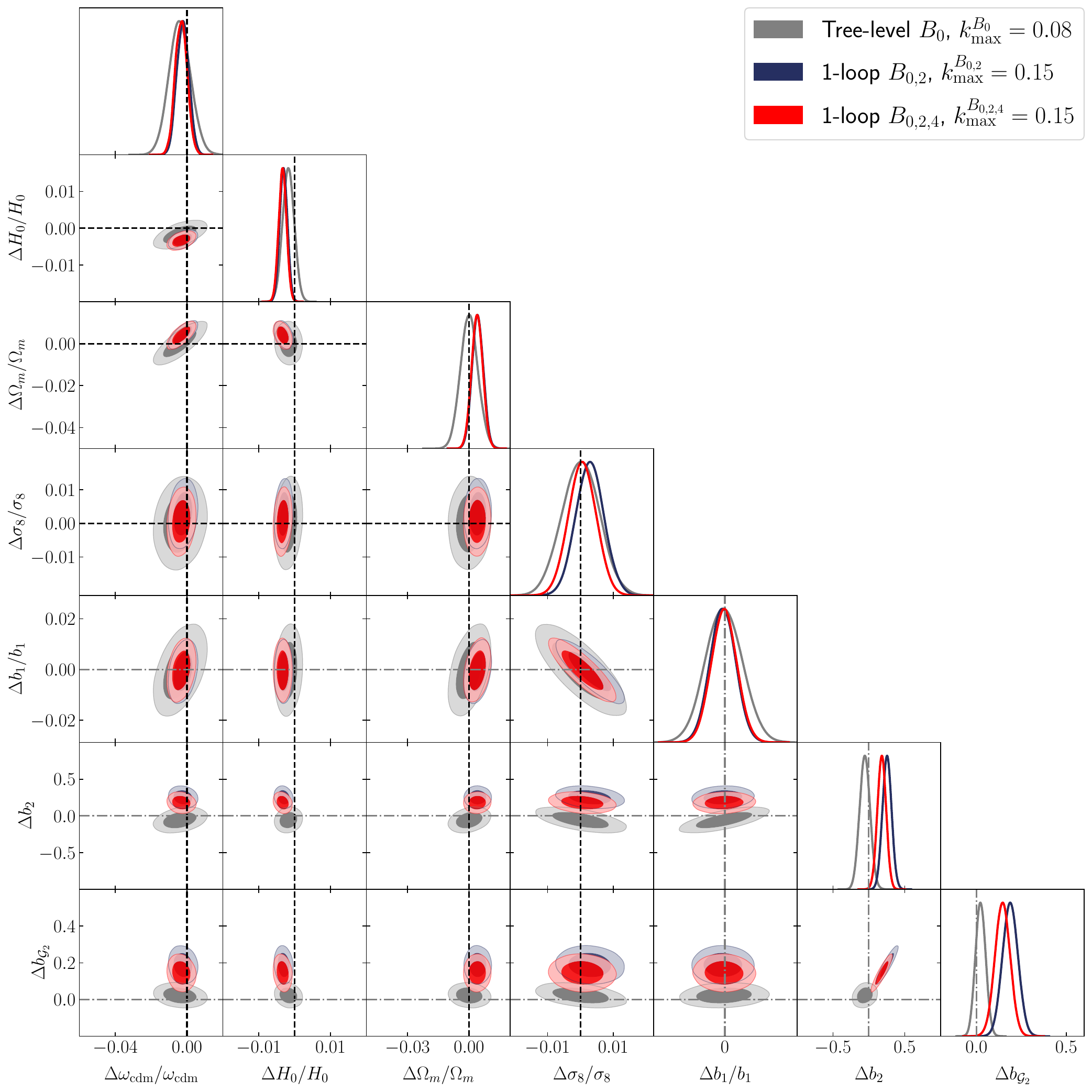}   \caption{Cosmological and EFT parameter posteriors obtained from analyzing the PTChallenge dataset using the power spectrum and one-loop bispectrum multipoles. 
All quoted $k_{\rm max}$ values in the legend are in the $~\hMpc$ units.
In contrast to Figure~\ref{fig:cosmo}, we add the bispectrum hexadecapole moments with $k_{\rm max}^{B_{4}}=0.15~\hMpc$ (red).
Results for the tree-level and one-loop
monopole plus quadrupole data vectors
with $k_{\rm max}^{B_{0,2}}=0.15~\hMpc$ are presented for comparison. Cosmological constraints do not tighten considerably as a result of adding 
the bispectrum hexadecapole. 
    } \label{fig:multi}
\end{figure*}

{
In our main analysis 
we fix $k_{\rm max}^{B_{4}}=0.15~\hMpc$ and $k_{\rm max}^{B_{0,2}}=0.15~\hMpc$, keeping the same power spectrum scale-cuts as before. This yields the results shown in Figure \ref{fig:multi} and Table \ref{tab:tab3}. 
For comparison, we also show 
the results for the tree-level 
bispectrum monopole analysis at $k_\text{max}^{B_0}=0.08~\hMpc$ (following \citep{Ivanov:2021kcd}), as well as our baseline
one-loop 
monopole-plus-quadrupole analysis discussed in Section \ref{subsec: monopole}. When adding the bispectrum hexadecapole, we find consistent results for all the parameters.
Importantly, for our baseline scale cut, we find only marginal  tightening of the errorbars:
$\Omega_m$, $\sigma_8$, 
and $H_0$ improve by $3.6\%$,
$2.3\%$, and $0\%$, respectively.
These marginal improvements are similar to those coming 
from the bispectrum multipoles analyzed at the tree level~\cite{Ivanov:2023qzb}.}

\begin{table*}[htb!]
 \begin{tabular}{l|c|c|cc}
\multicolumn{5}{c}{PTC $P_\ell+B^{\rm 1-loop}_\ell$, $k_{\rm max}^{B_{0,2,4}}=0.15~\hMpc$} \\
  \hline
& best-fit & mean$\pm\sigma$ & \multicolumn{2}{c}{95\% lower / upper} \\ \hline
$10^2\Delta \omega_{\rm cdm}/\omega_{\rm cdm}$ & $-0.31$ & $-0.32\pm 0.33$ & $-0.97$ & $0.34$ \\
$10^2\Delta H_0/H_0$ & $-0.35$ & $-0.34\pm 0.11$ & $-0.55$ & $-0.12$ \\
$10^2\Delta \sigma_8/\sigma_8$ & $0.02$ & $0.06\pm 0.42$ & $-0.77$ & $0.88$ \\
$10^2\Delta \Omega_m/\Omega_m$ & $0.44$ & $0.40\pm 0.27$ & $-0.14$ & $0.94$ \\
$10^2\Delta b_1/b_1$ & $0.02$ & $-0.03\pm 0.52$ & $-1.04$ & $0.99$ \\
$10^2\Delta b_{2}$ & $17$ & $18\pm 6.1$ & $6.0$ & $30$ \\
$10^2\Delta b_{\mathcal{G}_2}$ & $14$ & $14.5\pm 4.2$ & $6.2$ & $23$ \\
$ b_{\Gamma_3}$ &$0.41$ & $0.40\pm 0.16$ & $0.10$ & $0.71$ 
 \end{tabular}
 \caption{Cosmological and EFT parameters obtained from the PTChallenge power spectrum and bispectrum multipoles. These results can be compared to the monopole+quadrupole-only constraints given in Table~\ref{tab:tab1}. The inclusion of the hexadecapole 
 $B_4$ does not 
 substantially improve constraints on the cosmological parameters.}
 \label{tab:tab3}
 \end{table*}

{
All in all, our results show that the \cobra method
works well for the one-loop bispectrum hexadecapole up to $k^{B_{4}}_{\rm max}=0.15~\hMpc$, 
but the inclusion of this moment
does not lead to an appreciable 
improvement of constraining power. 
Given that the theoretical modeling of 
$B_4$ requires substantial discreteness weights, whose computation is quite costly, 
our results suggest that 
an analysis based on 
the first two 
moments ($B_0,B_2$) is an 
optimal way to extract the 
cosmological information
from  the bispectrum
multipole data. 
We note however, that this conclusion is strictly speaking 
true only in $\Lambda$CDM.
The $B_4$
data may be important for 
tests of hypothetical 
new physics models that produce 
a significant signal 
in the anisotropic 
part of the bispectrum.}

\section{Conclusions}\label{sec:concl}
\noindent In this work, we have applied the \cobra formalism introduced in \citep{Bakx:2024zgu} to the one-loop bispectrum and demonstrated its use in galaxy clustering analyses. In particular, we have demonstrated how the efficient numerical basis allows the one-loop bispectrum integrals to be expressed as low-dimensional tensor products that can be evaluated in just $\mathcal{O}(1)$ second with negligible approximation error ($\lesssim 0.01\%$ for the monopole, cf. Figure \ref{fig: cobra-bispectrum-comparison}). 
We supplemented this improved numerical treatment with significant developments on the theory side, deriving new mixed one-loop stochastic contributions in real-space (cf. \eqref{eq:mixed1loopstoch_r}) and in redshift space (cf. \eqref{eq:mixed1loopstoch_s}). 
Furthermore, we have applied the model to the high-fidelity PTChallenge simulation data, demonstrating that \cobra provides highly accurate predictions for the redshift-space 
bispectrum on mildly non-linear scales, which lead
to an unbiased recovery of $\Lambda$CDM cosmological
parameters with precision 
adequate even for the next 
generation of galaxy surveys. 

For luminous red galaxies at $z=0.61$ (appropriate for BOSS- and DESI-like samples), the addition of the one-loop bispectrum  monopole and quadrupole likelihood 
at $\kmax=0.15~\hMpc$ leads to significantly sharper cosmological parameter constraints than those possible with the analogous tree-level  monopole likelihood. In particular, the bounds on $\Omega_m$, $H_0$,
and $\sigma_8$ tighten by {$35\%$, $32\%$ and $25\%$},
respectively, with those on equilateral primordial non-Gaussianity improving by $\approx 10\%$. We caution that these results are specific to the redshift and galaxy sample analyzed, and other choices may require different scale-cuts and find different parameter
improvements. In particular, one may expect an increased information yield for emission line galaxies and Lyman-alpha emitters, 
which exhibit weaker non-linear
redshift-space distortion
signatures (`fingers-of-God'~\cite{Jackson:2008yv}) 
and are thus more perturbative than the luminous red galaxies studied herein~\cite{Ivanov:2021zmi,Ivanov:2024dgv,Sullivan:2025eei}. 
In addition, the constraining 
power of the one-loop bispectrum is expected to increase
in analyses of beyond-$\Lambda$CDM models \cite{spaar,braganca,Tsedrik:2022cri,Creminelli:2013nua,Lewandowski:2019txi,Crisostomi:2019vhj}.
We leave a detailed exploration
of \cobra-based one-loop
bispectra in the context of other 
galaxy samples and 
cosmological models 
for future work, see~\cite{Chudaykin:2025vdh,Ivanov:2026dvl} for applications
to DESI~DR1.

This work has also examined the impact of the higher-order bispectrum hexadecapole $B_{4}$ at the one-loop level. 
Unlike $B_0$ and $B_2$
the bispectrum hexadecapole
requires substantial 
discreteness corrections 
and carries very little 
signal, which raises questions about its inclusion
in the analysis. 
As demonstrated in Figure \ref{fig:multi}, this moment does not source improved cosmological parameter constraints. These results 
suggest that the bulk of the cosmological information
in the redshift-space multipole
moments in captured by the 
monopole and quadrupole moments
only.

We conclude that modeling the bispectrum beyond the tree-level approximation harbors significant potential improvements on cosmological parameters and primordial non-Gaussianity. Whilst computing the one-loop bispectra is expensive using conventional methods \citep[cf.,][]{Philcox:2022frc}, we have demonstrated that it is perfectly feasible using the fast \cobra scheme. 
The above conclusions clearly motivate several other developments besides the theoretical systematics we have addressed here. To wit, efficient convolution with the window function is more complicated than for the power spectrum and has been a topic of recent investigation \cite{Sugiyama:2018yzo,Pardede:2022udo,wang_window}, though other works have suggested using `window-free' estimators instead, or computing the window numerically \cite{Philcox:2020vbm,Philcox:2021ukg,Philcox:2024rqr}. The latter technique has been successfully applied recently to DESI data in \cite{chudaykin_desi}. Moreover, it would be useful to conduct a comparison of $N$-body codes along the lines of \cite{grove_comp} to assess the possible systematics associated to measuring the redshift-space bispectrum multipoles in $N$-body simulations. These endeavors are key to extracting all possible cosmological information from the large-scale structure of the Universe. 

\acknowledgements
\noindent
{\footnotesize
We thank Takahiro Nishimichi for providing the PTChallenge bispectra, as well as Guido D'Amico and Matthew Lewandowski for useful discussions. OHEP was a Junior Fellow of the Simons Society of Fellows. This publication is part of the project `A rising tide: Galaxy intrinsic alignments as a new probe of cosmology and galaxy evolution' (with project number VI.Vidi.203.011) of the Talent programme Vidi which is (partly) financed by the Dutch Research Council (NWO). For the purpose of open access, a CC BY public copyright license is applied to any Author Accepted Manuscript version arising from this submission. ZV acknowledges the support of the Kavli Foundation. Part of the computations in this work were run at facilities supported by the Scientific Computing Core at the Flatiron Institute, a division of the Simons Foundation. Other computations were performed on the Sherlock cluster. We would like to thank Stanford University and Stanford Research Computing for providing computational resources and support that contributed to these research results.}

\appendix 

\section{Alternative treatment 
of the dependence on the Hubble constant}
\label{app:APtest}

\begin{figure*}[htb!]
\centering
\includegraphics[width=0.49\textwidth]{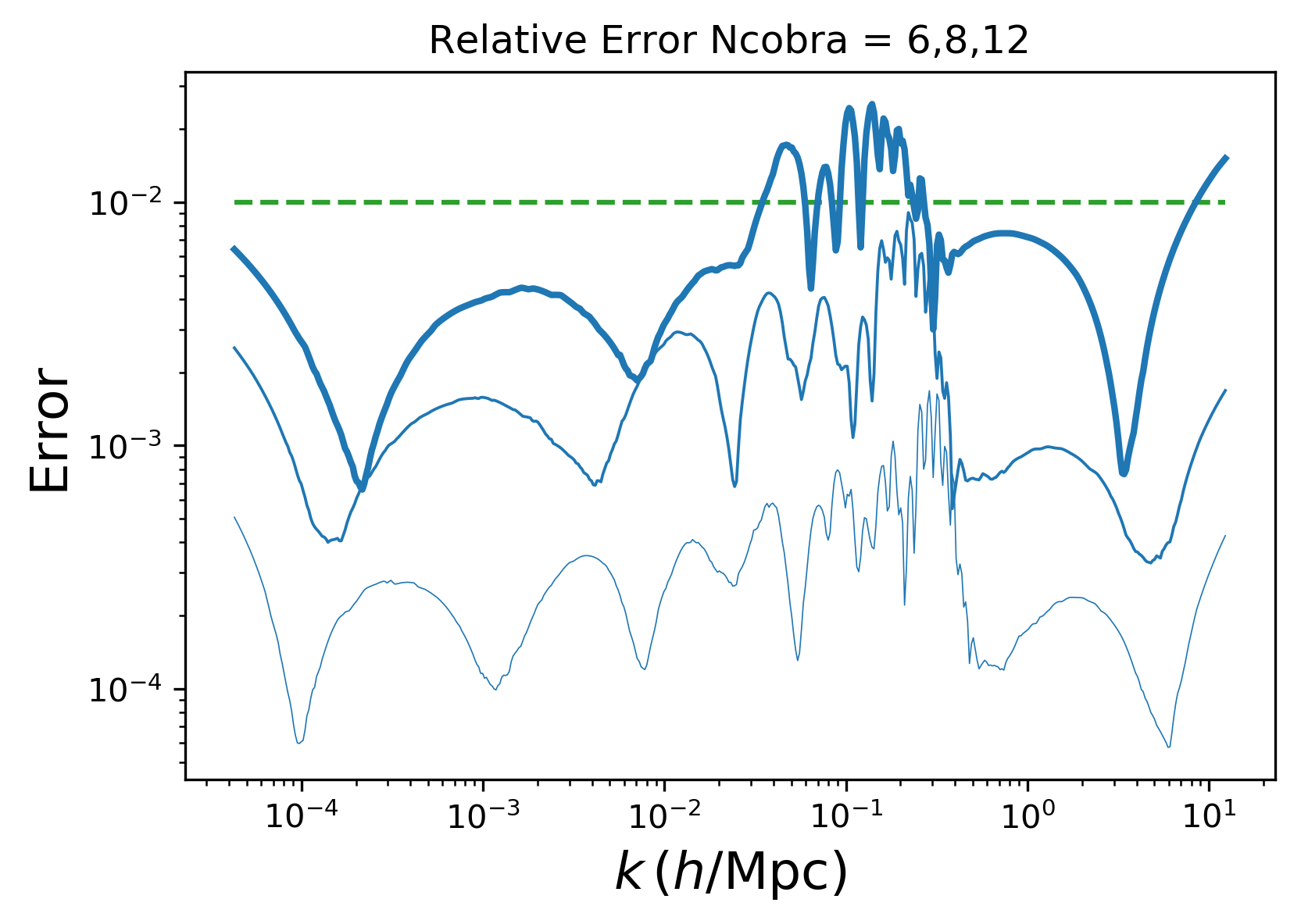}  
\includegraphics[width=0.49\textwidth]{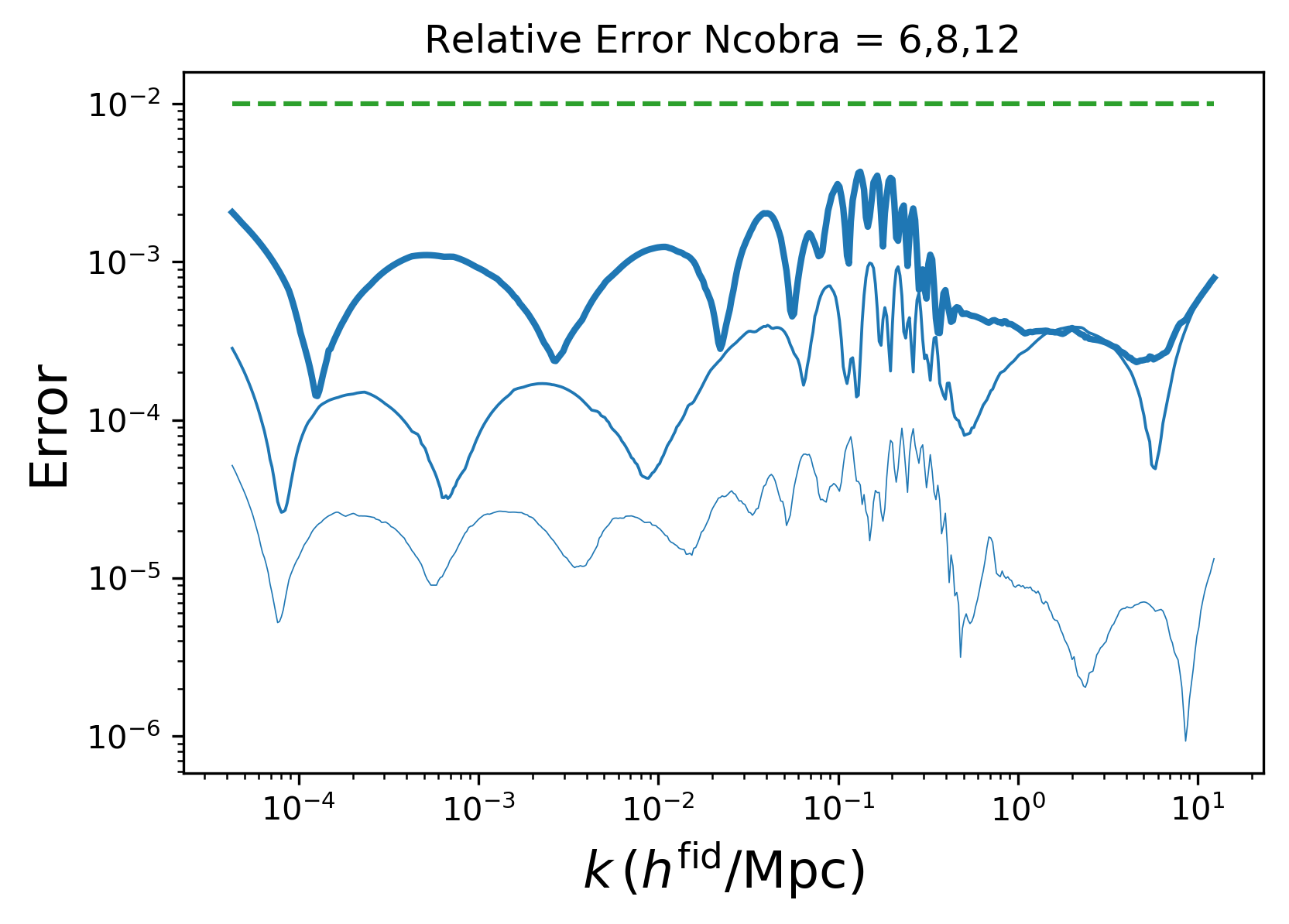}  
\caption{{Accuracy of the linear power spectrum reconstruction with COBRA: 
using the basis where the templates are computed in units 
of $h$/Mpc, where $h$ is explicitly sampled (left panel), 
and the templates computed in  units of 
$h^{\rm fid}$/Mpc (right panel). The former requires more SVD modes for $1\%$
recovery of the PTChallenge linear matter power spectrum, resulting in a more costly analysis.}} \label{fig:varyingh}
\end{figure*}

\noindent 
In this Appendix we present results of the alternative treatment of the dependence on $h$ 
in \cobra. We prepare the template bank by sampling cosmological
parameters in ranges specified in 
Sec~\ref{sec:cobra}
as well as the reduced Hubble constant
$h\in [0.55,0.8]$.
All other settings remain the same
as in our baseline 
analysis. The accuracy 
of the linear power spectrum
reconstruction with the new and the old template banks
are shown in Fig.~\ref{fig:varyingh}.
Due to the additional parameter, the explicit $h$-sampling method requires more SVD modes for a $1\%$ accuracy reconstruction
of the true linear matter power spectrum of the PTChallenge simulation, making this approach considerably lossier than working in $h^{\rm fid}/\mathrm{Mpc}$ units and accounting for the $h$-dependence via the AP rescaling. Nevertheless, since the one-loop bispectrum represents a small correction to the total bispectrum amplitude,
one may expect that 
even $\mathcal{O}(1\%)$ accuracy 
on the linear power spectrum should be sufficient for accurate bispectrum predictions.
Moreover, explicitly including the $h$-dependence 
removes any uncertainty
potentially present in our perturbative treatment of the AP effects, bracketing the accuracy of the $h$ inference with \cobra.

\begin{figure*}[htb!]
\centering
\includegraphics[width=0.99\textwidth]{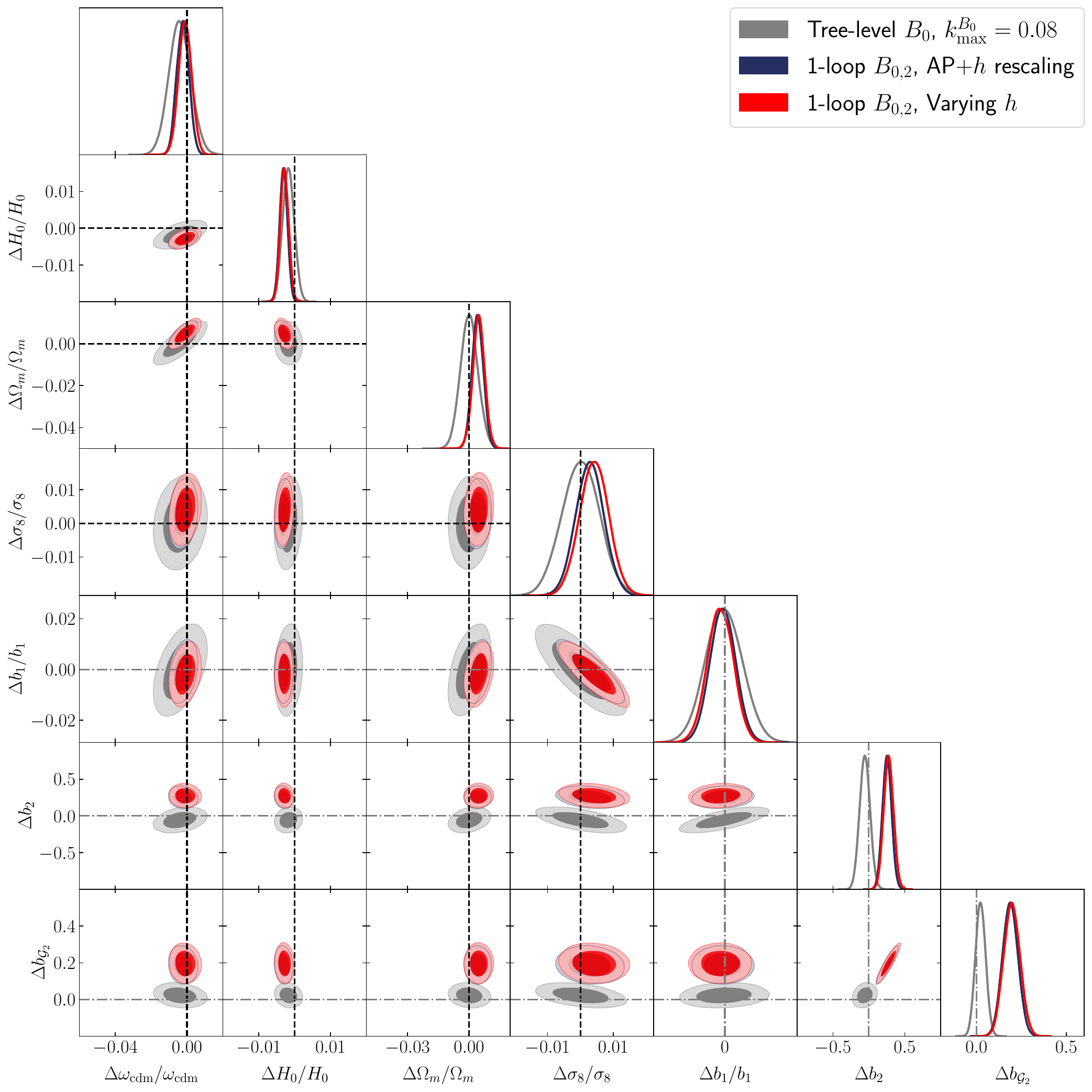}   \caption{{Cosmological and EFT parameter posteriors obtained from the PTChallenge dataset using the power spectrum and one-loop bispectrum multipoles for two different 
treatments of the dependence on 
$h$. We display results for our 
baseline analysis, obtained
by absorbing the $h$-dependence
into the Alcock-Pacynski 
rescaling  (AP+$h$ method, blue contours),
and by utilizing a \cobra basis
with an explicit $h$-dependence (Varying $h$ method, red contours). 
The quoted $k_{\rm max}$ value in the legend is in the $~\hMpc$ units.
In addition, we display the benchmark tree-level monopole results. Both methods give virtually identical results validataing our approach.}
    } \label{fig:cosmo_varyingh}
\end{figure*}

Using the same methods as before, we compute the one-loop bispectrum in the \cobra basis, explicitly varying $h$ (setting $N_{\rm cobra}=6$). We repeat the cosmological parameter inference from the PTChallenge power spectrum and bispectrum using the same settings as our baseline analysis, with results displayed in Fig.~\ref{fig:cosmo_varyingh}. We see that the results are virtually identical to those of the baseline study, 
with the varying $h$ method giving $\sim 1\%$ larger errors, which we interpret as a result of the more lossy compression scheme.
We conclude that our baseline results are stable with respect to the treatment
of $h$ in the one-loop bispectrum templates.

\bibliographystyle{apsrev4-1}
\bibliography{short}

\end{document}